\documentclass[rmp,aps,nofootinbib,twocolumn,10pt]{revtex4-1}
\usepackage[T1]{fontenc}

\usepackage{graphics}
\usepackage{epsfig}
\usepackage{amssymb}
\usepackage{graphicx}
\usepackage{graphics}
\usepackage[english]{babel}
\usepackage{amsmath}        
\usepackage{amsfonts}       
\usepackage{amsbsy,amssymb}     
\usepackage{xcolor}
\definecolor{red}{rgb}{0.7,0,0}
\definecolor{green}{rgb}{0.,0.35,0.}
\definecolor{blue}{rgb}{0.2,0.2,0.7}
\usepackage{times}
\usepackage{latexsym}
\usepackage{fancyhdr}
\usepackage{verbatim}
\usepackage{epsf}
\usepackage{subfigure}
\usepackage{bm}
\usepackage{epsfig}
\usepackage{psfrag}
\usepackage{dcolumn}
\usepackage{makeidx}
\usepackage[T1]{fontenc}

\def\inbar{\,\vrule height1.5ex width.4pt depth0pt}
\def\IR{\relax{\rm I\kern-.18em R}}
\def\IC{\relax\hbox{$\inbar\kern-.3em{\rm C}$}}

\newcommand{\bs}[1]{\boldsymbol{#1}}

\newcommand{\ket}[1]{\left|#1\right\rangle}
\newcommand{\bra}[1]{\left\langle#1\right|}
\newcommand{\braket}[2]{\bigl\langle#1\bigl|\bigr.#2\bigr\rangle}

\newcommand{\pa}{\partial}
\newcommand{\up}{\uparrow}
\newcommand{\dw}{\downarrow}
\newcommand{\pd}{{\phantom{\dagger}}}

\def\ie{\emph{i.e.},\ }
\def\eg{\emph{e.g.}\ }
\def\ea{\emph{et al.}}

\definecolor{red}{rgb}{0.7,0,0}
\definecolor{green}{rgb}{0.,0.35,0.}
\definecolor{blue}{rgb}{0.2,0.2,0.7}

\newcommand{\beq}{\begin{equation}}
\newcommand{\eeq}{\end{equation}}

\newcommand{\be}{\begin{equation}}
\newcommand{\ee}{\end{equation}}
\newcommand{\bea}{\begin{eqnarray}}
\newcommand{\eea}{\end{eqnarray}}

\newcommand{\gdD}[1]{g}
\newcommand{\UdD}[1]{U}

\newcommand{\ba}{{\bf a}}

\def\ee{\mathord{\rm e}}

\def\bs#1{\boldsymbol{#1}}

\def\be{\begin{equation}}
\def\ba{\begin{align}}
\def\enda{\end{align}}
\def\bi{\begin{itemize}}
\def\ei{\end{itemize}}

\def\Rb87{^{87}\rm{Rb}}                 
\def\Li6{^{6}\rm{Li}}                   
 \usepackage{color}
\usepackage[normalem]{ulem} 

\definecolor{color_dl}{rgb}{1 0 0} 

\begin{document}

\title{Interacting topological insulators: a review}
\author{Stephan Rachel}
\affiliation{School of Physics, University of Melbourne, Parkville, VIC 3010, Australia}

\date{\today}

\begin{abstract}
The discovery of the quantum spin Hall effect and topological insulators more than a decade ago has revolutionized modern condensed matter physics. Today, the field of topological states of matter is one of the most active and fruitful research areas for both experimentalists and theorists. The physics of topological insulators is typically well described by band theory and systems of non-interacting fermions.
In contrast, 
several of the most fascinating effects in condensed matter physics merely exist due to electron-electron interactions, examples include unconventional superconductivity, the Kondo effect, and the Mott--Hubbard transition.

The aim of this review article is to give an overview of the manifold directions which emerge when topological bandstructures and correlation physics interfere and compete. These include the study of the stability of topological bandstructures and correlated topological insulators. Interaction-induced topological phases such as the topological Kondo insulator provide another exciting topic. More exotic states of matter such as topological Mott insulator and fractional Chern insulators only exist due to the interplay of topology and strong interactions and do not have any bandstructure analogue. Eventually the relation between topological bandstructures and frustrated quantum magnetism in certain transition metal oxides is emphasized.
\end{abstract}
\maketitle
\tableofcontents

\section{Introduction}\label{sec:intro}

Condensed matter physics is the broad field of physics research which is devoted to understand solids and liquids. In the last century, condensed matter physics was a success story for both theorists and experimentalists 
due to the Landau--Ginzburg paradigm\,\cite{ginzburg-50zetf1064} and the concept of spontaneous symmetry breaking. Most aspects of quantum magnetism, superconductivity, and superfluidity can be well understood in this context, just to mention a few. 
All these examples have in common that at high temperatures the physical system is in a {\it disordered} phase. Upon decreasing the temperature, the system acquires {\it order} due to spontaneous symmetry breaking below a critical temperature $T_c$. For instance, the spins of localized electrons in a magnet will randomly point in arbitrary directions at high temperatures, but are aligned along a certain direction at low temperatures leading to a finite magnetization acting as the local order parameter.
Despite the fact that other important paradigms such as Anderson localization\,\cite{anderson58pr1492} were available, physics associated with spontaneous symmetry breaking dominated the condensed matter research. 

A mentionable exception is provided by the Berezinsky--Kosterlitz--Thouless transition\,\cite{berezinskii72spjetp610,kosterlitz-73jpc1181} in two spatial dimensions: below a critical temperature 
the system does not order as expected but forms a quasi-ordered state with vortex-antivortex pairs instead. This exception is particular worth mentioning since vortices themselves are regarded as topologically stable objects. The different phases of $^3$He\,\cite{vollhardt-90} provide another exception, and that line of research partially anticipated the rich field of topological Weyl semimetals\,\cite{bevan-97n689,Volovik03,wan-11prb205101,xu-15s613} (which is, however, not the focus of this review). 
A third exception is the Su--Schrieffer--Heger (SSH) model intended to describe solitons in polyacetylene\,\cite{su-79prl1698}. It was acknowledged early that these solitons are ``topological objects'' and the SSH model is an ideal test ground for {\it fractionalization}, \ie that a particle carries parts of an elementary quantum number\,\cite{jackiw-76prd3398,laughlin99rmp863}. What was not recognized is that the SSH model realizes a one-dimensional (1D) topological insulator characterized by non-local invariants. The simple reason is that the aforementioned concepts were not available at the time.

Certainly the most striking event in the context of ``non-Landau--Ginzburg physics'' was the discovery of the integer quantum Hall effect (QHE) featuring the exact quantization of the Hall conductivity (1980)\,\cite{klitzing-80prl494}.
In the quantum Hall experiments, a two-dimensional electron gas as realized in gallium arsenide heterostructures is placed in a strong perpendicular magnetic field. While Drude theory predicts a linear dependence of the transverse resistivity on the magnetic field strength, at sufficiently low temperatures and strong magnetic fields plateaus were observed instead\,\cite{klitzing-80prl494}. This result is even more striking when it is expressed in the inverse quantity, the transverse (Hall) conductivity, and in units of $e^2/h$: the plateaus correspond to integer numbers with an unprecedented accuracy. Thouless and coworkers showed that this integer number is in fact a topological invariant\,\cite{thouless-82prl405}, the first Chern number, which also correspond to the number of chiral edge states being responsible for the transport in this otherwise insulating system. At the same time, various experimental groups performed QHE experiments which eventually led to the discovery of the {\it fractional} quantum Hall effect (FQHE)\,\cite{tsui-82prl1559} featuring a subtile interplay of non-trivial topology of the Landau levels, disorder, and electron-electron interactions. The elementary excitations, quasi-holes, carry charge $e/3$ (for the first observed plateau at $\nu=1/3$\,\cite{tsui-82prl1559}) as demonstrated in the famous shot noise experiments more than 20 years ago\,\cite{de-picciotto-97n162,saminadayar-97prl2526}. The FQHE with its quasi-hole excitations provides one of the most impressive examples of electron fractionalization\,\cite{laughlin83prl1395,arovas-84prl722} in nature. 
Today, we also consider the FQHE as the only widely accepted realization of a system with {\it intrinsic topological order} (see Sec.\,\ref{sec:SPTvsTO}). 

It was more than 20 years later, when C.\ Kane and E.\ Mele\,\cite{kane-05prl146802,kane-05prl226801} and independently B.\ A.\ Bernevig and S.-C.\ Zhang\,\cite{bernevig-06prl106802} predicted a new state of matter (2005) dubbed {\it quantum spin Hall} (QSH) effect which has revolutionized
the  field of topological phases and made it to one of the main research directions of modern condensed matter physics: topological condensed matter physics. QSH insulators are spinful extensions of integer QHE systems and protected by time-reversal (TR) symmetry. First proposed to be realized in two-dimensional (2D) graphene monolayers\,\cite{kane-05prl226801} and strained zinc blende semiconductors\,\cite{bernevig-06prl106802}, it was shortly after generalized to three spatial dimensions\,\cite{roy09prb195322,moore-07prb121306,fu-07prl106803}, today known as ``Topological Insulators''. Due to the vanishingly small amount of spin-orbit coupling present in graphene\,\cite{min-06prb165310,yao-07prb041401}, the QSH effect was eventually observed in HgTe/CdTe heterostructures for the first time\,\cite{bernevig-06s1757,koenig-07s766}. Unlike the QHEs, topological insulators do not require any magnetic field. Instead the intrinsic spin-orbit coupling leads to the non-trivial band topology. Moreover, instead of having chiral edge modes, topological insulators exhibit spin-filtered counterpropagating pairs of chiral edge modes, so-called {\it helical} edge modes. These are only two out of various reasons why topological insulators are believed to have a bright future regarding applications similar to spintronics (sometimes referred to as ``topotronics''). 

Both the integer QHE as well as topological insulators can be described as free fermion models (in contrast to the FQHE). This raises the question to what extent are such topological band structures
stable with respect to electron-electron interactions? And if they are not stable, what happens otherwise?
As we well know  Coulomb interactions are always present in solids, sometimes screened and weak, sometimes strong and causing new phenomena such as the metal-to-insulator transition\,\cite{mott68rmp677} or the Kondo effect\,\cite{kondo64ptp37}. 

In addition to condensed matter systems, tremendous progress has been made in the past years in realizing topological matter using quantum simulators\,\cite{ozawa-18arXiv1802.04173,cooper-18arXiv1803.00249,goldman-14rpp126401,lu-16np626}. Most notably, ultra-cold quantum gases also allow to tune interactions\,\cite{goldman-14rpp126401,cooper-18arXiv1803.00249}, thus providing an alternative and controlled approach to investigate interacting topological insulators.

%
%
\subsection{Outline}\label{sec:outline}

This review aims to shed some light on the various aspects of the interplay of topological band structures and (strong) electron-electron interactions, simply referred to as {\it Interacting Topological Insulators}. In the literature, there are already several excellent review articles including the general reviews about topological insulators\,\cite{qi-11rmp1057,hasan-10rmp3045} as well as the insightful
textbook {\it Topological Insulators and Topological Superconductors}\,\cite{bernevig13}. Several slightly more specialised review articles about TIs are available\,\cite{maciejko-11arcmp31,zhang-13pss72,hasan-11arcmp55,wehling-14ap1} as well as reviews about topological materials\,\cite{yan-12rpp096501,ren-16rpp066501}, about correlation effects in topological bandstructures\,\cite{hohenadler-13jpcm143201}, about the edge physics of 2D topological insulators\,\cite{dolcetto-16rnc113}, and about fractional Chern insulators\,\cite{bergholtz-13ijmpb1330017,neupert15ps014005} as well as reviews about spin-orbit coupling in quantum gases covering the development of topological phases in cold atoms\,\cite{galitski-13n49,goldman-14rpp126401,cooper-18arXiv1803.00249}.
Also G.\ E.\ Volovik's textbook addresses many aspects of to\-po\-lo\-gy which are directly relevant for
the physics reviewed in this article\,\cite{Volovik03} although it was written before the discovery of topological insulators.
There are several points which might come to the reader's mind when thinking about {\it topology} and {\it interactions}. Some of them are discussed in the following.

In Section \ref{sec:TI} the most important topological insulator (TI) models, Chern insulators and $\mathbb{Z}_2$ TIs, are reviewed. 
A short discussion of the effective theory and the spin texture of helical edge states of $\mathbb{Z}_2$ TIs is provided and the influence of a broken axial spin symmetry emphasized. The section is completed by briefly reviewing and comparing the concepts {\it symmetry protected topological order} and {\it intrinsic topological order}.
In Section \ref{sec:corr-TI}, various paradigmatic models for interacting TIs at half filling are discussed: 
examples of 1D correlated TIs (Sec.\,\ref{sec:SSHH}),
the Haldane-Hubbard model as a prototype of an interacting Chern insulator (Sec.\,\ref{sec:corr-CI}), the Kane--Mele--Hubbard model as a prototype of a correlated $\mathbb{Z}_2$ TI (Sec.\,\ref{sec:KMH}), its bond-dependent, multi-directional generalization dubbed sodium--iridate--Hubbard model (Sec.\,\ref{sec:SIH}), the Bernevig--Hughes--Zhang--Hubbard model (Sec.\,\ref{sec:BHZH}),  the TR invariant Hofstadter--Hubbard model (Sec.\,\ref{sec:TRI-HH}) and other correlated topological insulators in higher dimensions (Sec.\,\ref{sec:3D-corrTI}).  For all of these models the stability towards electron correlations and, eventually, the breakdown of the TI phases into conventionally ordered or sometimes even into other exotic phases are con\-sidered. 
A completely different perspective is discussed in Section \ref{sec:IITI} which reviews interaction-induced topological insulator phases. The idea is to find scenarios where a topologically trivial band-structure experiences a phase transition into a topological insulator phase driven by electron-electron interactions. The considered mechanisms contain mere bandstructure renormalization effects, fluctuation induced phases (Sec.\,\ref{sec:renormalize}),
and spontaneous symmetry breaking (Sec.\,\ref{sec:raghu}). Eventually, a brief discussion about the most promising class of candidate materials, topological Kondo insulators, is presented (Sec.\,\ref{sec:TKI}).
In Section \ref{sec:stronglycorr-TI} topological Mott insulators as well as fractional Chern insulators and frational topological insulators are discussed as prototype systems which merely exists due the interplay of topo\-lo\-gy and strong interactions and do not have any band structure analogue.
%
In Section \ref{sec:TIsurface} the physics of strongly interacting surface states
of a three-dimensional strong TI is reviewed. 
Eventually in Section \ref{sec:iridates} the connection between TI physics and the honeycomb iridates A$_2$IrO$_3$ (A=Na, Li)  
and other so-called Kitaev materials, which have been proposed to host various topological states of matter such as QSH insulators, strong TIs, and quantum spin liquids, is emphasized.
Section \ref{sec:conclusion} provides a summary and an outlook.

%
%
\section{Non-Interacting Topological Insulators}
\label{sec:TI}

Given the enormous interest in topological band insulators there are already several notable reviews\,\cite{zhang-13pss72,hasan-10rmp3045,bernevig13,maciejko-11arcmp31,hasan-11arcmp55,yan-12rpp096501,ren-16rpp066501}. This section about non-interacting TIs is not meant to replace them but rather give the necessary introduction to render this review self-contained.
First, we will briefly introduce the integer quantum Hall effect. Then we will review the most important band insulators with a non-trivial topology, Chern insulator and $\mathbb{Z}_2$ topological insulators. Both have in common that they can be formulated as Bloch Hamiltonians, \ie tight-binding models of free particles on a lattice.
This formulation has the benefit that the models can be easily extended to Hubbard-type models perfectly suited for the investigation of electron-electron interactions discussed in the following sections.
At the end of this section, we further elaborate on the role of ``topology''. We point out that there are different concepts of topology floating around in the literature. Moreover, we briefly review the most important characterization schemes for topological phases and give a very brief overview of candidate materials.

A topological insulator (TI) is a band insulator in the bulk featuring metallic edge or surface states which are topologically protected against elastic single-particle back-scattering\,\cite{qi-11rmp1057}.
The term ``topological insulator'' is often used or meant synonymously with ``time-reversal invariant $\mathbb{Z}_2$ TI''. We will use, however, the more relaxed definition that any band insulator featuring a ``non-trivial topology'' is a TI, and that also includes Chern and quantum Hall insulators.

\subsection{Integer Quantum Hall Effect}
\label{sec:QHE}

The integer QHE\,\cite{klitzing-80prl494} can be considered as the mother state of all topological states of matter. The QHE is realized in a two-dimensional electron gas which is subject to a perpendicular, homogeneous magnetic field. The resulting system is a bulk-insulator, being reflected in a vanishing longitudinal conductivity $\sigma^{xx}=\sigma^{yy}=0$ where $\underline{\sigma}$ is the conductivity tensor. At the same time the transverse or Hall conductivity $\sigma^{xy}$ is finite and quantized,
\begin{equation}\label{sigma_xy}
\sigma^{xy} = C \frac{e^2}{h}
\end{equation}
with $C\in\mathbb{Z}$, the first Chern number, which will be discussed below. Since the system is bulk-insulating the finite Hall conductivity must be associated with edge transport. In fact, the integer number $C$ corresponds to the number of chiral edge modes where each mode carries a unit of conductance $e^2/h$. This intuitively explains why $\sigma^{xy}$ shows a step-like behavior as a function of magnetic field: as long as a bulk gap is present $C$ is integer-valued and constant; only due to a fine-tuned magnetic field value the bulk might become metallic allowing $C$ to switch to another value which must be again integer-valued when the bulk-gap reopens.

Thouless, Kohmoto, Nightingale, and de\,Nijs (TKNN)\,\cite{thouless-82prl405} used the Kubo-formula to compute the Hall conductivity and indeed found the result \eqref{sigma_xy} with
\begin{equation}\label{chern-number}
C=\frac{1}{2\pi} \int_{\rm BZ} d\bs{k} \left( \frac{\pa A_y(\bs{k})}{\pa k_x} - \frac{\pa A_x(\bs{k})}{\pa k_y}\right)
\end{equation}
and
\begin{equation}
A_\mu = -i \sum_{{\rm occ.\,bands}\,\alpha} \bra{\alpha\bs{k}} \frac{\pa}{\pa k_\mu}\ket{\alpha\bs{k}}
\end{equation}
where ``BZ'' stands for the first Brillouin zone and ``occ.\,bands'' refers here to filled Landau levels or energy bands, respectively. The vector $\bs{A}$ is the Berry connection or Berry vector potential of the Bloch state $\ket{\alpha\bs{k}}$ and the integrand in \eqref{chern-number} is the corresponding Berry curvature $\mathcal{F}_{xy}(\bs{k})$. The expression \eqref{chern-number} is formally equivalent to the Berry  phase\,\cite{berry84prsl45} [see also \cite{zak89prl2747,niu-85prb3372}] with the difference that the latter can have a range of integration which is a smaller subset of the BZ. $C$ is the first Chern number, sometimes also called TKNN invariant. 
Efficient methods to numerically compute the Chern number have been proposed\,\cite{fukui-05jpsj1674}.
Chern numbers appear in algebraic topology and differential geometry in the context of Chern classes which are characteristic classes associated with complex vector bundles. In \eqref{chern-number} the range of integration covers the full BZ, \ie $C$ is a highly non-local object and some people find it helpful to consider the Chern number as a non-local order parameter. This provides a complete explanation of the integer QHE. Each Landau level carries a Chern number $C=1$. The magnetic field strength controls how many Landau levels are filled. As long as the magnetic field strength is such that the Fermi level is in the gap between two Landau levels, the Chern number corresponds to this number of occupied Landau levels and stays constant (explaining the quantized and perfectly flat Hall plateaus). And only at a certain fine-tuned magnetic field strength when the Fermi level hits a Landau level, the system becomes metallic (reflected in finite $\sigma^{xx}$) and looses its quantized Hall conductivity.

The second important physical effect is the fractional quantum Hall effect (FQHE)\,\cite{tsui-82prl1559,laughlin83prl1395} which, unlike the integer effect, results in a strongly correlated state of matter: 
additional plateaus can be observed which correspond to fractions of the Hall conductivity in units of $e^2/h$, 
\ie rational numbers such as 1/3, 1/5, 2/7 etc., while the system is incompressible. Since the Chern number $C$ needs to be integer-valued, Eq.\,\eqref{sigma_xy} in conjunction with \eqref{chern-number} cannot provide an explanation for the observed phenomenon\,\cite{girvinFQHE}. In order to observe FQHE one has to tune the magnetic field strength such that the Fermi level lies inside a Landau level. Instead of showing metallic behavior strong electron-electron interactions drive the system into a gapped, incompressible phase which is not adiabatically connected with the physics of the integer QHE. The reason why strong interactions can be present at all in a 2D electron gas stems from the flatness of the Landau levels. Since the Landau levels are essentially dispersionless, the kinetic energy is heavily quenched and even the smallest electron-electron interactions, which are always present, turn out to be effectively large. That is, not the interactions but the ratio ``interactions / kinetic energy'' is large. We note that the FQHE is a state of matter which exists due to a subtle interplay between the topologically non-trivial Landau levels and strong interactions. It is worth mentioning that also disorder is a necessary ingredient to provide full understanding of this effect\,\cite{girvinFQHE}. It furthermore turns out that the elementary excitations in FQHE systems are exotic quasi particles which carry only fractions of the electron charge $-e$. To give a famous example: when the magnetic field is tuned such that the transverse conductivity takes the value $\sigma^{xy}=\frac{1}{3}\frac{e^2}{h}$ then the elementary excitations are quasi-holes with charge $-\frac{e}{3}$\,\cite{laughlin83prl1395}. One should keep in mind that every FQH state will be the groundstate of a corresponding many-body Hamiltonian involving some type of Coulomb interactions\,\cite{haldane83prl605} and a single-particle description is insufficient. Later we will discuss the lattice version of FQH systems, the so-called frational Chern insulators in Sec.\,\ref{sec:fci+fti}. In contrast to the FQHE, the integer QHE is described by a free fermion theory.

\begin{figure}[t!]
\centering
\includegraphics[scale=0.53]{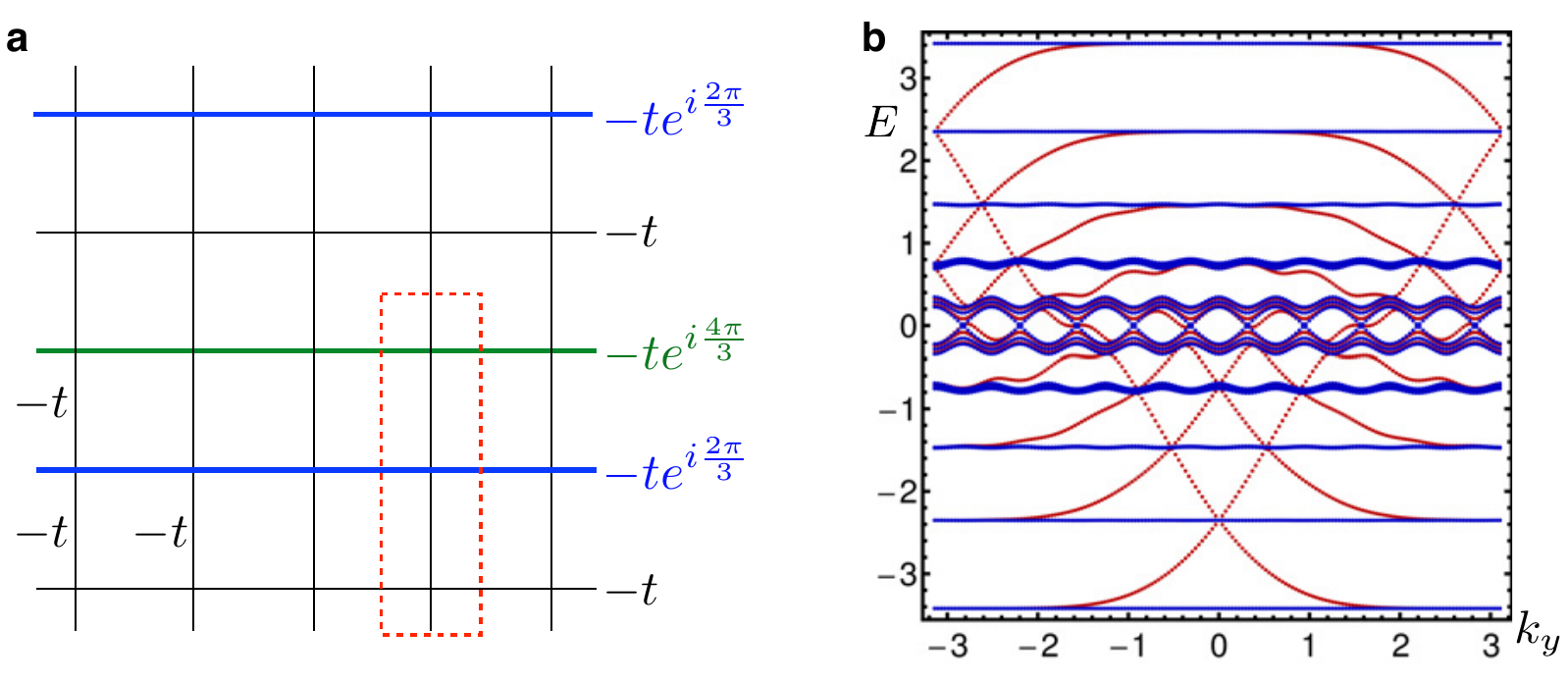}
\caption{\small The Hofstadter problem. (a) Square lattice with modified hoppings for vertical bonds according to \eqref{peierls-hopp} for a magnetic field of strength $\alpha=1/3$ in units of $\phi_0$. The red box indicates the unit cell. (b) Spectrum for $\alpha=1/10$ computed on a cylinder geometry. Bulk bands are shown in blue, chiral edge modes localized at both cylinder edges are shown in red\,\cite{orth-13jpb134004}.}
\label{fig:hofstadter}
\end{figure}

The formation of Landau levels in a 2D electron gas assumes free fermionic particles in a magnetic field with kinetic energy $T=\sum_i\frac{p_i^2}{2m}$.
Now we will address the interesting  question how non-interacting {\it lattice} fermions are affected by a strong (orbital) magnetic field. This problem was first successfully analyzed by Hofstadter in 1976\,\cite{hofstadter76prb2239}, even a few years before the discovery of the integer QHE. Here we briefly illustrate the {\it Hofstadter problem} on the square lattice which will be the starting point for the study of interaction effects in Hofstadter bands, the Hofstadter-Hubbard model\,\cite{cocks-12prl205303,goldman-10prl255302}, and fractionalization in Hofstadter bands\,\cite{moeller-15prl126401}. Spinless fermions on the square lattice are described by a hopping Hamiltonian $H=-\sum_{\langle ij \rangle} \big( t c_i^\dag c_j^\pd + {\rm h.c.}\big)$. In order to include a finite magnetic field we choose Landau gauge and modify the hopping term using Peierls substitution,
\begin{equation}\label{peierls-hopp}
t \, c_i^\dag  c_j^\pd \longrightarrow t \, c_i^\dag \exp{\left(i \frac{2\pi}{\phi_0}\oint_{\mathcal{S}} \bs{A} d\ell\right)} c_j^\pd\ ,
\end{equation}
with the Dirac flux quantum $\phi_0=hc/e$. For a vector potential of strength $\alpha=p/q$ (we do not consider irrational values here) being the number of flux quanta per lattice cell yields
\begin{equation}
\bs{A}(\bs{r}) = \frac{p}{q}\frac{\phi_0}{A_{LC}}y \hat x\ .
\end{equation}
The area of an elementary lattice cell is $A_{LC}=a^2$ for  a square lattice as considered in the following ($a$ being the lattice spacing). For a square lattice we find in the chosen gauge the hopping pattern, which is shown in Fig.\,\ref{fig:hofstadter}\,(a) for $\alpha=1/3$, a typical spectrum solved on a cylinder for $\alpha=1/10$ is shown in panel (b).

Note that for $\alpha=p/q$ Peierls phase factors $\exp{(m i 2\pi/ q)}$ appear in the $m$th row or line, respectively. When $m=q$, the phase becomes (a multiple of) $2\pi$ which corresponds to the field strength of one flux quantum $\phi_0$. The corresponding phase factor yields $+1$\footnote{That is the reason why people often say ``one flux quantum is no flux quantum''.}. It is further worth mentioning that 
translational symmetry is broken in the Hofstadter problem. This might appear to be counterintuitive since the applied magnetic field is homogeneous. One can show, however, that the applied vector potential explicitly breaks translation symmetry\,\cite{kohmoto-85ap343,bernevig13}. The system remains invariant under translation of $q$ sites along the direction of increased unit cell, see for instance the $y$ direction in  Fig.\,\ref{fig:hofstadter}\,(a).

For $\alpha=p/q$ there are $q$ bands; each of the isolated bands carries a Chern number $\Delta C=1$.
According to the TKNN formula, in order to obtain the Hall conductance or total Chern number, associated with a given energy gap, one has to sum up the Chern numbers of the occupied bands. For instance, let us consider the case $q=10$ where the lowest gap exhibits a Chern number $C=1$, the second lowest gap  $C=2$ etc.
As mentioned previously, in each gap there must be $C$ chiral edge modes.
This is illustrated in Fig.\ref{fig:hofstadter}\,(b) where a cylinder geometry has been used; thus there are even $C$ chiral edge modes {\it per edge}, which are shown in red for the case $q=10$.

Note that for any $\alpha \not= 1/2$ time-reversal symmetry is broken due to the orbital magnetic field. The Hofstadter problem is a realization of the quantum Hall effect on a lattice. When plotting the energy spectrum (without any momentum quantum numbers) vs.\ the magnetic field strength $\alpha \in [0,1]$ it shows for the square lattice a remarkable fractal structure having the shape of a moth or a butterfly\,\cite{hofstadter76prb2239}. On other lattices similar fractal structures appear.

%
%
\subsection{Chern Insulators}
\label{sec:CI}

A Chern insulator is defined as a band insulator which possesses a quantized Hall conductivity $\sigma^{xy}$ but no net-magnetic field. Alternatively, one can think of a translation invariant band insulator with  quantized Hall conductivity $\sigma^{xy}$. Behind the concept of a Chern insulator stands F.\,D.\,M.\ Haldane's insight that not the orbital magnetic field but the broken time-reversal symmetry is essential for the realization of QHE\,\cite{haldane88prl2015}. Such a QHE without orbital magnetic field or Landau levels is usually referred to as 
quantum anomalous  Hall effect (QAHE).  
Haldane proposed a simple tight-binding model on the honeycomb lattice with real nearest-neighbor hoppings and complex-valued second-neighbor hoppings\,\cite{haldane88prl2015}. This is accomplished by a staggered flux pattern for the different sublattices $A$ and $B$ such that the total flux enclosed by a hexagon, the smallest lattice cell, is zero. But the second neighbor hoppings acquire magnetic phases which leads to breaking of time-reversal symmetry and a QAHE phase with Chern numbers $C=+1$ or $-1$, respectively. Most remarkably, recent progress in ultracold quantum gases and optical lattices has led to the experimental realization of Haldane's model\,\cite{jotzu-14n237}.

\begin{figure}[h!]
\centering
\includegraphics[scale=0.63]{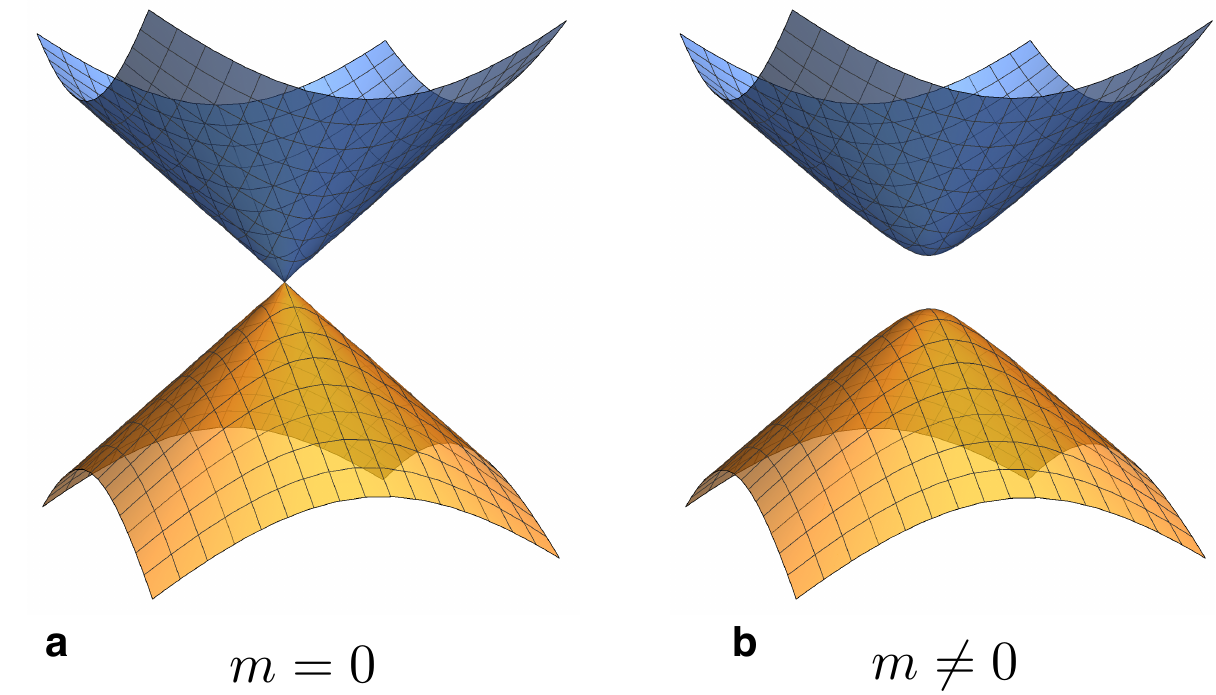}
\caption{\small Spectrum of (a) a massless ($m=0$) and (b) a massive Dirac Hamiltonian ($m\not= 0$). 
}
\label{fig:massive-dirac}
\end{figure}

To understand Haldane's seminal paper it is most instructive to consider the physics of a Dirac Hamiltonian in 2+1 dimensions. For two states $\ket{+}$ and $\ket{-}$ (which might be spin, isospin, orbital degree of freedom, etc.) a Dirac theory is governed by the Hamiltonian
\begin{equation}
\mathcal{H} = \sum_{\bs{k}} \left( c_{\bs{k}.+}^\dag ~~ c_{\bs{k},-}^\dag \right) h(k_x, k_y) \left(\begin{array}{c} c_{\bs{k},+}\\ c_{\bs{k},-} \end{array}\right)
\end{equation}
where the summation runs over $\bs{k}=(k_x, k_y)$ corresponding to momentum quantum numbers accounting for the two-dimensionality of the system. The $2\times 2$ matrix $h$ expressed in terms of Pauli matrices $\sigma^i$ is then given by
\begin{equation}\label{massless-dirac}
h(k_x, k_y) = \vec k \cdot \vec \sigma \equiv k_x \sigma^x + k_y \sigma^y
\end{equation}
leading to a linearly dispersing spectrum $\epsilon_\pm = \pm | \bs{k} |$ with band-touching at $\bs{k}=0$ at zero energy, see Fig.\,\ref{fig:massive-dirac}\,(a).

By adding a mass term to \eqref{massless-dirac}, $h_m = m \sigma^z$, the Dirac node opens a gap of size $2m$, see Fig.\,\ref{fig:massive-dirac}\,(b). Such a gapped or {\it massive} Dirac theory is always associated with a Chern number $C={\rm sgn}(m)/2$ displaying half-quantum Hall effect. The result can be readily derived\,\cite{volovik88zetf123,bernevig13}. The Berry connection reads
\begin{equation}
A_x = \frac{-k_y}{2\sqrt{k^2 + m^2}[\sqrt{k^2 + m^2} -m]}
\end{equation}
and
\begin{equation}
A_y = \frac{k_x}{2\sqrt{k^2+m^2}[\sqrt{k^2+m^2}-m]}
\end{equation}
leading to the Berry curvature
\begin{equation}
\mathcal{F}_{xy} = \frac{m}{2(m^2 + k^2)^{3/2}}
\end{equation}
which agrees with the general expression $\mathcal{F}_{xy}=\epsilon_{abc}d_a\pa_x d_b  \pa_y d_c /(2d^3)$ if the Hamiltonian is written as $h(k_x,k_y) = \vec d \cdot \vec \sigma$ with $\vec d^{\,T} = (k_x, k_y, m)$. Integrating $\mathcal{F}_{xy}$ over the infinite plane eventually leads to\,\cite{volovik88zetf123,bernevig13}
\begin{equation}\label{examples-chern}
C = \frac{1}{2\pi} \int d^2k \mathcal{F}_{xy} = \frac{m}{2} \int_0^\infty \frac{k dk}{(m^2 + k^2)^{3/2}} = 
\frac{m}{2} \frac{1}{\sqrt{m^2}}\ .
\end{equation}
This half-QHE stems from the continuum description of the Dirac fermion. On a lattice, the bands which have finite band width need to bend down at some point; these contributions of the band will then add 
another half of a Chern number leading to either a trivial (\ie total $C=0$) or a topological phase (\ie total $C=\pm 1$). Apparently the continuum is not sufficient to determine whether or not a phase is topologically non-trivial; what can be determined, however, is that the Chern number will change by $\Delta C=\pm 1$ when $m$ changes it sign from $\pm m$ to $\mp m$ through a gap-closing transition at $m=0$. The presence of this half-QHE is sometimes also referred to as a {\it meron} which is in high-energy physics defined as half an instanton.

Regularizing the massive Dirac Hamiltonian and defining it on a lattice brings down the high-energy modes which contain ``the other half of a Chern number''. Depending on the sign (\ie depending on the topology) the two halves of the Chern number cancel or add up to an integer Chern number. Apparently the realization of a topological phase of a lattice Dirac Hamiltonian depends on the model details;  nonetheless the two most important lattice realizations of the massive Dirac Hamiltonian are certainly the Haldane model\,\cite{haldane88prl2015} and the two-orbital square lattice Chern insulator\,\cite{qi-08prb195424} which corresponds to a single spin-channel of the Bernevig--Hughes--Zhang model\,\cite{bernevig-06s1757}. In the literature these models are referred to as {\it lattice Chern insulators} or simply Chern insulators.

The two-orbital square lattice model\,\cite{qi-08prb195424} is a straight-forward regularization of $h=k_x \sigma^x + k_y \sigma^y + m \sigma^z$ and reads
\begin{equation}
\label{ham:ci-squarelattice}
\mathcal{H}_{2{\rm -orb.}} = \sum_{\bs{k}} \left( c_{s,\bs{k}}^\dag~c_{p,\bs{k}}^\dag\right)
h(\bs{k})_{2{\rm -orb.}} \left(\begin{array}{c} c_{s,\bs{k}}\\ c_{p,\bs{k}}\end{array}\right)
\end{equation}
with the orbital indices $s$ and $p$ and the Bloch matrix
\begin{equation}
\label{bloch:ci-squarelattice}
h(\bs{k})_{2{\rm -orb.}} \!=\! t \sin{k_x} \sigma^x + t \sin{k_y} \sigma^y  + \big(m+\cos{k_x}+\cos{k_y}\big) \sigma^z.
\end{equation} 
%
\begin{figure}[t!]
\centering
\includegraphics[scale=0.47]{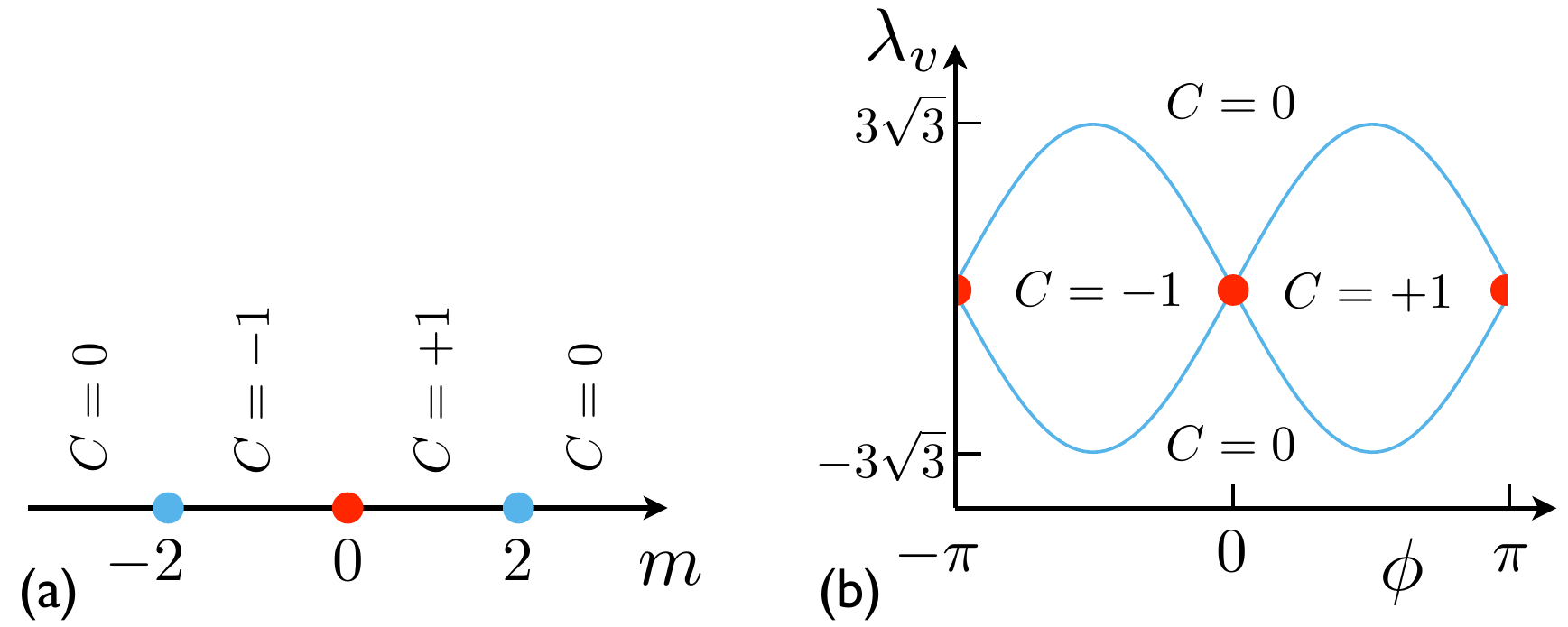}
\caption{\small Phase diagrams of the Chern insulator models on the (a) square\,\cite{qi-08prb195424} and
(b) honeycomb lattices\,\cite{haldane88prl2015}. The blue dots (lines) correspond to gap-closing points (lines)  where the Chern number changes by $|\Delta C|=1$. At the red dots, at two points in the Brillouin zone the gap closes simultaneously associated with a change of Chern number $|\Delta C|=2$.}
\label{fig:CI-phasediagrams}
\end{figure}
%
Here in this simple form there is only one free parameter $m$ (assuming $t=1$).
Possible gap-closing positions in the Brillouin zone are restricted to momenta $k_x=0, \pi$ and $k_y=0,\pi$. Assuming that the system is topologically trivial for $m\to\pm\infty$ (``atomic limit'') one readily derives the phase diagram shown in Fig.\,\ref{fig:CI-phasediagrams}\,(a). At $m=0$, two diabolic points (\ie Dirac points) appear at $\bs{k}_0=(0,\pi)$ and $(\pi,0)$ which allows the system to change the Chern number by $|\Delta C|=2$. The other gap-closing points at $m=\pm 2$ involve only a single Dirac point changing the Chern number only by $|\Delta C|=1$.

To regularize a massive Dirac Hamiltonian on a honeycomb lattice is rather simple because already a real nearest-neighbor hopping model realizes a spectrum with two band-touching points effectively described by the Dirac Hamiltonian. These band-touching points are denoted as $\bs{K}$ and $\bs{K}'$ located at the corners of the hexagonal Brillouin zone. The honeycomb lattice features a two-atomic basis where we label the different basis sites as $A$ and $B$, respectively. Adding a potential imbalance between $A$ and $B$ sites correspond to a mass term $\lambda_v \sigma^z$\,\cite{semenoff84prl2449}. It turns out, however, that the gapped phase caused by $\lambda_v \sigma^z$ is topologically trivial. Duncan Haldane pointed out in his famous paper from 1988 that the mass term leading to the Chern insulator phase must contain not only a sign-change between the sublattices $A$ and $B$ but also between the two Dirac points (or ``valleys'') $\bs{K}$ and $\bs{K}'$, $\lambda \sigma^z \tau^z$, where $\tau$ corresponds to valley degree of freedom. These conditions can be realized on a honeycomb lattice due to imaginary second-neighbor hopping breaking time-reversal  but preserving translation symmetry.

The Haldane model\,\cite{haldane88prl2015} is governed by the Hamiltonian
\begin{equation}
\label{ham:haldane}
\mathcal{H}_{{\rm H}} = t\sum_{\langle ij \rangle} c_i^\dag c_j^\pd + \lambda \sum_{\langle\!\langle ij \rangle\!\rangle} c_i^\dag e^{i\phi \nu_ij} c_j^\pd + \lambda_v \sum_i \xi_i c_i^\dag c_i^\pd\ .
\end{equation}
Here $t$ denotes the amplitude for real nearest-neighbor hopping, $\lambda$ for second neighbor hopping and $\phi$ the phase which mixes real and imaginary second neighbor hopping, $\lambda_v$ the Semenoff mass, and $\xi_i=\pm1$ on sublattice $A$ or $B$, respectively. The three parameters $\lambda$, $\lambda_v$, and $\phi$ lead to a richer phase diagram than that of the square lattice CI. 
The phase $\phi$ essentially mixes real and imaginary second-neighbor hopping; while the former breaks particle-hole symmetry but leaves the Dirac points unchanged otherwise, the latter breaks TR symmetry and opens the gap into the topological phase. In contrast, $\lambda_v$ opens the gap into a trivial phase. The interplay of $\phi$ and $\lambda_v$ leads to the phase diagram shown in Fig.\,\ref{fig:CI-phasediagrams}\,(b). 
The Bloch matrix corresponding to \eqref{ham:haldane}
contains close to the Dirac points $\bs{K}$ and $\bs{K}'$ the  mass term $d_z \sigma^z$ with
\begin{equation}
d_z = M \pm 3\sqrt{3}\lambda \sin{\phi}\ .
\end{equation}
Plus and minus signs correspond to valley $\bs{K}$ and $\bs{K}'$, respectively. In analogy to the limit $m\to\infty$ in case of the square lattice CI, here we consider the atomic limit $\lambda_v\to\infty$ where all particles are localized on one of the sublattices, clearly being topologically trivial ($C=0$). This phase remains insulating for $3\sqrt{3}\lambda\sin{\phi} < M < \infty$ and due to adiabaticity the Chern number cannot change. Further decreasing $M$ closes the gap for $M=3\sqrt{3}\lambda\sin{\phi}$ at $\bs{K}'$ in the Brillouin zone. The sign change of the mass at the Dirac point $\bs{K}'$ is again associated with a change of Chern number $|\Delta C|=1$. Upon further decreasing $M$ one passes another gap-closing for $M=-3\sqrt{3}\lambda\sin{\phi}$ at $\bs{K}$ and the Chern number changes back to $C=0$. Along the blue lines in the phase diagram Fig.\,\ref{fig:CI-phasediagrams}\,(b) only a single gapless Dirac cone is present. This is by no means a violation of the {\it fermion doubling theorem}\,\cite{nielsen-81npb173} since TR symmetry is broken for $\phi\not= 0,\pm\pi$. Only at $\phi=0$ or $\phi=\pm\pi$ (marked by the red dots in the phase diagram in Fig.\,\ref{fig:CI-phasediagrams}), TR is intact and for these parameters the upper and lower bands touch simultaneously at both $\bs{K}$ and $\bs{K}'$ points in the Brillouin zone. By analogy, we see that only for $m=0$  the square lattice Chern insulator model \eqref{ham:ci-squarelattice} can be TR invariant. For a deeper discussion about the relationship of TR, fermion doubling, and lattice Dirac fermions we refer the reader to \cite{herbut-11prb245445}.

\begin{figure}[t!]
\centering
\includegraphics[scale=0.44]{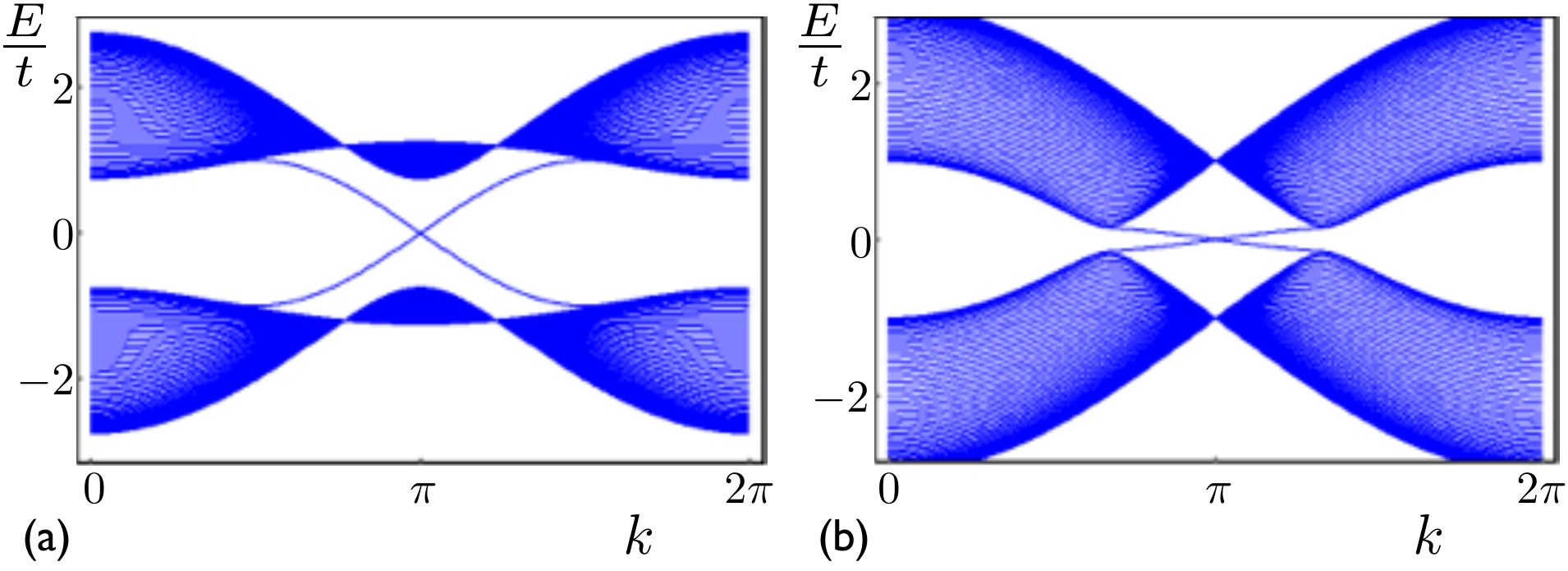}
\caption{
\small Spectra computed on a cylinder geometry (``nano-ribbon''). (a) 
Two-orbital square lattice model\,\cite{qi-08prb195424} for parameter $m=0.75\,t$.
(b) Haldane model\,\cite{haldane88prl2015} for parameters $\lambda=0.03\,t$, $\phi=\pi/2$, and $\lambda_v=0$. Both spectra correspond to the $C=+1$ phase and feature a single chiral edge mode per edge. Both spectra were computed for a cylinder containing 48 unit cells.}
\label{fig:CI-cylinderspectra}
\end{figure}

Chern insulator models were also discussed on other lattices -- mainly motivated by the search for fractional Chern insulators (see Sec.\,\ref{sec:fci+fti}) -- including the kagome lattice\,\cite{tang-11prl236802,bergholtz-13ijmpb1330017}, the triangular lattice\,\cite{venderbos-12prl126405}, the checkerboard lattice\,\cite{sun-11prl236803,neupert-11prl236804}, the dice lattice\,\cite{wang-11prb241103}, and the ruby lattice\,\cite{hu-11prb155116}. 

In the previous section, the first Chern number as a topological invariant was introduced for systems with broken time-reversal symmetry (note that the Chern number {\it must} vanish whenever time-reversal invariance is enforced). The ``quantization'' of the Chern number and its stability is guaranteed by the bulk gap: since the band above the gap carries a topological defect (\ie a monopole in its Berry curvature) and the band below the gap another one (an antimonopole) there is no way to remove these defects (this is a rephrasing of the earlier statement that the sign of the mass in the  Dirac Hamiltonian cannot change without gap-closing). Now let us consider a heterostructure consisting of an insulator with finite Chern number and another insulator with zero Chern number (\eg vacuum). At the interface the topological defects of the Chern bands (\ie the monopoles in the Berry curvature) must recombine. This is accomplished by {\it locally} closing the gap at the interface (\ie the edge of the Chern insulator). This ``local gap-closing'' corresponds to the {\it edge states} traversing the bulk gap in the energy spectrum.
Whenever a system carries a finite Chern number $C$, there must be $C$ chiral edge modes at its boundaries to trivial insulators or vacuum. That is the {\it bulk boundary correspondence}. In Fig.\,\ref{fig:CI-cylinderspectra} we show the energy spectra of both discussed CI models for parameters corresponding to the $C=+1$ phase computed on a cylinder geometry. Since a cylinder exhibits two edges the spectra feature two chiral edge modes which cross at $k=\pi$ where $k$ is the momentum quantum number due to translation invariance along the circumference of the cylinder. Analyzing the corresponding eigenvectors clearly reveals that the two edge modes are localized on opposite edges of the cylinder (not shown here). A chiral edge mode is considered to be stable against not too strong disorder. Loosely speaking the inherent chirality of the system forbids the edge mode to stop or even ``turn around''.

In principle, the quantum Hall  and Chern insulators are systems which do not require any symmetry, only TR must be broken. 
Apart from the fact the TR is a discrete symmetry, note that TR does not need to be spontaneously broken.
Moreover, a Chern insulator does not possess a {\it local} order parameter. The Chern number which is sometimes considered a ``topological order parameter'' is a highly {\it non-local} object as it involves summation of the full Brillouin zone, see Eq.\,\eqref{chern-number}. Eventually one might ask whether the U(1) symmetry associated with particle number or charge conservation is required which we refer to in the following as U(1)$_{\rm charge}$. While the quantized Hall conductivity $\sigma^{xy}$ looses its quantized value when U(1)$_{\rm charge}$ is broken\,\cite{volovik88zetf123} the Chern number does not change and the chiral edge mode does not acquire a gap. Note that the part of the Hall conductance related to the edge states is in topological superconductors generally expected to be quantized\,\cite{volovik92jetp368}. The persistence of the chiral edge mode (and thus the Chern number) in the presence of a superconducting pairing term has explicitly been shown\,\cite{rachel16jpcm405502}.

%
%
\subsection{$\mathbb{Z}_2$ Topological Insulators}
\label{sec:Z2TI}

\begin{figure}[h!]
\centering
\includegraphics[scale=0.5]{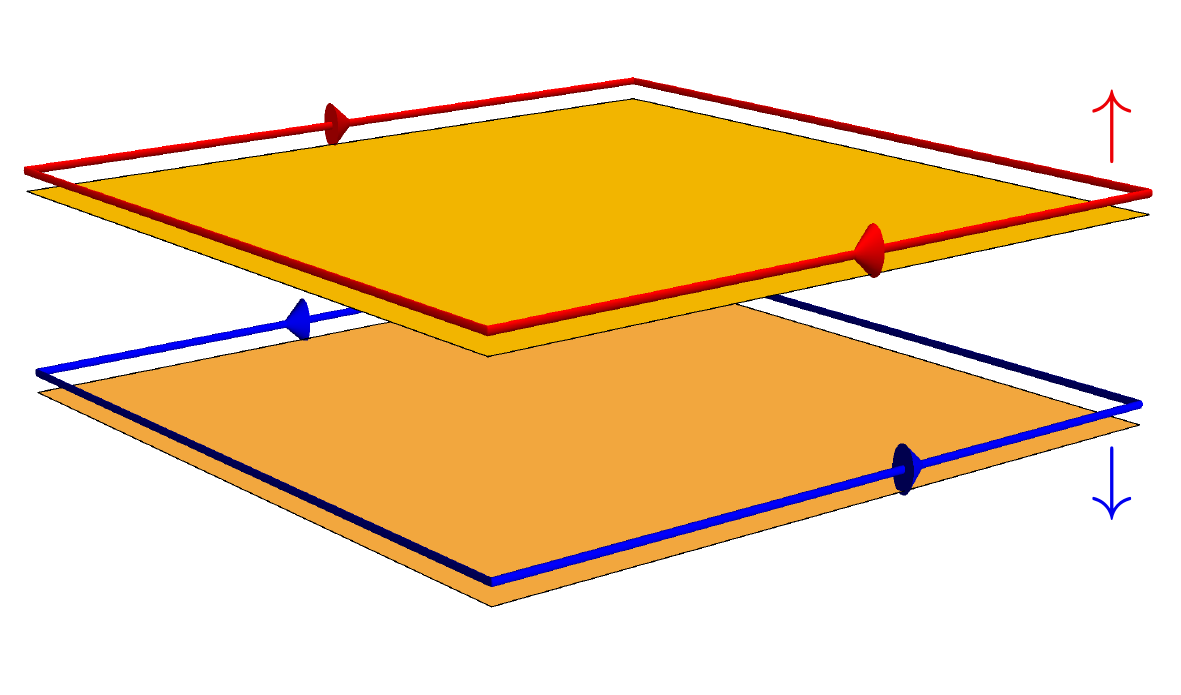}
\caption{The QSH system as a merger of two QH layers for the opposite spin components with opposite chirality.}
\label{fig:qsh-construction}
\end{figure}

In 2004 and 2005, Kane and Mele\,\cite{kane-05prl146802,kane-05prl226801} and independently Bernevig and Zhang\,\cite{bernevig-06prl106802} proposed the {\it Quantum Spin Hall} (QSH) effect in two spatial dimensions. The QSH effect is very similar to the integer QHE with the major differences that no external field is required and time-reversal symmetry is {\it not} broken. The usual way most people introduce the QSH effect is by considering Bloch electrons with spin-1/2 degree of freedom: while the $\up$-spins are described by an integer QHE with positive chirality ($C=+1$), the $\dw$-spins by an integer QHE with negative chirality ($C=-1$) which is the time-reversal conjugate copy of the $\up$-spin system.
This construction is manifest in the Bloch-matrix \eqref{bloch-QSH} of a typical QSH system: the $2\times 2$ block describing the $\up$-spin system is odd under TR, and the $2\times 2$ block of the $\dw$-spin system is its TR conjugate.
The Bloch matrix thus reads
\begin{equation}\label{bloch-QSH}
H(\bs{k})_{\rm QSH} = \left( \begin{array}{cc} {\color{red}H(\bs{k})_{\rm QHE}} & 0 \\ 0 & {\color{blue}H^\ast(-\bs{k})_{\rm QHE}} \end{array}\right)\ 
\end{equation}
where the basis is defined via the spinor $\Psi^\dag = \{ c^\dag_{\up, O_1}, c^\dag_{\up, O_2}, c^\dag_{\dw, O_1}, c^\dag_{\dw, O_2}\}$ and $O_{1/2}$ refers to different orbitals or sublattices, respectively.
This can be visualized as two quantum Hall layers with opposite chirality, see Fig.\,\ref{fig:qsh-construction}.
By construction, the resulting system is TR invariant and while the bulk still is insulating the $\up$-spins circulate, say, clockwise and the $\dw$-spins counter-clockwise around the sample edge: The QSH effect features {\it helical}, spin-filtered metallic edge modes\,\cite{murakami11jpcs012019}.

Before discussing more of the phenomenology of the QSH effect and its microscopic realizations let us take a different perspective. Following Kane and Mele, in two spatial dimensions and in the presence of time-reversal symmetry, there are precisely two different types of band insulators\,\cite{kane-05prl146802}: topologically trivial and topologically non-trivial ones, the latter displaying the QSH effect. This is a very strong and a rather unexpected result; these two phases are distinguished or classified by a $\mathbb{Z}_2$ invariant originally introduced by Kane and Mele. Note that this insight in conjunction with the $\mathbb{Z}_2$ invariant directly leads to  the finding that in three spatial dimensions there are 16 different band insulators when TR symmetry is conserved\,\cite{roy09prb195322,moore-07prb121306,fu-07prl106803}, see below. At that time, Moore and Balents also coined the name ``topological insulator''. Today it is also common in two spatial dimensions, with the names ``QSH insulator'' and ``$\mathbb{Z}_2$ topological insulator'' often used as synonyms.

Kane and Mele introduced the QSH effect as a possible ground state for graphene as they realized  that TR invariant ``intrinsic'' spin-orbit coupling (SOC) of the type $\vec L\cdot\vec S$ is allowed by symmetry. This SOC leads to a  gap in the energy spectrum of  graphene. It is instructive to consider only its $z$ component, $L^z\cdot S^z$, as it reveals that the tight-binding version of $L^z\cdot S^z$ corresponds to a spinful version of the Haldane mass term\,\cite{kane-05prl226801,konschuh-10prb245412}. The Kane-Mele SOC reads
\begin{equation}\label{KM-SOC}
H_{\rm SOC} = i\lambda \sum_{\langle\!\langle ij \rangle\!\rangle} \sum_{\alpha\beta=\up,\dw} c_{i,\alpha}^\dag  \,\nu_{ij} \, s^z_{\alpha\beta}\, c_{j\beta}^\pd\ .
\end{equation}
The minimal version of the Kane--Mele (KM) model is complemented by the Semenoff mass term $H_v$\,\cite{semenoff84prl2449} and by a Rashba SOC $H_{\rm R}$ which might be caused by an external field or substrate\,\cite{kane-05prl226801},
\begin{equation}\label{semenoff+rashba}
H_v =\lambda_v \sum_{i\sigma} \xi_i c_{i\sigma}^\dag c_{i\sigma}^\pd~,\qquad\quad H_{\rm R} = \lambda_R \sum_{\langle ij \rangle} c_i^\dag (\bs{s}\times \bs{d})_z c_j^\pd\ .
\end{equation}
%
\begin{figure}[t!]
\centering
\includegraphics[scale=0.95]{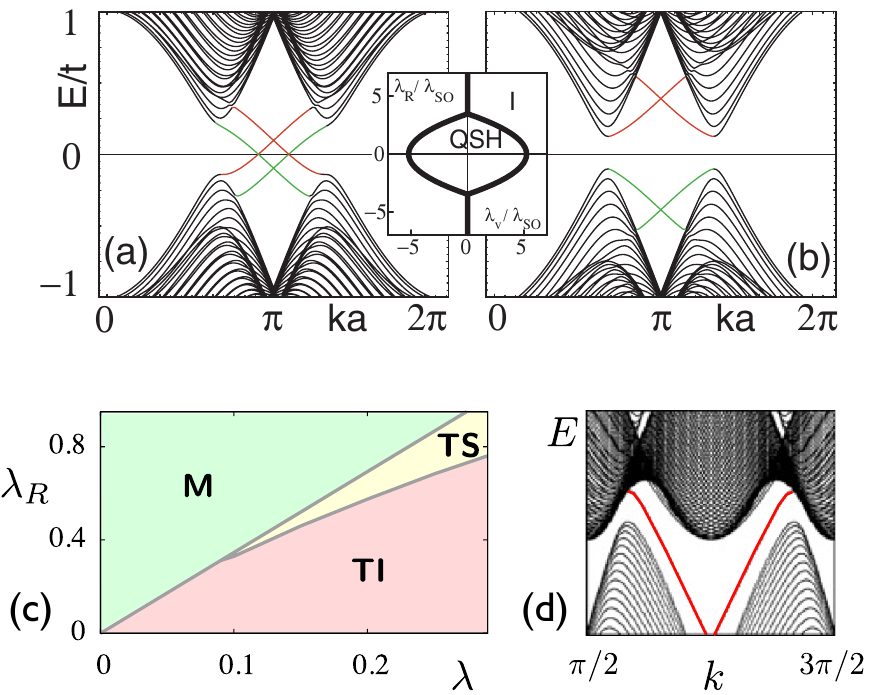}   
\caption{ 
\small (a) Ribbon spectrum of the Kane-Mele model for finite $\lambda$, $\lambda_R$, and $\lambda_v$ within the QSH phase. (b) Same model but with parameters corresponding to the topologically trivial phase. Inset: $\lambda_R$--$\lambda_v$ phase diagram with trivial insulator (I) and QSH phases; thick black lines indicate (semi-) metallic regions. 
Reprinted with permission from \cite{kane-05prl146802}. Copyright (2005) by the American Physical Society.
(c) At $\lambda_v=0$, an additional topological semiconductor (TS) phase appears between QSH and metallic phases for sufficiently large $\lambda$. (d) Ribbon spectra within the TS phase: topological edge states exist within the indirect bulk gap.
Panels (c)+(d) are reprinted from \cite{laubach-14prb165136}.}
\label{fig:km+rashba}
\end{figure}
%
Here $\xi_i=+1$ for $i\in$ sublattice A and $\xi_i=-1$ for $i\in$ sublattice B. The vector $\bs{d}$ points from site $i$ to $j$. 
More importantly, while $H_{\rm SOC}$ and $H_v$ preserve the U(1)$_{\rm spin}$ symmetry associated with spin conservation, $H_{\rm R}$ breaks it down to $\mathbb{Z}_2$; it further breaks particle-hole symmetry and the mirror $z\to -z$ symmetry. Kane and Mele  showed that the QSH effect remains stable for finite Rashba SOC; that is, the QSH effect is not only a singular point where the U(1)$_{\rm spin}$ symmetry protects two ``non-communicating'' quantum Hall systems which are fully characterized by two Chern numbers for the different spin channels, $C_\up$ and $C_\dw=-C_\up$. Instead, the QSH phase remains intact when the U(1)$_{\rm spin}$ symmetry is broken due to $H_{\rm R}$ (see the phase diagram as a function of $\lambda_R$ and $\lambda_v$ as inset between Fig.\,\ref{fig:km+rashba}\,(a) and (b)\,\cite{kane-05prl146802}). In addition, Kane and Mele introduced a new $\mathbb{Z}_2$ invariant (previously mentioned) which does not rely on spin-symmetry and distinguishes the QSH phase from the trivial insulating phase. The helical edge states of the QSH phase remain stable as long as the bulk gap remains finite. Note that the KM phase diagram [inset between Figs.\,\ref{fig:km+rashba}\,(a) and (b)] is valid only for small $\lambda < 0.1\,t$. Increasing $\lambda$ further stabilizes an additional phase between TI and the (semi-) metallic phase (at $\lambda_v=0$) where due to bending of the bulk bands only an indirect gap with helical edge states still persists\,\cite{laubach-14prb165136}, see Fig.\,\ref{fig:km+rashba}\,(c) and (d).

As beautiful the Kane-Mele proposal for QSH effect in graphene might be, subsequent {\it ab initio} calculations predicted that the intrinsic SOC is tiny (too small to be experimentally observable) and graphene remains a semi-metal\,\cite{min-06prb165310,yao-07prb041401,konschuh-10prb245412}. Ever since then material scientists are searching for other honeycomb lattice materials with possibly heavier elements to guarantee a stronger SOC necessary to realize the Kane-Mele scenario. A brief discussion of candidate materials is added below.

Shortly after the initial proposal of Kane and Mele, also Bernevig and Zhang proposed the QSH effect in zinc blende semiconductors under strain\,\cite{bernevig-06prl106802}. In a subsequent paper, Bernevig, Hughes, and Zhang (BHZ) predicted that the QSH effect should be realized in HgTe/CdTe quantum wells\,\cite{bernevig-06s1757}. The bands in vicinity of the Fermi level become inverted leading to a change of the sign of the mass gap (as it happens for SOC). Using $\bs{k}\cdot\bs{p}$ theory BHZ showed that the Bloch matrix around the points in the Brillouin zone where the band inversion occurs is effectively described by a spinful and TR invariant version of the two-orbital CI model\,\eqref{ham:ci-squarelattice} in agreement with the QSH structure \eqref{bloch-QSH}. They further proposed that the most likely source of U(1)$_{\rm spin}$ breaking is due to a bulk inversion asymmetry. Rashba SOC has been considered as another possible source\,\cite{rothe-10njp065012}. 

In 2007, a quantized and sample-width independent conductance of $2e^2/h$ in HgTe/CdTe quantum wells 
was measured in the group of L.\ Molenkamp\,\cite{koenig-07s766}, in agreement with the predictions of BHZ\,\cite{bernevig-06s1757}. 
Since the quantized conductance was independent of sample width it was attributed to edge transport, 
and today it is widely accepted as the first observation of the QSH effect.
%
This short sequence of theoretical prediction and experimental confirmation is certainly one of the main reasons why the field of topological insulators became so popular.
In the meantime, QSH effect was also proposed in InAs/GaSb quantum wells\,\cite{knez-11prl136603} and is today considered as a candidate material\,\cite{knez-14prl026602,nichele-16njp083005}.
Topological insulating band structures can be defined on any lattice, early examples include the Lieb and perovskite lattices\,\cite{weeks-10prb085310} and the ruby lattice\,\cite{hu-11prb155116}.

The QSH effect is today considered as a paradigm for a {\it symmetry protected topological} (SPT) phase. Although it was clear that the QSH effect is protected by TR symmetry (\ie in the absence of TR symmetry backscattering at the helical edges is possible) it was not anticipated at the time that QSH insulators belong to a much bigger class of physical systems. In the meantime it is common knowledge that QSH insulators are protected by TR and U(1)$_{\rm charge}$ symmetry.
If either of these symmetries is broken the QSH phase can be adiabatically connected to a trivial band insulator (assuming no other symmetry such as U(1)$_{\rm spin}$ or point group symmetries are present). Usually the absence of topological protection is signalled by a gap opening at the helical edge states.
This can be explicitly tested by applying an antiferromagnetic Zeeman field or a superconducting term to the BHZ or KM models. In all cases the edge modes start to gap out immediately; an exception is the KM model for $\lambda_{\rm R}=0$ when the staggered Zeeman field points in the $z$ direction. Due to the particular choice of the Zeeman field U(1)$_{\rm spin}$ is still  preserved and takes over the topological protection of the QSH phase\,\cite{rachel16jpcm405502}. This state has been dubbed ``spin Chern insulator''\,\cite{ezawa-13sr2790}. These ideas can be applied to situations where the symmetry breaking perturbations are acting more locally on the QSH system. For instance, only half of the sample or only the sample edges or even a few edge sites can be exposed to such a perturbation and making the edge states disappear accordingly\,\cite{rachel-14prb195303}. 

Inhomogeneous and locally varying perturbations have been used to predict applications and devices, see \eg\,\cite{rachel-14prb195303,qi-08np273,timm12prb155456}; the underlying fundamental physics base, however, on the concept of SPT phases which will be discussed in Sec.\,\ref{sec:SPTvsTO}

The edge states of a QSH insulator can be seen as two counterpropagating spin-filtered chiral edge modes and provide a new symmetry class of one-dimensional liquids: the ``helical Luttinger liquid''\,\cite{wu-06prl106401}. In a conventional spinful Luttinger liquid right- and left-moving electrons carrying $\up$-spin or $\dw$-spin, respectively, are the constituents of the system. In case of the helical edge theory of a QSH insulator where U(1)$_{\rm spin}$ symmetry is preserved the right-mover is a pure spin-$\up$ state while the left-mover is a pure spin-$\dw$ state. In the literature this phenomena has been dubbed {\it spin filtering} or {\it spin-momentum locking}. Since spin and momentum are not independent degrees of freedom anymore, the helical liquid is reminiscent of a spinless Luttinger liquid despite its spin index.
In a helical liquid, the topological protection due to TR symmetry is reflected in the absence of all terms related to elastic single-particle back-scattering\,\cite{wu-06prl106401}. This can intuitively be understood in a simplified picture where we denote the right-moving (left-moving) edge state as $\ket{R,\up}$ ($\ket{L,\dw}$). Any back-scattering term should contain a matrix element of the form $\braket{L,\dw}{R,\up}$ which is identically zero if both states are fully polarized in opposite spin-projection. Assuming that both $\ket{L,\dw}$ and $\ket{R,\up}$ are eigenstates to the same energy, $\braket{L,\dw}{R,\up}=0$ even when the axial spin symmetry U(1)$_{\rm spin}$ is broken; this follows from Kramers theorem.

\begin{figure}[t!]
\centering
\includegraphics[scale=0.72]{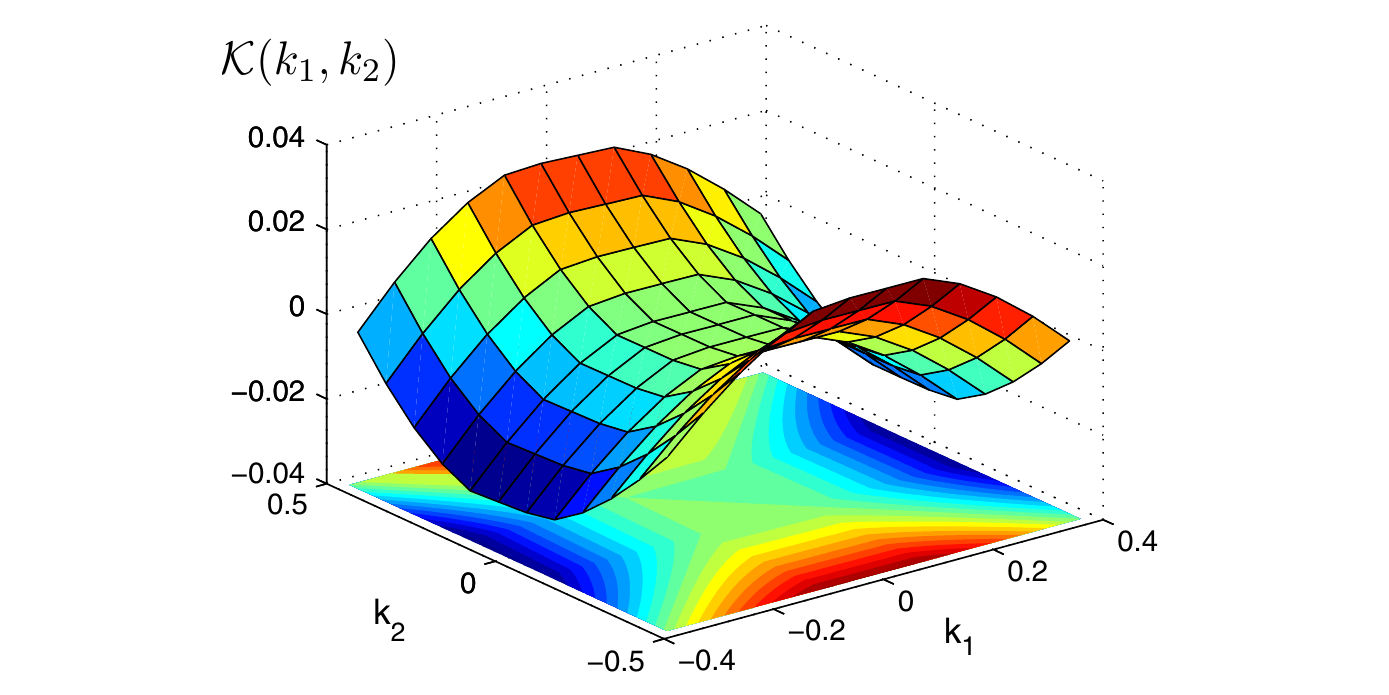}
\caption{\small Rotation of spin quantization axis as function of momenta, $\mathcal{K}(k_1,k_2)$ (reprinted from\,\cite{schmidt-12prl156402}). The plot corresponds to the BHZ model with finite bulk inversion asymmetry which breaks the U(1)$_{\rm spin}$ symmetry.}
\label{fig:RSQA}
\end{figure}

It is worthwhile to further elaborate on the role of the axial spin symmetry, U(1)$_{\rm spin}$. In experimental situations it will usually be broken since external electrical fields and the effect of a substrate are unavoidable\,\cite{kane-05prl146802}.
In the absence of the U(1)$_{\rm spin}$ symmetry the spin quantization axes will be tilted away from the $\hat z$ direction. Moreover, as demonstrated explicitly in \cite{schmidt-12prl156402,rod-15prb245112}, the spin quantization axes rotates as a function of momentum (see Fig.\,\ref{fig:RSQA}):
\begin{equation}
\mathcal{K}(k_1,k_2) = \int dx\, \psi^\dag_{-,k_2}(x,y_0) \psi^\pd_{+,k_1}(x,y_0)\ .
\end{equation}
Here $\psi_{\pm,k}(x,y)$ denotes the wave function of the right (left) moving edge state with momentum $k$, evaluated at real space position $(x,y)$ ($x$ is the coordinate perpendicular to the edge). At long wavelengths one finds $\mathcal{K}(k_1,k_2)  \approx k_0^{-2}\left( k_1^2 - k_2^2\right)$ with the material-specific parameter $k_0$.
 The corresponding edge theory of a QSH insulator where the U(1)$_{\rm spin}$ symmetry is broken has been named {\it generic helical liquid}\,\cite{schmidt-12prl156402,kainaris-14prb075118}.
 The physics of TR invariant backscattering potentials with broken U(1)$_{\rm spin}$ symmetry 
or a ``Rashba impurity'' have been extensively discussed\,\cite{stroem-10prl256804,crepin-12prb121106,lezmy-12prb235304,geissler-14prb235136,xie-16prl086603,kharitonov-17prb155134}.
For an elastic scattering process, \ie initial and final state have the same energy, left- and right-mover 
still have orthogonal spin components due to Kramer's theorem. In contrast, in case of an inelastic process right- and left-mover are not orthogonal anymore and the matrix element which enters the back scattering Hamiltonian will in general be finite, $\braket{L}{R}\not= 0$. In the presence of electron-electron interactions and an impurity, it has been shown that the combinations of these ingredients give rise to a temperature dependent correction of the conductance, $\delta G \propto T^4$. It turned out that this is at low temperatures the strongest correction to the otherwise quantized spin Hall conductance\,\cite{schmidt-12prl156402,kainaris-14prb075118}.
To derive the dependence of the spin quantization axes on momentum we originally considered a cylindrical geometry where momentum is associated with translation symmetry around the circumference. That is, the Brillouin zone is effectively one-dimensional. In order to approach geometries as used in real experiments, \eg Hall bars, one can also consider circular disks where the rotation of spin quantization axes can be obtained as a function of angular momentum\,\cite{rod-15prb245112}. Since the energy spectrum is in the vicinity of the crossing point of the edge modes approximately linear the (angular) momentum is locked to the energy values. This allows to define the rotation of spin quantization axes depending on the energy of the initial and final states in the absence of {\it any} translational or rotational symmetries\,\cite{rod-15prb245112}.
As discussed in Sec.\,\ref{sec:SIH}, the broken U(1)$_{\rm spin}$ symmetry does not only alter the spin texture of the helical edge states but it also influences the phase diagrams of the topological Hubbard models drastically.

Helical Luttinger liquids\,\cite{wu-06prl106401} represent an important subject as it opens the research topic ``topological insulators'' to the field of one-dimensional electron systems and the powerful methods developed therein. Also, helical li\-quids allow to study interesting scenarios such as helical quantum edge gears\,\cite{chou-15prl186404} or, in combination with disorder and interactions, the emergence of gapless glassy edge states spontaneously breaking time-reversal symmetry\,\cite{chou-17arXiv1710.04232}, just to mention a few. A recent review article discusses the edge physics of 2D TIs and helical Luttinger liquids\,\cite{dolcetto-16rnc113}.

%

Now let us consider the three-dimensional (3D) case: while the integer quantum Hall effect does not exist in odd space dimensions, shortly after the prediction of the 2D QSH insulating state, three groups indepedently introduced the 3D generalization\,\cite{roy09prb195322,moore-07prb121306,fu-07prl106803}. The central idea is to apply the $\mathbb{Z}_2$ invariant previously introduced for 2D systems to 3D Bloch matrices $H(k_x, k_y, k_z)$.
By fixing one of the momentum quantum numbers $k_x$, $k_y$, or $k_z$ to take the constant value 0 or $\pi$, the remaining Bloch matrix is effectively two-dimensional. Let us name the new invariants $x_0 := \mathbb{Z}_2[H(0, k_y, k_z)]$, $x_\pi := \mathbb{Z}_2[H(\pi, k_y, k_z)]$, $y_0 := \mathbb{Z}_2[H(k_x, 0, k_z)]$ etc. Each of these six new invariants can  take the values even and odd, but from the $2^6=64$ possible combinations only 16 are inequivalent. Eventually we define the ``strong'' invariant $\nu = x_0 \cdot x_\pi = y_0 \cdot y_\pi = z_0 \cdot z_\pi$\footnote{This ``multiplication'' works according to even $\cdot$ odd = odd $\cdot$ even = odd, odd $\cdot$ odd = even,  even $\cdot$ even = even.} and simply write $(\nu; x_0, y_0, z_0)$ in order to fully characterize the phase. 
%
%
%
Thus there are 16 time-reversal invariant and insulating phases in 3D\,\cite{roy09prb195322,moore-07prb121306,fu-07prl106803}. Eight of them are so-called {\it strong} topological insulators  (STIs) with $\nu=1$ and seven are {\it weak} topological insulators  (WTIs) with $\nu=0$; the phase with $(0;000)$ is the trivial insulating state. 
An efficient way to compute these invariants numerically has been derived\,\cite{fukui-07jpsp053702}.
STIs possess at the surface or interface to a topologically distinct phase an odd number of surface states which realize a 2+1--dimensional Dirac theory\,\cite{qi-08prb195424}. In contrast, WTIs feature surfaces with an even number (including zero) of such surface states.
In a topological quantum field theory there is only the distinction between topological insulators with $\theta=\pi$ and trivial insulators with $\theta=0$\,\cite{qi-08prb195424}. While the latter corresponds to STIs, the former to topologically trivial states of matter including WTIs. That is the main reason why WTIs did not attract much attention in the early years of topological insulators. First detailed studies of WTIs in the presence of disorder or interactions revealed, however, that WTIs are not much less robust than their strong cousins and similarly interesting\,\cite{mong-12prl076804,kobayashi-13prl236803,liu-12pe906}. Recently, also the topological terms of interacting topological insulators were derived\,\cite{wang-15prl031601,guo-18ap244} consistent with Witten's approach\,\cite{witten16rmp035001}. 

After the prediction and discovery of 2D and 3D TIs about ten years ago, several extensions and generalizations of these states have attracted much attention: for instance, topological crystalline insulators\,\cite{fu11prl106802}, Weyl semimetals\,\cite{bevan-97n689,Volovik03,wan-11prb205101,xu-15s613} (sometimes referred to as gapless topological insulators), and ``new fermions''\,\cite{alexandradinata-14prl116403}. In addition, a full classification of topological bandstructures based on quantum chemistry considerations has been derived\,\cite{bradlyn-17n257}.


In the following, we will give a very short overview of candidate materials for 2D and 3D TI phases. A more complete and thorough discussion of materials can be found in the relevant review articles\,\cite{hasan-10rmp3045,qi-11rmp1057,hasan-11arcmp55,ando-15arcmp361,yan-12rpp096501,ren-16rpp066501}.

In two spatial dimensions, besides the previously discussed quantum well systems HgTe/CdTe\,\cite{koenig-07s766} and InAs/GaSb\,\cite{knez-14prl026602} the main focus has been on other honeycomb-lattice materials involving sufficiently heavy atoms to produce a significant spin-orbit gap. Most interestingly is the family of silicene, germanene, and stanene\,\cite{liu-11prb195430,xu-13prl136804,geissler-15njp119401,rachel-14prb195303,xu-18prb035122}, where the latter is the most promising system. Also molecular graphene\,\cite{ghaemi-12prb201406} has been suggested and cold-atom settings\,\cite{goldman-10prl255302} which might feature the QSH effect.
Recently, the adatom system bismuthene on a SiC substrate\,\cite{reis-17s287} has been reported to feature a signficantly large bulk gap suggesting even  the presence of a room temperature QSH phase.

In three spatial dimensions, after the discovery of the first STI material Bi$_{1-x}$Sb$_x$\,\cite{hsieh-08n970} a countless number of candidate materials has been proposed and many of them were claimed to be successfully measured and identified as TIs. Here we only review the first few experiments.
The discovery of the first STI phase was reported in the compound Bi$_{0.9}$Sb$_{0.1}$\,\cite{hsieh-08n970,teo-08prb045426}. Angle-resolved photoemission spectroscopy (ARPES) measurements on the (111) surface of Bi$_{0.9}$Sb$_{0.1}$ demonstrated the presence of unusual surface states which cross the Fermi energy five times between $\bar\Gamma$ and $\bar M$. Another feature of the topological surface state is their inherent $\pi$ Berry phase since spin and momentum are locked. Within spin resolved ARPES also this feature could be observed\,\cite{hsieh-09s919}.

The second generation of materials includes the prominent systems Bi$_2$Se$_3$, Bi$_2$Te$_3$, and Sb$_2$Te$_3$. In contrast to the experimental challenges to handle BiSb compounds, in particular Bi$_2$Se$_3$ shows topological insulating behavior even at room temperature and in the absence of magnetic fields\,\cite{hasan-10rmp3045}. Moreover, Bi$_2$Se$_3$ features a single Dirac cone on its surface\,\cite{hsieh-09n1101}. At the same time, pioneering {\it ab initio} calculations identified Bi$_2$Se$_3$, Bi$_2$Te$_3$, and Sb$_2$Te$_3$ as single-Dirac-cone materials\,\cite{zhang-09np438}.

Further important TI materials are HgTe and $\alpha$-Sn, as proposed already in 2007\,\cite{fu-07prb045302}. Strained HgTe is a strong TI as shown through magneto transport and ARPES measurements\,\cite{bruene-11prl126803}. Similarly, strained $\alpha$-Sn realized a strong 3D TI phase as revealed through ARPES measurement combined with {\it ab initio} calculations; $\alpha$-Sn  represents the first elemental TI which is promising for engineering future devices\,\cite{barfuss-13prl157205,ohtsubo-13prl216401}.

Today, topological insulators represent a major research direction for the fields of spectroscopy, material growth, and transport measurements\,\cite{hasan-10rmp3045,qi-11rmp1057,hasan-11arcmp55,ando-15arcmp361,yan-12rpp096501,ren-16rpp066501}.

%
%
\subsection{Symmetry Protection vs.\ Topological Order}
\label{sec:SPTvsTO}


The discovery of the quantum spin Hall effect and topological insulators accelerated the field of topological phases and turned it into one of the most active fields of contemporary condensed matter research. As a consequence, a large amount of scientific publications related to condensed matter physics contain nowadays the word ``topological''. Unfortunately, there are (at least) two important concepts which are often 
mixed up
in the literature: {\it symmetry protected topological} (SPT) phases and {\it topologically ordered} phases. 
In the following, we will briefly define them, list the most important examples, and emphasize the difference between them. Note that both concepts are very active fields of research and this section only aims to give a brief overview but neither a detailed review nor a complete list of references. For further reading we refer to recent review articles\,\cite{senthil15arcmp299,wen17rmp041004}.

An SPT phase\,\cite{chen-10prb155138,Volovik03,chen-12s1604,wen14prb035147,pollmann-10prb064439,wen12prb085103,wang-15prl031601} has short-range entanglement, is protected by one or several symmetries, and carries gapless modes at its edges or boundaries, respectively. The central idea is that such a phase is stable against (small) perturbations as long as the protecting symmetry is preserved. When stronger perturbations are applied, a change of the ground state is only possible via closing of the bulk gap. We previously discussed the QSH effect as a paradigm for an SPT phase (the discussion also applies to 3D TIs).
 Note that free fermion phases in arbitrary dimensions were classified even before SPT phases became popular: Schnyder, Ryu, Furusaki, and Ludwig\,\cite{schnyder-08prb195125} and independently Kitaev\,\cite{kitaev09} proposed a table (``ten-fold way'') corresponding to the random matrix classification of Altland and Zirnbauer\,\cite{zirnbauer96jmp4986,altland-97prb1142}. This classification uses anti-unitary symmetries, time-reversal and particle-hole,  to distinguish the different phases. The TR invariant $\mathbb{Z}_2$ topological insulators belong to class AII in this table, the integer QHE to class A (the class which does not require any symmetry agreeing with our previous statement). Topological insulators which rely on other symmetries such as inversion -- and thus fall out of this classification scheme -- were proposed\,\cite{hughes-11prb245132}. Also topological crystalline insulators -- TIs which are protected by a space group symmetry instead of TR -- were proposed\,\cite{fu11prl106802} and subsequently discovered experimentally\,\cite{junwei-14nm178}, representing today another exciting research direction\,\cite{slager-13np98,kargarian-13prl156403,xu-12nc1192,liu-13prb241303,junwei-14nm178}.

Another instructive example of an SPT phase is the spin 1 antiferromagnetic chain (aka Haldane chain)\,\cite{haldane83prl1153} including the exactly soluble AKLT model\,\cite{affleck-87prl799}. Here the protecting symmetries are TR, bond-inversion, and dihedral symmetries\,\cite{pollmann-10prb064439}. One needs to break all these symmetries in order to loose its character,
\ie only then it is possible to adiabatically transform it to the atomic limit, a trivial product state, without closing of the bulk gap. The spin 1 chain features gapless zero modes at its edges, the so-called dangling spins. Other SPT phases of interests are {\it bosonic} TIs\,\cite{liu-14prl267206}.

The concept of SPT phases has been extensively applied to classify quantum phases in the past years. In one spatial dimension, SPT phases were claimed to be completely classified by the elements in the second group cohomology class\,\cite{chen-11prb235128,pollmann-10prb064439}. Also for non-interacting fermionic systems in $d$ spatial dimensions a complete characterization has been derived based on both group cohomology theory and $K$-theory\,\cite{wen12prb085103,schnyder-08prb195125,kitaev09}. Classification based on SPT phases for interacting bosonic states of matter in $d > 1$ have recently been explored\,\cite{chen-12s1604}.

Note that the previously discussed example of the spin-Chern insulator demonstrates that also other symmetries can protect quantum phases. TR symmetry and particle hole (\ie charge conjugation)  are certainly the most robust symmetries -- because they are anti-unitary symmetries. But inversion, mirror or other point group symmetries can in principle lead to ``topological'' protection, the previously mentioned ``crystalline topological insulators'' constitute a prominent example\,\cite{fu11prl106802}.
In such cases, the presence of edge states is, however, not guaranteed\,\cite{hughes-11prb245132}. 
It is worth emphasizing that SPT phases can be realized in non-interacting theories, \ie systems of free fermions, but are not limited to them.

This is in stark contrast to systems exhibiting (intrinsic) topological order. First introduced by X.-G. Wen\,\cite{wen90ijmp239} in the context of the fractional QHE, today several families of strongly correlated electron systems are known to be {\it topologically ordered}. In principle, no symmetry is required to be preserved for this type of order -- but no symmetry is required to be spontaneously broken either\footnote{Nonetheless we know topologically ordered states of matter which break a symmetry (\eg  fractional quantum Hall states break (explicitly) TR symmetry) or which preserve a symmetry (\eg the toric code\,\cite{kitaev03ap2} possesses a TR invariant ground state manifold).}. Topologically ordered states possess long-range entanglement (which immediately rules out free fermion theories), exhibit a ground state degeneracy when defined on a torus but no degeneracy when defined on a sphere, and the elementary excitations of such systems are anyons, exotic quantum particles obeying fractional statistics. Examples are the $\nu=1/m$ Laughlin fractional quantum Hall states ($m$ being an odd integer) featuring an $m$-fold degenerate ground state on the torus and the toric code\,\cite{kitaev03ap2} as a prototype of an $\mathbb{Z}_2$ spin liquid featuring a four-fold degenerate ground state on the torus geometry. Different topological orders can be characterized either by their topological quantum field theories or by their modular matrices\,\cite{zhu-14prl096803}. 
Recently, a hierarchy construction which allows deriving all topological orders in two spatial dimensions was developed\,\cite{lan-17prl040403}.
In this review, we are only touching topologically ordered phases when discussing the CI$^\star$ and QSH$^\star$ phases in Sec.\,\ref{sec:corr-CI} and in Sec.\,\ref{sec:SIH}, respectively, topological Mott insulators (in Sec.\,\ref{sec:stronglycorr-TI}), fractional Chern and topological insulators (Sec.\,\ref{sec:fci+fti}), Kitaev materials (Sec.\,\ref{sec:iridates}) and when discussing the effect of electron-electron interactions on the surface of a strong TI in Sec.\,\ref{sec:TIsurface}.

Several comments are in order.\\
(i) Topologically ordered phases always involve (strong) electron-electron interactions.\\
(ii) In some cases in the literature, topological insulators are referred to as systems with ``topological order'' -- which has nothing to do with the previously discussed concept of intrinsic topological order.\\
(iii) The role of interactions in the context of SPT phases is somewhat ambivalent -- on the one hand they can stabilize SPT phases which do not have a free fermionic analog, a nice example is provided by parafermionic chains\,\cite{motruk-13prb085115,alicea-16arcmp119}; on the other hand they can destroy SPT phases, best illustrated by the reduction of the BDI symmetry class to the $\mathbb{Z}_8$ classification in one spatial dimension\,\cite{turner-11prb075102,fidkowski-11prb075103}.
To test this reduction experimentally was suggested both in the superlattice material CeCoIn$_5$/YbCoIn$_5$\,\cite{yoshida-17prl147001} and in a cold atom system\,\cite{yoshida-17arXiv1711.09538}.
\\
(iv) It is possible to construct states of matter which exhibit topological order {\it and} possess a protecting symmetry simultaneously. Such states of matter are named {\it symmetry enriched topological} (SET) phases\,\cite{lu-16prb155121}. A straight-forward example are TR-invariant fractional topological insulators\,\cite{levin-09prl196803}, see Sec.\,\ref{sec:fci+fti}.\\
(v) The integer QHE is neither topologically ordered (because it is a free fermion theory)
nor protected by any symmetry.
On the other side, it is clearly the {\it mother state} of all topological phases. The QHE  simply realizes a chiral phase. The example of the QHE nicely demonstrates the dilemma of characterizing {\it all} topological phases in a simple fashion. This issue is, however, beyond the scope of this review.

%
%
\section{Interacting Topological Insulators}
\label{sec:interacting-TI}

The previously discussed Chern and topological insulators  are fascinating states of matter in all the varieties they appear. Most notably, they are all described by theories of free fermions. In condensed matter physics, many of the striking phenomena are due to (strong) electron--electron interactions. Examples include unconventional superconductivity\,\cite{steglich-79prl1892,bednorz-86zpb189,stewart17ap75}, the Kondo-effect\,\cite{kondo64ptp37}, or the Mott-Hubbard transition\,\cite{mott68rmp677}, just to mention a few. Furthermore, in all solids Coulomb interactions are unavoidable, sometimes they might be screened and weak, but sometimes they are strong and can even dominate the physical properties of a material. Therefore it is both a natural and important question to ask what the effect of interactions in these topological insulators might be. One can group the most relevant aspects into the following list of questions:
\begin{itemize}
\item Is a topological bandstructure stable with respect to electron-electron interactions and what happens when the TI phase breaks down?
\item The conceptually inverse question is whether electron-electron interactions can turn a topologically trivial ground state into a non-trivial one? With other words, are interaction-induced topological phases possible?
\item  Can more exotic states of matter emerge when the non-trivial band topology competes with strong electron-electron interactions? How does the band filling affect this competition?
\item How are the topological edge or surface states influenced by strong interactions?
\item Eventually one can ask to what extent the topological properties of non-interacting bandstructures 
become relevant for the corresponding strong coupling phases of certain frustrated quantum magnets?
\end{itemize}

In the following, we will focus on all these aspects -- some of them are discussed in more detail, others are briefly mentioned.

%
%
\subsection{Early Considerations of Interacting Topological Insulators}
\label{sec:earlyTI}

Already in the first Kane-Mele paper\,\cite{kane-05prl226801} the effect of long-ranged Coulomb interactions in graphene is considered: in order to estimate the value of the spin-orbit potential 
Kane and Mele took into account the renormalization of the spin-orbit gap $\Delta_{\rm SO}$ due to the interaction of electrons with the exchange potential induced by $\Delta_{\rm SO}$. They claimed that Coulomb interactions  may increase the energy gap. The renormalization causes a divergent correction to  $\Delta_{\rm SO}$ which together with other contributions can be summed using the renormalization group. Once the renormalization group equations are derived and solved one can indeed see that the spin-orbit potential is enhanced due to Coulomb interactions. Note that these considerations provide an argument for the stability of the TI phase with respect to electron-electron interactions but they are limited to the perturbative regime where Coulomb interactions are weak.

Most other works focussing on correlation effects included the interactions exclusively for the edge states on the level of helical Luttinger liquids\,\cite{wu-06prl106401}. Another notable early work addresses the effects of interactions in combination with disorder for the topological edge states of TIs\,\cite{xu-06prb045322}.

Another direction where interactions were considered enclose the attempts to generalize or extend the Kane-Mele invariant\,\cite{kane-05prl146802} to interacting systems\,\cite{wang-10prl256803,gurarie-11prb85426,wang-12prb165126,wang-14prx011006}; several of these papers are inspired by earlier work of Volovik\,\cite{Volovik03}. An interesting discussion can be found in Ref.\,\cite{budich-13pss109}.
 The main idea is to express the invariant in terms of the single-particle Green's function and its derivatives. It can be shown that these formulations are equivalent to the Kane-Mele\,\cite{kane-05prl146802} or Fu-Kane\,\cite{fu-07prb045302} invariants in the absence of interactions. They allow, however, easily to implement electron correlations encoded in the self-energy\,\cite{wang-10prl256803,gurarie-11prb85426,wang-14prx011006} and thus presumably extend the classification of non-interacting topological bandstructures to their correlated counterparts. Another interesting concept is the {\it topological Hamiltonian}\,\cite{wang-13jpcm155601} which is defined as the Bloch matrix of the noninteracting system plus the self-energy evaluated at zero frequency, $h_{\rm topo} = h(\bs{k}) + \Sigma(\omega=0, \bs{k})$. In particular for numerical methods which rely on strategies to efficiently compute the full Green's function (\eg dynamical mean-field theory, quantum cluster appraoches etc.) it has the major advantage to be easily accessible in contrast to other invariants.

%
%
\subsection{Correlated Topological Insulators}
\label{sec:corr-TI}

Testing the stability of topological bandstructures and deriving interacting phase diagrams of model Hamiltonians has stimulated a large amount of research in the past years.
Here we consider several paradigmatic two-dimensional TI models supplemented by Coulomb interactions. In order to simplify the discussion we always consider local Hubbard interactions for spinful models,
\begin{equation}\label{hubbard}
\mathcal{H}_I = U \sum_i n_{i\up} n_{i\dw}\ ,
\end{equation}
and nearest-neighbor interactions for spinless models,
\begin{equation}\label{NN-coulomb}
\mathcal{H}_I^{(1)} = V_1 \sum_{\langle ij \rangle} n_{i} n_{j}\ ,
\end{equation}
if not mentioned otherwise. Although the Hubbard term \eqref{hubbard} represents the simplest interaction term for spinful electrons, in two and three spatial dimensions there is no exact solution for Hubbard-type models.
The focus of this section will be mainly on the stability of the TI phase, on the emergence of conventional orders due to spontaneous symmetry breaking, but also on novel, exotic phases. 
There is  an earlier review\,\cite{hohenadler-13jpcm143201} about correlated TIs covering some of the material discussed in the following.

%
%
\subsubsection{One-dimensional Correlated Topological Insulators}
\label{sec:SSHH}

As discussed earlier, topological (Chern) insulators are lattice realizations of the Dirac Hamiltonian. Originally, theoretical and experimental interest was primarily focussed on 2D and 3D systems. In the meantime, also 1D systems have been considered with and without inclusion of electron-electron interactions and experimental setups proposed and even realized. Details can be found in the brief review article\,\cite{guo-16scpma637401}.

The simplest model realizing a 1D TI is the Su-Schrieffer-Heger (SSH) model which was introduced in 1979 to describe polyacetylene\,\cite{su-79prl1698,heeger-88rmp781}. It is a Peierls-type model with alternating weak and strong hoppings governed by the Hamiltonian
\begin{equation}\label{ham:ssh}
H_{\rm SSH} = \sum_i \left( t - (-1)^i\delta t \right) [ c_{i+1}^\dag c_i^\pd + {\rm H.c.}]\ .
\end{equation}
For $\delta t=0$ the model is gapless and describes free fermions with $\cos{k}$ dispersion. For finite $\delta t$, the energy spectrum is gapped being topologically non-trivial (trivial) for $\delta t<0$ ($\delta t > 0$). Often also the spinful version of \eqref{ham:ssh} is discussed. Topological invariants are available, either based on Berry phase, single-particle Green's functions, or entanglement entropy\,\cite{guo-11prb195107,manmana-12prb205119,wang-15prb115118}. Other 1D TIs are based on superlattices or Creutz-type models\,\cite{hetenyi-18jpcm10LT01,gholizadeh-18epl27001}. As mentioned before, the topological properties can be best understood from the perspective of gapped Dirac Hamiltonians\,\cite{shen-11spin33}.

Several of these models have been discussed in the presence of electron-electron interactions: spinless SSH Hubbard model\,\cite{sirker-14jsm10032}, spinful SSH Hubbard model\,\cite{manmana-12prb205119,wang-15prb115118,barbiero-18prb201115} as well as other 1D topological Hubbard models\,\cite{guo-11prb195107,junemann-17prx031057}. All of them turn out to be robust against weak or moderate interaction strength. Also a bosonic SSH model with repulsive interactions has been analyzed including topologically non-trivial Mott phases\,\cite{grusdt-13prl260405}.

Experimental progress in realizing 1D TIs has been made using ultra-cold atoms or photonic crystals\,\cite{atala-13np795,kraus-12prl106402,verbin-13prl076403}. At least ultra-cold quantum gases allow to tune finite interactions between the atoms and thus realize interacting TIs in 1D.

%
%
\subsubsection{Correlated Chern Insulators and Haldane-Hubbard model}
\label{sec:corr-CI}

The first paper addressing the effect of electron-electron interactions in the spinless Haldane model was an exact diagonalization study using the Lanczos method\,\cite{varney-10prb115125}. Triggered by the nearest-neighbor Coulomb term \eqref{NN-coulomb} 
the transition from the weakly-correlated Chern insulator phase with circulating edge modes into the Mott phase with charge density wave order was investigated. The ana\-lo\-gous study for hardcore bosons did not show any topological features\,\cite{varney-10prb115125}.  Later, the bosonic Haldane-Hubbard model was reconsidered; superfluid and Mott insulating phases supporting local plaquette currents were found\,\cite{vasic-15prb094502}.

Motivated by the experimental realization of the Haldane Chern insulator in an optical lattice\,\cite{jotzu-14n237}, ``spinful'' versions of the Haldane model were investigated where both spin channels feature a Chern number leading to a Chern insulator phase with $C=\pm 2$ (``doubled Haldane model''). The effect of interactions was studied using slave-rotor method\,\cite{hickey-15prb134414} as well as slave-spin technique\,\cite{prychynenko-15pb53,maciejko-13prb241101}. Besides the variety of  magnetically ordered phases, also correlated CI and Chern metal phases were found; most remarkably, it was suggested that an exotic CI$^\star$ phase could possibly be realized given that the topological as well as the interaction term are sufficiently large\,\cite{maciejko-13prb241101}. This phase corresponds to the ``$\mathbb{Z}_2$ fractionalized Chern insulator" found in an exactly solvable model\,\cite{zhong-13prb045109}.
Properties of the exotic phase include topological order, chiral edge states, a quantized Hall conductance. It can be seen as the chiral, time-reversal broken version of the QSH$^\star$ phase or as an ``orthogonal Chern insulator'', further discussed in Sec.\,\ref{sec:SIH}.
In contrast to the CI$^\star$ phase found within slave-spin method, slave-rotor theory  led to the prediction of chiral spin liquid phases for sufficiently strong interactions\,\cite{hickey-15prb134414}.

In the meantime, also numerical studies for the spinful Haldane-Hubbard model are available. Within Dynamical Cluster Approach a direct transition from the the Chern insulator phase into an antiferromagnet was found which is likely to be of first order\,\cite{imriska-16prb035109}. Using dynamical mean-field theory, the model has been studied with additional staggered sublattice potential $H_v$ as defined in Eq.\,\eqref{semenoff+rashba}. Besides the trivial band insulating and trival (Mott) insulating phases (magnetism was not considered), an intermediate phase with $C=1$ was found\,\cite{vanhala-16prl225305}.

Another example of a correlated Chern insulator was discussed in Ref.\,\cite{doretto-15prb245124} where a staggered $\pi$-flux lattice leads to explicit time-reversal breaking (the same bandstructure was the starting point in Ref.\,\cite{neupert-11prl236804}). Using a bosonization formalism, the quarter filled system can be mapped to an interacting boson model; interestingly, the spin-wave excitation spectrum above the correlated Chern insulator remains gapless while for the analogous TR-invariant version it features a gap.

%
%
\subsubsection{Kane-Mele-Hubbard Model}
\label{sec:KMH}

\begin{figure}[t!]
\centering
\includegraphics[scale=0.80]{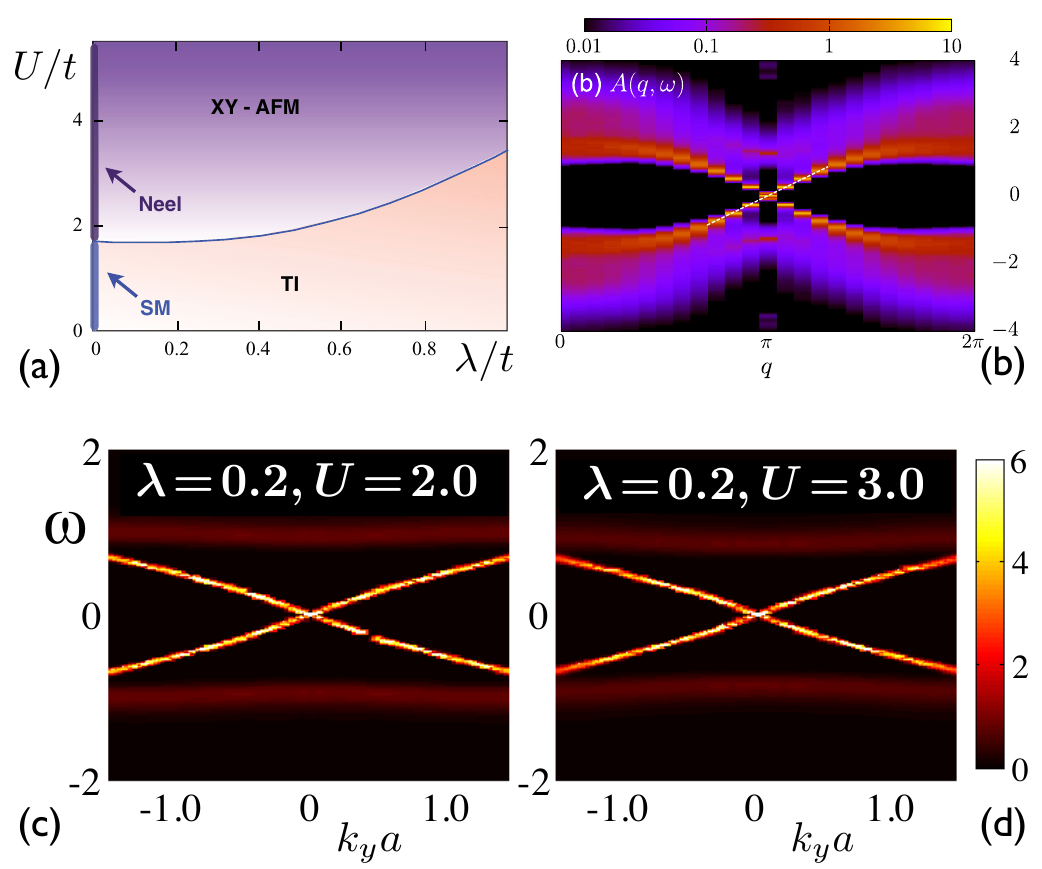}
\caption{\small (a) Phase diagram of the Kane-Mele-Hubbard model as a function of $U/t$ and $\lambda/t$ (figure taken from Ref.\,\cite{reuther-12prb155127}). (b) Single-particle spectral function $A(q,\omega)$ of the same model for $U/t=2$ and $\lambda/t=0.25$ obtained within Quantum Monte Carlo.
Reprinted with permission from \cite{hohenadler-12prl229902,hohenadler-11prl100403}. Copyright (2012) by the American Physical Society.
(c)+(d) Single particle spectral function $\mathcal{A}(k_y, \omega)$ computed within cluster dynamical mean-field theory for $\lambda/t=0.2$ and  (c) $U/t=2$ and (d)  $U/t=3$ (figures are taken from Ref.\,\cite{wu-12prb205102}).}
\label{fig:kmh+cdmft}
\end{figure}

The most important model to capture both non-trivial band topology and correlation physics in 2D is the Kane-Mele-Hubbard (KMH) model\,\cite{rachel-12prb075106} governed by the Hamiltonian
\begin{equation}\label{ham:KMH}
\begin{split}
\mathcal{H}_{\rm KMH} =&  -t \sum_{\langle ij \rangle} \sum_{\sigma} c_{i\sigma}^\dag c_{j\sigma}^\pd  + i\lambda \sum_{\langle\!\langle ij \rangle\!\rangle} \sum_{\alpha\beta=\up,\dw} c_{i,\alpha}^\dag  \,\nu_{ij} \, s^z_{\alpha\beta}\, c_{j\beta}^\pd \\
&+U \sum_i n_{i\up} n_{i\dw}
\end{split}\end{equation}
consisting of a real hopping term with amplitude $t$, the beforementioned SOC \eqref{KM-SOC} with amplitude $\lambda$, and the Hubbard term \eqref{hubbard} with amplitude $U$.
In the meantime, the KMH model has been extensively studied both 
 analytically\,\cite{rachel-12prb075106,soriano-10prb161302,lee11prl166806,griset-12prb045123,mardani-11arXiv1111.5980,hamad-16prb205113}   and 
 numerically\,\cite{wu-12prb205102,laubach-14prb165136,hohenadler-11prl100403,hohenadler-12prb115132,
 zheng-11prb205121,yamaji-11prb205122,yu-11prl010401,hung-13prb121113,meng-14mplb143001,zeng-17prb195118}
and its phase diagram is well-established [see Fig.\,\ref{fig:kmh+cdmft}\,(a)]. It contains in the weak-coupling regime up to moderate interactions $U \overset{>}{\sim} t$ the topological band insulator phase; at some critical $U_c$ a phase transition into a magnetically ordered phase occurs. We will see below that this is a rather generic behavior for TIs. The KMH model in the absence of any Rashba SOC $\lambda_R =0$ features above $U_c$ an XY antiferromagnet with in-plane magnetization.

The presence of an antiferromagnetic XY phase can be best seen by deriving the strong coupling limit of \eqref{ham:KMH} leading to the Kane-Mele spin model\,\cite{rachel-12prb075106},
\begin{equation}
H = \frac{4t^2}{U}\sum_{\langle ij \rangle} \bs{S}_i \bs{S}_j + \frac{4\lambda^2}{U}\sum_{\langle\!\langle ij \rangle\!\rangle} \left( -S^x_i S^x_j - S^y_i S^y_j + S_i^z S_j^z \right)\ .
\end{equation}
The antiferromagnetic nearest-neighbor Heisenberg model on the honeycomb lattice exhibits a Neel ordered ground state since the lattice is bipartite.
While the second-neighbor spin exchange for $x$ and $y$ components is compatible with the antiferromagnetic nearest-neighbor exchange, the second-neighbor exchange $S_i^z S_j^z$ is competing with the nearest-neighbor contributions. Instead of resulting in a situation of frustrated magnetism, the system finds, however, an elegant way to circumvent it. By turning the magnetization vector into the XY plane (and avoiding any finite component in the $\hat z$ direction) the $S^z_i S^z_j$ exchange is neutralized. This argument is strictly valid only for large $U$ but mean-field treatment of different magnetizations shows that already at $U_c$ the in-plane magnetized AFM is energetically favorable compared to an easy axis order\,\cite{wu-12prb205102}. These findings are in agreement with numerically exact quantum Monte Carlo simulations\,\cite{hohenadler-11prl100403,hohenadler-12prb115132,zheng-11prb205121}, with pseudo-fermion functional RG analysis\,\cite{reuther-12prb155127}, cluster dynamical mean-field theory\,\cite{wu-12prb205102}, variational cluster approach\,\cite{laubach-14prb165136}, and density-matrix renormalization group\,\cite{zeng-17prb195118}. Also the slave-rotor analysis points towards an XY antiferromagnetically ordered phase\,\cite{rachel-12prb075106}, see Sec.\,\ref{sec:TMI}. In several works, it has been argued and shown that the phase transition from the TI into the in-plane antiferromagnet is of three-dimensional XY universality class\,\cite{hohenadler-11prl100403,wu-12prb205102,griset-12prb045123,zeng-17prb195118}.

While the minimal version of the KMH model \eqref{ham:KMH} has been studied by many authors, only few works\,\cite{laubach-14prb165136,hohenadler-14prb245148} investigated the additional effect of the Rashba term. While the physics for not too large Rashba-SOC is similar to that of the pure KMH model, the combination of larger values of intrinsic and Rashba SOC even changes the non-interacting phase diagram\,\cite{laubach-14prb165136} as mentioned in Sec.\,\ref{sec:Z2TI} [see Fig.\,\ref{fig:km+rashba}\,(c)]. In contrast to the $\lambda_v$--$\lambda_R$ phase diagram shown in the inset between Figs.\,\ref{fig:km+rashba}\,(a) and (b), between QSH phase and trivial insulator a metallic phase with topological edge states is present. Band bending due to large Rashba leads to an indirect band gap; locally in momentum space for each wavevector $\bs{k}$ the gap is still preserved. It was shown that this ``topological semiconductor'' phase persists for finite $U$\,\cite{laubach-14prb165136}.
Eventually the magnetism at large $U$ is influenced because the Rashba-term generates  Dzyaloshinsiki--Morya spin exchange which might favor spiral orders over collinear magnetic states.

That a topological {\it insulator} is stable with respect to weak interactions is certainly not surprising since it possesses an ener\-gy gap like any other insulator.  The interesting question one might ask is whether or not the metallic edge states persist in the presence of interactions. Apart from mean-field attempts, numerical methods including dynamical mean-field theory, its cluster extension, and variational cluster approach where  the full Green's function $\mathcal{G}(\bs{k},\omega)$ is (approximately) computed is particularly suited to answer this question. One can impose a cylinder geometry and compute the single-particle spectral function of the respective topological Hubbard model,
\begin{equation}
\mathcal{A}(k_y, \omega) = -\frac{1}{\pi} {\rm Im}\left\{ \mathcal{G}(k_y,\omega) \right\}
\end{equation}
where $k_y$ is the good momentum quantum number of the cylinder which is finite in the $x$ direction.
In Ref.\,\cite{hohenadler-11prl100403,wu-12prb205102,hohenadler-12prb081106,yu-11prl010401}, $\mathcal{A}(k_y, \omega)$ is computed in the correlated TI regime of the KMH model, see Fig.\,\ref{fig:kmh+cdmft}\,(c) and (d).

Amongst the manifold works on the KMH model, particular attention was given to the edges of the topological regime. While essentially all papers agree on the bulk phase diagram as shown in Fig.\,\ref{fig:kmh+cdmft}\,(a) (apart from actual numbers for the $U$ values where phase transitions occurs, which are somewhat method-dependent), the discussion of the edges turns out to be more diverse. Quantum Monte Carlo simulations\,\cite{zheng-11prb205121} predicted two different non-magnetic phases: for zero to weak interactions the TI phase with helical edge states, but for slightly stronger interactions an analogous phase where the helical edges become unstable. The edge states can spontaneously break TR symmetry and acquire magnetic order due to two-particle backscattering\,\cite{zheng-11prb205121}. A similar observation has been made using variational Monte Carlo. Increasing Hubbard interactions lead to a strong suppression of the charge Drude weight in the helical edge channels. This mechanism drives a phase transition from the TI phase to an edge-Mott insulator phase\,\cite{yamaji-11prb205122,yoshida-16prb085149}. Extensions of the KMH model were also applied to describe the topological and magnetic phase transitions of silicene nanoribbons\,\cite{lue-18njp043054}.

A bosonic version of the KMH model was recently shown to host an emergent chiral spin state\,\cite{plekhanov-18prl157201}.
Other correlated topological systems  similar to the (fermionic) KMH model have also been studied\,\cite{lado-14prl027203,parisen-15prb165108}. In particular, the interaction-driven phase transition from correlated topological insulator into an antiferromagnetic phase has been studied on the $\pi$-flux checkerboard lattice: in contrast to the KMH case discussed above, here the universality class of the transition seems to correspond to the one of the 2D Ising model\,\cite{zeng-17prb195118}.

%
%
\subsubsection{Sodium-Iridate-Hubbard Model}
\label{sec:SIH}

\begin{figure}[t!]
\centering
\includegraphics[scale=0.52]{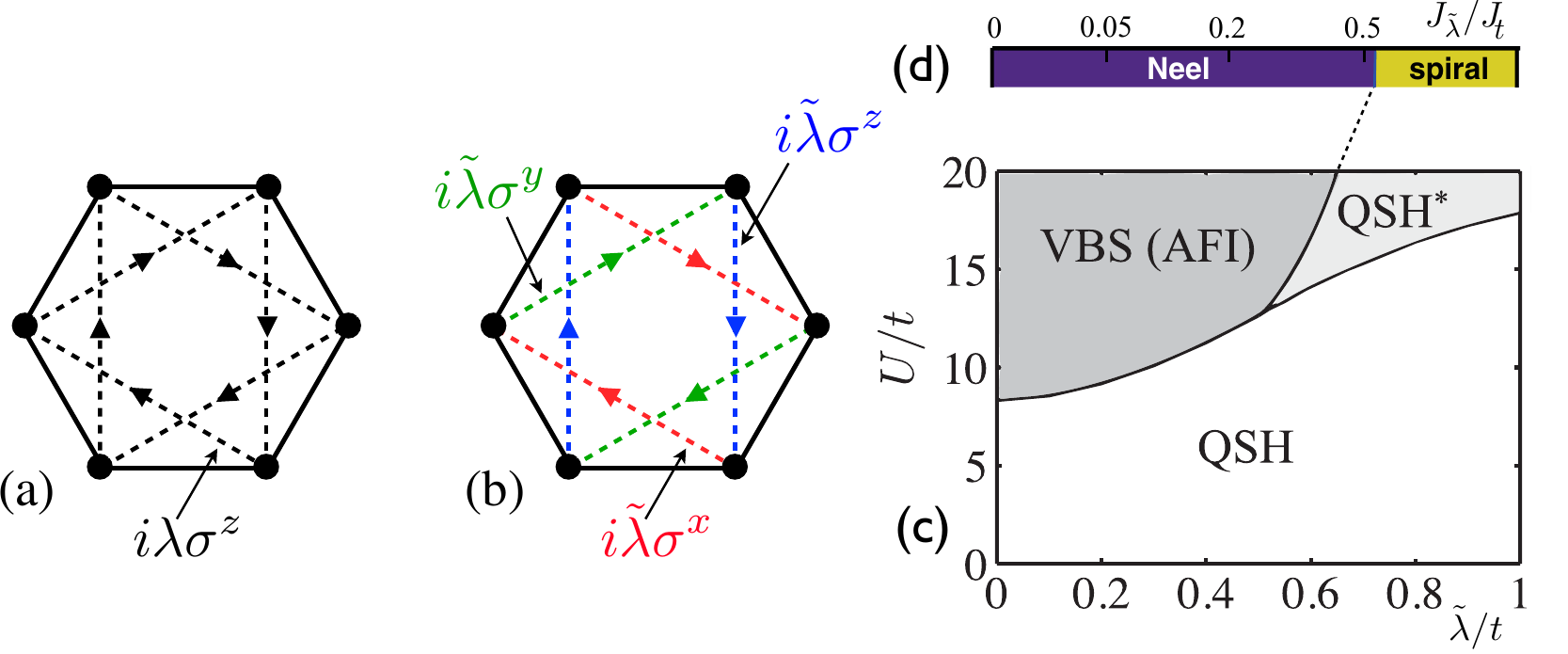} 
\caption{\small (a) Illustration of the nearest-neighbor hopping (solid lines) and second-neighbor spin-orbit hopping (dashed lines) with amplitude $i\lambda$ for the KM model. (b) Different second-neighbor bonds involve different Pauli matrices for the SIH model with amplitude $i\tilde\lambda$. (c) Phase diagram of the SIH model obtained within slave-spin method. Besides the QSH (TI) phase, an antiferromagnet (VBS / AFI) and the putative QSH$^\ast$ phase are present; see main text for details.
Reprinted with permission from \cite{ruegg-12prl046401}. Copyright (2012) by the American Physical Society.
(d) Phase diagram of the strong-coupling spin limit of the same model\,\cite{reuther-12prb155127}, with an AFM Neel phase at weak $\tilde\lambda$ and magnetic spiral phase at large $\tilde\lambda$. Remarkably, the phase boundary of the phase diagram in (c) is compatible with the one of the corresponding spin model in (d); magnetic exchange couplings are given by $J_x = 4x^2/U$ with $x=\tilde\lambda, t$. For details see main text. Panels (a), (b), and (d) are taken from\,\cite{reuther-12prb155127}.
}
\label{fig:shitade}
\end{figure}

Kane and Mele proposed the quantum spin Hall effect originally for graphene\,\cite{kane-05prl226801}. Subsequent {\it ab initio} calculations clarified, however, that the spin-orbit gap is far too small to be observed\,\cite{min-06prb165310,yao-07prb041401}. Ever since then material scientists searched for other honeycomb-lattice materials with possibly heavier elements. One of the first proposals along these lines is the work by Shitade\,\ea\,\cite{shitade-09prl256403} in which Na$_2$IrO$_3$, a $5d$ transition metal oxide, is proposed as a layered correlated QSH insulator. Based on {\it ab initio} calculations an effective hopping model was derived which is reminiscent of the Kane-Mele model. Instead of isotropic SOC with amplitude $i\lambda \sigma^z$ [for an illustration see Fig.\,\ref{fig:shitade}\,(a)] the three inequivalent second-neighbor hopping directions involve different Pauli matrices, see Fig.\,\ref{fig:shitade}\,(b). The band structure realizes a $\mathbb{Z}_2$ TI\,\cite{shitade-09prl256403} being in the same universality class as the Kane-Mele model; \ie when interpolating between both band structures the gap will not close and edge states (in case of a disc or cylinder geometry) will persist\,\cite{rachel16jpcm405502}. The authors of Ref.\,\cite{shitade-09prl256403} argued that a correlated TI phase is present if Coulomb interactions are not too strong. The experimental status of Na$_2$IrO$_3$ will be briefly discussed in Sec.\,\ref{sec:iridates}.

Within a {\it slave-spin approach}, R\"uegg and Fiete investigated the aforementioned band structure supplemented with a Hubbard interaction, in the following referred to as {\it sodium--iridate Hubbard} (SIH) model\,\cite{ruegg-12prl046401}. Its Hamiltonian is given by
\begin{equation}\label{ham:sih}
\begin{split}
H_{\rm SIH} =&   -t \sum_{\langle ij \rangle} \sum_{\sigma} c_{i\sigma}^\dag c_{j\sigma}^\pd  + i\tilde\lambda \sum_{\langle\!\langle ij \rangle\!\rangle_\gamma} \sum_{\alpha\beta=\up,\dw} c_{i,\alpha}^\dag  \,\nu_{ij} \, s^\gamma_{\alpha\beta}\, c_{j\beta}^\pd \\
&+U \sum_i n_{i\up} n_{i\dw}
\end{split}
\end{equation}
and $\gamma=x,y,z$ as illustrated in Fig.\,\ref{fig:shitade}\,(b). Most notably, the corresponding slave-spin analysis revealed for sufficiently large SOC $\tilde\lambda$ and Hubbard $U$ an additional phase, dubbed QSH$^\star$, which is absent in the KMH phase diagram. The phase diagram of the SIH model\,\cite{ruegg-12prl046401}
 is shown in Fig.\,\ref{fig:shitade}\,(c); therein, QSH refers to the quantum spin Hall phase (the 2D TI) and VBS (AFI) to ``valence bond solid (antiferromagnetic insulator)''. Due to a limitation of the used slave-spin approach, magnetic solutions cannot be found; instead, a valence bond solid is found as the state which comes energetically closest to the antiferromagnet. In other methods, this phase would clearly show up as the Neel ordered antiferromagnet.
The time-reversal invariant QSH$^\star$ phase possesses topological order associated with a topological degeneracy on a torus geometry. It was further claimed to be described by the same field theory than the toric code\,\cite{kitaev03ap2}; excitations were identified as Abelian anyons\,\cite{ruegg-12prl046401}. In Ref.\,\cite{zhong-13prb045109}, an exactly soluble model was proposed which features a fractionalized phase which is comparable to the QSH$^\star$ phase. 
It is worth emphasizing that one can interpret them also as an {\it orthogonal QSH} phase in analogy to the {\it orthogonal metal} phase\,\cite{nandkishore-12prb045128,zhong-12prb165134}. The orthogonal metal has the same transport and thermal signatures (\ie two-particle responses) as standard Fermi liquids but their single-particle spectrum is gapped. It seems that the slave-spin approach naturally generates phases such as QSH$^\star$ or CI$^\star$ (see Sec.\,\ref{sec:corr-CI}) which can be considered as ``orthogonal'' phases.

It turned out to be challenging to verify or further investigate the exotic QSH$^\star$ phase because one has to do nothing less than solving the 2D Hubbard model with broken spin symmetry at intermediate interaction strength. Due to the topological order, most mean-field type approaches cannot be used. At least there were two successful attempts to encircle the predicted QSH$^\star$ phase in the phase diagram: (i) by deriving the strong coupling limit, the ground state of the corresponding spin Hamiltonian has been solved\,\cite{reuther-12prb155127}; (ii) by deriving the limit of strong spin-orbit coupling the SIH model effectively decouples into two copies of triangular lattice Hubbard models with a peculiar band structure\,\cite{rachel-15prl167201}.

The intrinsic SOC as illustrated in Fig.\,\ref{fig:shitade}\,(b) gives rise to the same spin exchange for the vertical bonds as in the KMH case ($\propto\sigma^z$), but the red bonds ($\propto\sigma^x$) lead to $S_i^x S_j^x - S_i^y S_j^y - S_i^z S_j^z$ and the green bonds ($\propto\sigma^y$)  to $-S_i^x S_j^x +S_i^y S_j^y - S_i^z S_j^z$. In a compact form one can write the sodium-iridate spin Hamiltonian as
\begin{equation}\label{SI-SOC-spin}
\mathcal{H} = \frac{4t^2}{U}\sum_{\langle ij \rangle} \bs{S}_i \bs{S}_j -
\sum_{\langle\!\langle ij \rangle\!\rangle} \frac{{\tilde\lambda}^2}{4U} \bs{S}_i \bs{S}_j + 2  \sum_{\langle\!\langle ij \rangle\!\rangle_\gamma}  \frac{{\tilde\lambda}^2}{4U} S_i^\gamma S_j^\gamma\ .
\end{equation}
Note that $\gamma$ is defined as for $\mathcal{H}_{\rm SIH}$; the bond-dependent Ising exchange $S^\gamma_i S^\gamma_j$ is referred to as compass interactions\,\cite{nussinov-15rmp1} or Kitaev exchange\,\cite{Kitaev06ap2}. Both the antiferromagnetic nearest neighbor and the ferromagnetic second-neighbor exchange stabilizes Neel order on the honeycomb lattice. The non-trivial second-neighbor Kitaev exchange eventually turns the Neel state into an incommensurate spiral ordered state for sufficiently large $\tilde\lambda$\,\cite{reuther-12prb155127}. Interestingly, this magnetic phase diagram fits nicely to the mean-field phase boundaries of Ref.\,\cite{ruegg-12prl046401}, see the dashed line between panels (c) and (d); the combination of both phase diagrams is shown in Fig.\,\ref{fig:shitade}\,(c) and (d).

Taking the limit of infinitely large SOC is more subtile as it is equivalent to setting the nearest-neighbor hoppings to zero, $t\to 0$. That is, the honeycomb lattice decouples into two independent triangular lattices.
The ``new'' nearest--neighbor hopping on such a decoupled triangular lattice is governed by the Hamiltonian
\begin{equation}\label{imaginay-piflux}
\mathcal{H}_\triangle = i \lambda \sum_{\langle ij \rangle_\gamma} c_{i\alpha}^\dag \sigma_{\alpha\beta}^\gamma c_{j\beta}^\pd\ .
\end{equation}
In order to derive a more intuitive understanding it is helpful to apply Klein transformations to the creation and annihilation operators\,\cite{rachel-15prl167201}; Klein transformations are usually applied to spin operators\,\cite{chaloupka-10prl027204,kimchi-14prb014414,reuther-12prb155127}.

For a Klein map, the lattice is divided into four sublattices. All creation and annihilation operators defined on one of the sites remain untouched. The other three type of sites are transformed such that their spin component is rotated around the $x$, $y$, or $z$ axes by $\pi$. Surprisingly, the formerly spin- and bond-dependent imaginary hopping problem described by \eqref{imaginay-piflux} appears to be real and spin-independent after the Klein map. The difference to an ordinary real tight-binding model on the triangular lattice are a few randomly distributed minus-signs. Rewriting them in terms of $\pi$-flux Peierls phases reveals the underlying flux patterns consisting of alternating triangles threaded by $\pi$ and $0$ flux. Now an appropriate unit cell is easily chosen and the band structure computed: its low-energy theory is a 2+1 dimensional Dirac theory, very similar to graphene.

A real spin-independent hopping problem with hoppings $\pm t$ results in the strong coupling limit to isotropic Heisenberg spin exchange because $J=4(\pm t)^2/U$ is independent of the sign of $t$. The corresponding ground state, although frustrated, is known to 
be the 120-degree Neel state. This situation is  quite interesting: a semi-metallic groundstate in the weak coupling, and a frustrated magnetically ordered groundstate in the strong coupling regime. In order to reveal the nature of the intermediate region of the phase diagram, a {\it Variational Cluster Approach} study suggested that a non-magnetic insulator phase is present\,\cite{rachel-15prl167201}. The nature of this quantum paramagnetic phase was claimed to be a spin liquid because the energy gain due to the formation of short-range resonating valence bonds\,\cite{anderson73mrb153} was much higher compared to the ordinary triangular lattice Hubbard model. Since the latter is believed to host a spin liquid phase the same should be true for the triangular lattice $\pi$-flux Hubbard model\,\cite{rachel-15prl167201}. 

In order to close the ``loop'' between the $U\to\infty$ and the $\lambda\to\infty$ result, one has to apply the same Klein map to the strong coupling groundstate and transform backwards. The 120 degree Neel state is transformed into a commensurate spiral ordered state with a 12-site unit cell; by weakly coupling two triangular lattices into a honeycomb lattice where each sublattice hosts such a spiral state, the spiral order becomes incommensurate as previously found within pseudo-fermion functional RG\,\cite{reuther-12prb155127}. 
Still the question about possible phases at intermediate $U$ and intermediate spin-orbit coupling remains an open issue. Whether or not the QSH$^\ast$ phase\,\cite{ruegg-12prl046401} is realized remains an open question. Alternatively, it might be a magnetically ordered spiral  phase\,\cite{liu-13prb245119,reuther-12prb155127}) or there could even multiple phases  be hidden.

%
%
\subsubsection{Bernevig-Hughes-Zhang-Hubbard Model}
\label{sec:BHZH}

The BHZ model\,\cite{bernevig-06s1757} plays an important role as it effectively describes the physics of the HgTe quantum wells\,\cite{koenig-07s766}, the first (and for a long time only) physical system displaying the QSH effect. On the other side, HgTe/CdTe quantum wells are a heterostructure built of different materials which all realize a 3D zinc blende structure. Using $\bs{k}\cdot\bs{p}$ theory one can show that the BHZ model is indeed the true low-energy theory. In order to study interaction effects one need to regularize this model, usually it is done on a two-orbital square lattice.
The BHZ model can be thought of as a spinful extension of the two-orbital square lattice Chern insulator (Sec.\,\ref{sec:CI}).
The Hamiltonian can simply be written using the spinor $\Psi^\dag = \left( c_{E,\up}^\dag, c_{H,\up}^\dag, c_{E,\dw}^\dag, c_{H,\dw}^\dag\right)$ as
\begin{equation}
\label{ham:BHZ}
\mathcal{H}_{\rm BHZ} = \sum_{\bs{k}}  \Psi^\dag \!\!\left(\begin{array}{c|c}
h(\bs{k})_{2{\rm -orb.}}&
\begin{array}{cc}&~~~~-\Delta~~~~\\[5pt]~~~~\Delta~~~~&\end{array}
\\[10pt]
\hline
&\\[-8pt]
\begin{array}{cc}
&~~~~\Delta~~~~\\[5pt]~~~~-\Delta~~~~&\end{array}
&h^\star(\bs{-k})_{2{\rm -orb.}}
\end{array}\right) \!\!\Psi^\pd
\end{equation}
where $h(\bs{k})_{2{\rm -orb.}}$ is the $2\times 2$ Bloch matrix earlier introduced in Eq.\,\ref{bloch:ci-squarelattice} and $\Delta$ quantifies the bulk inversion asymmetry\,\cite{bernevig-06s1757,qi-11rmp1057}. A real space formulation of \eqref{ham:BHZ} can be found, for instance, in Ref.\,\cite{rod-15prb245112}.

This square lattice version of the BHZ model is then a good starting point to investigate the role of Hubbard interactions; it is, however, not directly relevant anymore for the HgTe/CdTe quantum wells. Nonetheless several authors studied the interplay of topology and interactions in this system\,\cite{yoshida-12prb125113,yoshida-13prb085134,miyakoshi-13prb195133,tada-12prb165138,budich-13prb235104}. While some of the basic results are similar to those of the KMH model, for instance that the TI phase breaks down when a spontaneous magnetization sets in, the BHZ Hubbard model features a first-order phase transition\,\cite{amaricci-15prl185701} not known for other topological Hubbard models. 
Moreover, the orbital structure nicely allows to study the effect of Hunds-coupling and intra-orbital interactions in addition to the inter-orbital Hubbard interactions\,\cite{budich-13prb235104,amaricci-15prl185701}:
\begin{equation}
H_{\rm int} = H_U + H_V + H_J
\end{equation}
with the inter-orbital Hubbard term, $H_U = U\sum_i ( n_{i,\up}^{(E)} n_{i,\dw}^{(E)} + n_{i,\up}^{(H)} n_{i,\dw}^{(H)})$, intra-orbital repulsion $H_V = V\sum_i ( n_{i,\up}^{(E)} n_{i,\dw}^{(H)} + n_{i,\up}^{(H)} n_{i,\dw}^{(E)})$, and the Hund's coupling $H_J = (V-J)\sum_i ( n_{i,\up}^{(H)} n_{i,\up}^{(E)} + n_{i,\dw}^{(H)} n_{i,\dw}^{(E)})$. These additional interaction terms do not qualitatively change the general competition between topological weak-coupling regime and the strongly coupled Mott regime, but clearly they lead to a richer phenomenology. Ultimately, for a multi-orbital scenario such as in the BHZ model these terms will be present in real materials. The paramagnetic $m$-$U$ phase diagram of the BHZ Hubbard model with $H_{\rm int}$ is shown in Fig.\,\ref{fig:budich} in Sec.\,\ref{sec:IITI} for the parameters $V=U-2J$, $J=0.25 U$ and $t=0.3$.
%

%
%
\subsubsection{Hofstadter-Hubbard Model}
\label{sec:TRI-HH}

The role of interactions in the BHZ model becomes relevant in a completely different branch of physics:
ultracold quantum gases and optical lattices instead of semiconductors.
The tremendous progress to realize synthetic magnetic fields and SOC\,\cite{dalibard-11rmp1523,hauke-12prl145301,aidelsburger-13prl185301,kennedy-13prl225301} for ultracold atoms made in the past years  allows to stabilize topological states of matter. 
Indeed, the Haldane Chern insulator model has recently been realized in an optical lattice\,\cite{jotzu-14n237}. Most ultracold atom systems have in common that a local particle-particle interaction can be tuned and controlled. With the help of a Feshbach resonance, Hubbard-$U$ can even be tuned from the attractive to the repulsive regime. Thus the combination of synthetic gauge fields and arbitrary interactions makes ultracold quantum gases predestined as a playground for interacting topological insulators.

Here we will briefly discuss a specific cold atom proposal\,\cite{goldman-10prl255302} where a spinful and time-reversal invariant Hofstadter system is discussed. The standard Hofstadter problem realized the QHE for spinless fermions on the square lattice as discussed in Sec.\,\ref{sec:TI}. When the vector potential is created artificially, it is possible to choose the vector potential for both spin species individually. In particular, they can be chosen to  possess opposite sign thus realizing a TR invariant magnetic field. The resulting state is a $\mathbb{Z}_2$ TI exhibiting Hofstadter bands which can be interpreted as Landau levels in the limit of small magnetic field $\alpha=p/q$. While this is expected for values of the Fermi energy within all ``Hofstadter-gaps'' with an odd Chern number per spin channel\footnote{Even values of the Chern number would result for the spinful case  in an even number of pairs of helical edge states. According to the Kane-Mele classification such insulators are topologically trivial, see Sec.\,\ref{sec:Z2TI}.}, additional staggered sublattice potential and a Rashba-type spin-orbit hopping make it possible to obtain a QSH phase even at half filling\,\cite{goldman-10prl255302}. The Hamiltonian can be defined on a square lattice as
%
\begin{equation}
\label{ham:HH}
\begin{split}
&H_{\rm TRI-HH} = \lambda \sum_j (-1)^{j_x} c_j^\dag c_j^\pd \\
&- \sum_j \Big( t c_{j+\hat e_x}^\dag e^{2\pi i \gamma \sigma^x} c_j^\pd + t c_{j+\hat e_y}^\dag e^{2\pi i \alpha j_x \sigma^z} c_j^\pd + {\rm H.c.}\Big)
\end{split}
\end{equation}
describing spin-1/2 fermions subject to a synthetic gauge field. The $\sigma^z$ Pauli matrix in the second term corresponds to the spin-dependent field and guarantees TR invariance. The first term induces spin flips  if the particles move along the $x$-direction. The $\lambda$ term causes a staggering of the lattice in the $x$ direction. Most importantly, all terms in \eqref{ham:HH} can be experimentally realized. This non-interacting model exhibits trivial and topological insulator phases as well as metallic regimes, depending on the filling.

Together with local Coulomb interactions the rich phase diagram has been explored\,\cite{cocks-12prl205303,orth-13jpb134004}. In particular, a variety of magnetically ordered phases has been found including different spiral states. In Ref.\,\cite{scheurer-15sr8386} it has been shown that an  anisotropic version of the BHZ model emerges as the low-energy $\bs{k}\cdot\bs{p}$ theory of this time-reversal invariant Hofstadter model at half filling. 

Furthermore, a 3D stacked version of this model can be tuned such that weak and strong TI phases  emerge out of the critical point between normal and topological insulator phases\,\cite{scheurer-15sr8386} when hoppings in $z$ direction are turned on.
Requirement is, however, that these hoppings in the third direction are attached to synthetic gauge fields.
This setup is later discussed as the starting point for the investigation of the {\it topological Mott insulator} phase in a cubic lattice, see Sec.\,\ref{sec:TMI}.

%
%
\subsubsection{Higher-dimensional Correlated Topological Insulators}
\label{sec:3D-corrTI}

Compared to the two-dimensional case, investigations of interaction effects in 3D bandstructures are rather rare. Notable exceptions are the interesting  studies on the pyrochlore lattice (partially motivated by pyrochlore oxides and by the search for topological Mott insulators)\,\cite{go-12prl066401,wan-11prb205101}. As a consequence of electron-electron interactions, topological Weyl semimetals were predicted but also the more elusive axion insulator phase -- essentially a 3D strong TI with broken TR symmetry -- were found.

The effect of Coulomb interactions and Hund's exchange coupling was studied in a cubic bandstructure with strong and weak TI phases\,\cite{amaricci-16prb235112}. 

Most other works are focussed on interaction-induced topological phases such as the topological Kondo insulators (as discussed below). Other examples of interaction-induced topological phases are discussed in Refs.\,\cite{zhang-09prb245331,kurita-11jpsp044708}. And a linear response theory of interacting topological insulators was derived in Ref.\,\cite{culcer-11prb235411}.

Recent developments in the field of ultracold quantum gases in optical lattices has established the possibility of so-called ``synthetic'' dimensions\,\cite{lohse-18n55}. This led to the realization of the four-dimensional integer quantum Hall effect\,\cite{kraus-13prl226401,lohse-18n55}. 
The idea is that  an internal  degree of freedom is used as if it was another spatial degree of freedom. This generally allows to investigate high-dimensional systems and 
led to the recent study of interacting topological insulators in $d$ dimensional systems\,\cite{jian-18arXiv1804.03658}.

%
%
\subsection{Interaction-Induced Topological Insulators}
\label{sec:IITI}

In this section, we briefly consider systems where electron--electron interactions induce a topologically non-trivial phase. In the following we discuss three scenarios where this is realized. To begin with, we illustrate the simplest scenario possible. Consider the Kane-Mele model with staggered sublattice potential $H_v$ as defined in Eq.\,\eqref{semenoff+rashba}. The phase diagram coincides with the one of the Haldane model for $\phi=\pm\pi/2$, Fig.\,\ref{fig:CI-phasediagrams}, except that both CI phases are replaced by QSH phases. For sufficiently large $\lambda_v$ the system is in a trivial insulating phase and the majority of particles occupies $B$ sites rather than $A$ sites. At half filling, that implies that several $A$ sites are doubly occupied while $B$ sites are empty. 

\begin{figure}[t!]
\centering
\includegraphics[scale=0.89]{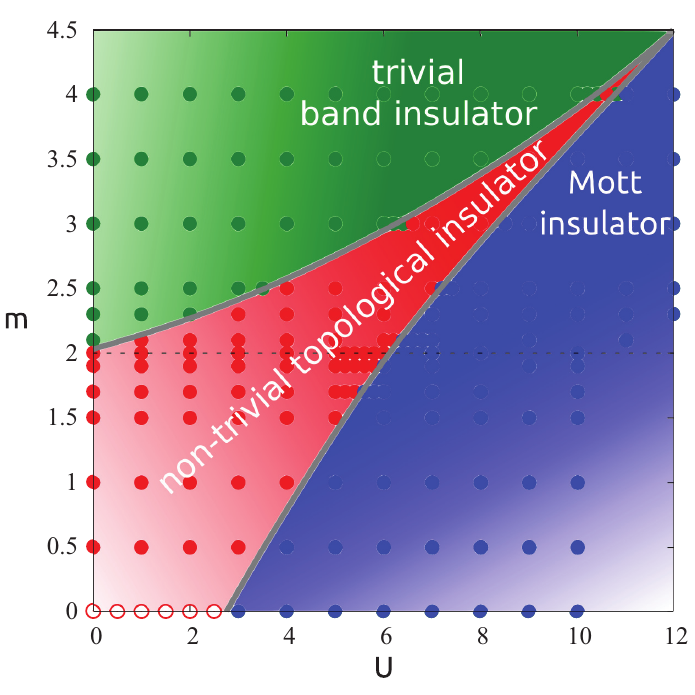}
\caption{\small Phase diagram of the BHZ Hubbard model with additional Hunds coupling and intra-orbital repulsion. The balance between $m$ causing a trivial band insulator and Hubbard $U$ causing a Mott insulator leads to the survival of the topological band insulator up to rather large values of $U$.
Reprinted with permission from \cite{budich-13prb235104}. Copyright (2013) by the American Physical Society.
}
\label{fig:budich}
\end{figure}

Now let us add local Coulomb interactions \eqref{hubbard}: as a consequence, doubly occupied sites cost an energy penalty $U$ while singly occupied and empty sites do not. If $U$ is sufficiently large, it will reverse the effect of $\lambda_v$ (for the sake of simplicity, we ignore the possible formation of magnetism due to $U$). Reversing the effect of $\lambda_v$ drives the ground state back into the topologically non-trivial phase: this is the simplest example of an interaction-induced topological phase. Similarly, Hubbard-$U$ can ``reverse'' or renormalize the ``mass'' term of the BHZ model; simultaneous increase of the mass term $m$ leading to the orbital imbalance and Hubbard $U$ forcing equal orbital occupancy can lead to a stable balance up to large energy scales. This has been explicitely shown for the BHZ Hubbard model (already discussed in Sec.\,\ref{sec:BHZH}), see Fig.\,\ref{fig:budich}\,\cite{budich-13prb235104}.

%
%
\subsubsection{Fluctuation-Induced Topological Phases}
\label{sec:renormalize}

\begin{figure}[b!]
\centering
\includegraphics[scale=0.81]{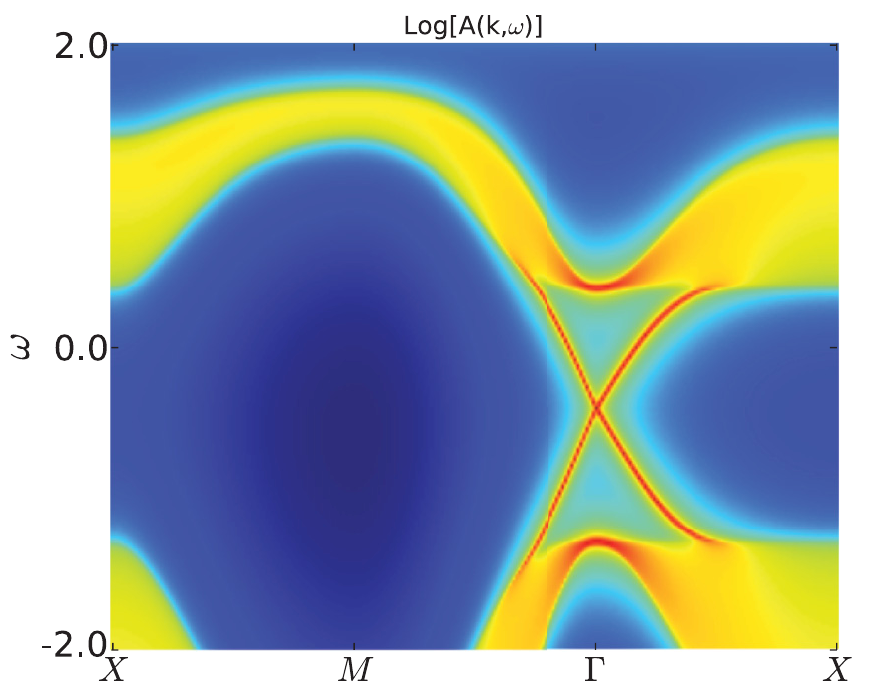}
\caption{\small The 3D bandstructure, originally prepared in a topologically trivial phase, is subject to an $\omega$-dependent self-energy; the resulting system is a strong TI. Shown is the corresponding surface spectral function with the chemical potential at zero. 
Reprinted with permission from \cite{wang-12prb235135}. Copyright (2012) by the American Physical Society.
}
\label{fig:fluct-induced}
\end{figure}

Quite generally, Coulomb interactions can ``tune'' topological insulator band structures from the trivial to the topological domain. This can already be understood in simplified mean-field pictures. Not only a ``true'' self-energy $\Sigma(\omega, \bs{k})$, but in principle also $\bs{k}$--dependent (static mean-fields) or purely $\omega$--dependent self-energies (fluctuation induced) can be responsible\,\cite{budich-12prb201407,wang-12epl57001,wang-12prb235135}. 
Examples of the former case\,\cite{cocks-12prl205303,orth-13jpb134004} but, in particular, also of the latter are nicely demonstrated within DMFT studies\,\cite{wang-12epl57001,wang-12prb235135}. 
Using the insight the additional poles in the self-energy can change the topology of the system's ground state motivated several works to study fluctuation-induced topological phases as an application of the pole-expansion method. A typical self-energy is given by $\Sigma(\omega) = \frac{V^2}{i\omega+P} + \frac{V^2}{i\omega- P}$ with $V$, $P$ being free parameters. It has been shown\,\cite{wang-12prb235135} that the topological index is trivial for $P=0$ and non-trivial otherwise. In Fig.\,\ref{fig:fluct-induced} the surface spectral function of a strong TI is shown for the case $V=1$ and $P=2$ of a system which was initially prepared in a trivial phase.

A different example of a topologically trivial system which enters a topological Chern insulator phase due to interactions has recently been discussed in the context of a large-$N$ exactly solvable model\,\cite{zhang-18arXiv1803.01411}.

%
%
\subsubsection{Topological Phases through Spontaneous Symmetry Breaking}
\label{sec:raghu}

A completely different scenario of interaction induced topological phases was put forward by Raghu \ea\,\cite{raghu-08prl156401}\footnote{The authors named their system ``topological Mott insulator'' but the reader should note that it is very different from the topological Mott insulators discussed in Sec.\,\ref{sec:TMI}. Instead, it corresponds to a state having a non-interacting analog.}. Additional longer-ranged Coulomb interactions shall spontaneously generate second-neighbor terms which either break TR symmetry (corresponding to a (doubled) Haldane term for spinless (spinful) fermions) or SU(2) spin symmetry (corresponding to the Kane-Mele term for spinful fermions only). In Ref.\,\cite{raghu-08prl156401} both nearest and next-nearest neighbor repulsions were considered, but the main player clearly is  the second-neighbor repulsion,
\begin{equation}
\mathcal{H}_{{I}}^{(2)} = V_2 \sum_{\langle\!\langle ij \rangle\!\rangle}  (n_i -1)(n_j-1)\ ,
\end{equation}
where for spinful fermions $n_i = n_{i\up} + n_{i\dw}$. For the spinful case, one can define the operator 
$\chi_{ij}^\mu = c_{i\alpha}^\dag \sigma_{\alpha\beta}^\mu c_{j\beta}^\pd$, $\mu=0,1,2,3$ with $\sigma^\mu = (\mathbf{1}, \bs{\sigma})$. The $\chi$ operators are related to the $V_2$ term by the identity
\begin{equation}
(n_i -1)(n_j-1) = 1 - \frac{1}{2} \left(\chi_{ij}^\mu\right)^\dag \chi_{ij}^\mu\ .
\end{equation}
A standard mean-field decoupling leads to order parameters $\langle\chi^0\rangle$ (Chern insulator phase) and $\langle\chi^i\rangle$, $i=1,2,3$ (QSH phase). Raghu\,\ea\,\cite{raghu-08prl156401} showed using mean-field and functional RG methods that for sufficiently large $V_2$ a topological phase is stabilized with a slight tendency towards the QSH phase. 

\begin{figure}[t!]
\centering
\includegraphics[scale=0.44]{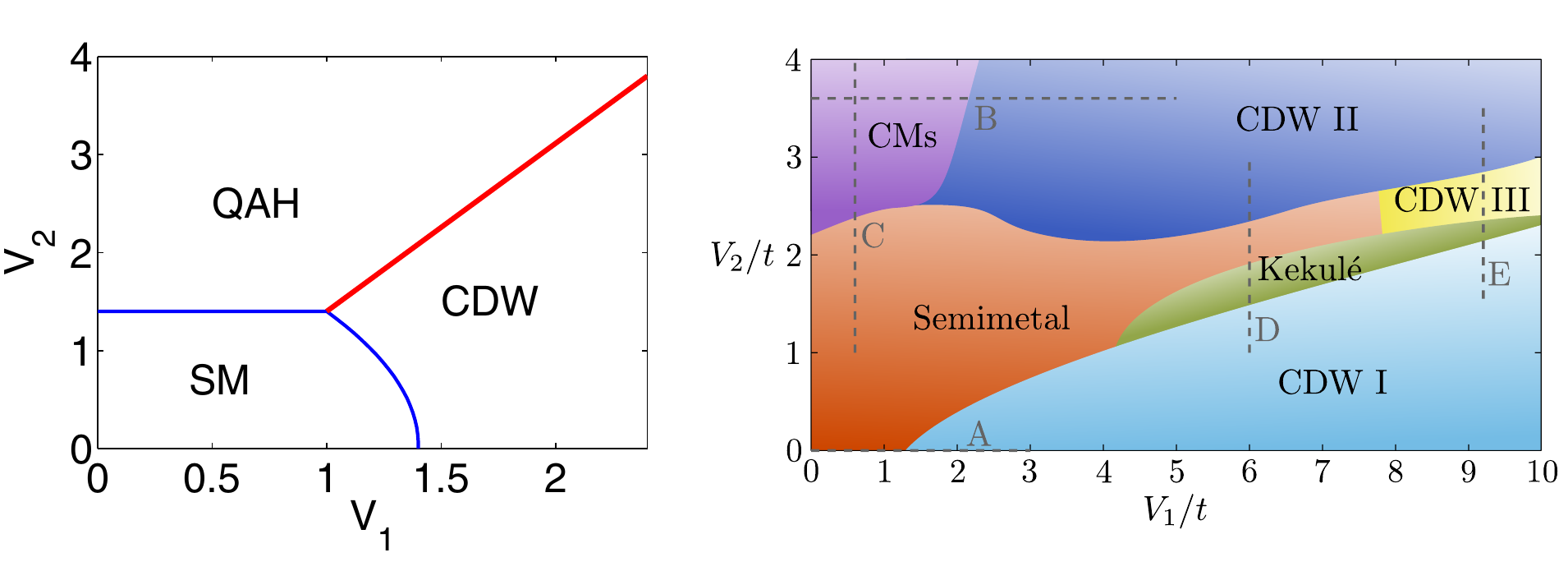}
\caption{\small (Left) Mean-field phase diagram for spinless fermions on the honeycomb lattice as originally proposed containing interaction-induced Chern insulator (QAH) and charge-ordered phases (CDW).
Reprinted with permission from \cite{raghu-08prl156401}. Copyright (2008) by the American Physical Society.
(Right) Density-matrix renormalization group phase diagram for the same model containing a variety of charge-ordered and modulated phases but not a topological phase.
Reprinted with permission from \cite{motruk-15prb085147}. Copyright (2015) by the American Physical Society.
}
\label{fig:raghu}
\end{figure}

For the spinless case, a similar procedure led to the proposal that a Chern insulator phase should be realized when $V_2$ is sufficiently large\,\cite{raghu-08prl156401}, see the original phase diagram in Fig.\,\ref{fig:raghu}\,(left). While the spinful problem has been too challenging to be attacked by numerical methods so far, many authors studied the spinless case\,\cite{wen-10prb075125,weeks-10prb085105,daghofer-14prb035103,garcia-13prb245123,grushin-13prb085136,duric-14prb165123,motruk-15prb085147,capponi-15prb085146,liu-16prb195153,venderbos-16prb195126,dauphin-12pra053618}. In most of these works a topological Chern insulator phase has not been found. Instead, charge density wave orders were stabilized. For instance, the phase diagram obtained within the density matrix renormalization group is shown in Fig.\,\ref{fig:raghu}\,(right); a topological phase is absent. 
For details and references see the recent review\,\cite{capponi1609.01161}. An analogous study for spinless Dirac fermions on the $\pi$-flux (square) lattice led to comparable results: while a topological phase was present within the mean-field treatment, exact diagonalization found charge order instead\,\cite{jia-13prb075101}. The Lieb lattice represents a third setup where the Raghu-idea can be tested; it might have the advantage that a cold-atom realization seems to be feasible\,\cite{dauphin-16pra043611}.

A somewhat analogous idea to generate  topological phases by opening the gap at a ``Dirac crossing'' has been proposed by Sun \ea\,\cite{sun-09prl046811}: a quadratic band crossing (QBC) point instead has the major advantage of a drastically increased density of states. Short-range repulsive interactions turn out to be marginally relevant in the renormalization group sense, \ie that the QBC point will acquire a gap even for arbitrarily weak interactions. Such weak interactions will be unable to generate charge order as it happens for the case of Dirac fermions. In the spinless case, a quantum anomalous Hall phase was found; in the spinful case, both quantum anomalous Hall and QSH phases can be realized. In addition, also nematic phases might  emerge for stronger interactions\,\cite{sun-09prl046811}. All these predictions are in agreement with subsequent analytical and numerical work\,\cite{ueblacker-11prb205122,vafek-10prb205106}. 
While the Chern insulator phase could never been detected within exact diagonalization or density matrix renormalization group for the case of Dirac crossings (\eg on the honeycomb lattice)\,\cite{motruk-15prb085147,capponi-15prb085146}, recently this has been accomplished for the case of QBC points on the kagome\,\cite{zhu-16prl096402} and on the checkerboard lattice\,\cite{wu-16prl066403,zeng-18arXiv1805.01101}. For the latter, also an intermediate bond-ordered phase sandwiched between the Chern insulator and nematic charge ordered phase has been found\,\cite{zeng-18arXiv1805.01101}.

One should not conclude, however, that the Raghu proposal is wrong; it rather seems that the honeycomb lattice with its linear density of states is a difficult terrain to search for this kind of physics.
The success in finding the interaction-induced phase for the case of QBC points demonstrates that the intuition of Raghu \ea\,\cite{raghu-08prl156401} was correct, only the energetics on the honeycomb lattice were unfavorable to stabilize the topological phase by virtue of interactions. As a side remark, it might be interesting to consider  the Raghu proposal for spinless fermions on an anisotropic honeycomb lattice: lattice anisotropies could be chosen such that the charge density wave order becomes energetically less favorable. It would be interesting to test whether this is sufficient to eventually stabilize the topological phase.

Ultimately, it would be desirable to probe the physical mechanism elucidated in this subsection experimentally.
Fermionic optical lattices might provide the proper environment to test these predictions experimentally\,\cite{sun-12np67}. Motivated by such proposals, it was realized that QBC points can also lead to interaction-driven topological phases of bosons: the bands which form the QBC point exhibit both non-negative curvature. Onsite-interactions induce a quantum anomalous Hall phase for the Bogoluibov quasiparticle spectrum  characterized by a finite Chern number\,\cite{diliberto-16prl163001}.

Other systems where interaction-induced topological phases were proposed are certain heterostructures, so-called digital transition metal oxide heterostructures\,\cite{xiao-11nc596,ruegg-11prb201103,ruegg-12prb245131}. The involved transition metal oxides have a cubic environment but are grown in the (111) direction which results in a ``buckled'' honeycomb net. The interplay of the complex orbital order and electron-electron interactions leads to either Chern insulator or topological insulator phases if interactions are not too strong. Otherwise Mott-type physics dominates\,\cite{okamoto-14prb195121} which might even lead to spin liquid groundstates\,\cite{okamoto13prl066403}.

%
%
\subsubsection{Topological Kondo Insulators}
\label{sec:TKI}

The topological Kondo insulator\,\cite{dzero-10prl106408,alexandrov-13prl226403,dzero-16arcmp249} has been proposed to be realized in SmB$_6$ and is the only interaction-induced topological insulator phase present in a real material so far. 
Other material proposals such as YbB$_{12}$ have been suggested\,\cite{hagiwara-16nc12690,weng-14prl016403}. For a detailed review about topological Kondo insulators we refer the reader to\,\cite{dzero-16arcmp249,sun-17rpp112501}.
Note that the Kondo effect\,\cite{kondo64ptp37} is an interaction-driven low temperature phenomenon. The so-called Kondo insulators or {\it heavy fermion} semiconductors are materials with a narrow band gap which is caused by the Kondo effect, the hybridization between conduction electrons and (localized) $f$ electrons. 

The main idea of topological Kondo insulators is explained in the following. Kondo insulators such as SmB$_6$ exist for more than 50 years and some of them are well-studied. The difference between a conventional and topological Kondo insulator can be summarized in two main points: (i) the relevance of strong spin-orbit coupling of $f$ electrons (much larger than the typical Kondo gap) and (ii) the different parity of $f$ and $d$ electrons, where the latter exhibits odd and the former even parity. That is, each time there is a band crossing between $f$ and $d$ electrons the $\mathbb{Z}_2$ index changes, which leads under the right circumstances to a TI phase.

The experimental situation is far more complicated; alternatively, one could say that the topological insulator proposal for SmB$_6$ has proven to be sufficiently exciting that it has stimulated a variety of experiments performed by many groups. While several experiments are compatible with or proof for the topological insulator picture, others are not. In the following, we only mention a few relevant experiments; a detailed discussion is given in the reviews by Dzero \ea\,\cite{dzero-16arcmp249} and by Sun and Wu\,\cite{sun-17rpp112501}; note that the latter proposes the scenario of an ``accompany-type valence fluctuation state'' which possibly coexists with the Kondo ground state of SmB$_6$.
Large single crystals of SmB$_6$ are available\,\cite{hatnean-sr3071} and many experiments find evidence of a surface state at low temperatures: combined ARPES and {\it ab initio} studies\,\cite{kim-13sr3150,wolgast-13prb180405,zhu-13prl216402,neupane-13nc2991,frantzeskakis-13prx041024,jiang-13nc3010}, point-contact spectroscopy\,\cite{zhang-13prx011011}, thickness-dependent transport measurements\,\cite{kim-13nm466}, magnetoresistance measurements\,\cite{chen-15prb205133}, and -- possibly most convincingly -- quantum oscillation measurements\,\cite{tan-15s287}. 2D Fermi surfaces have also been observed using torque magnetometry\,\cite{li-14s1208,xiang-17prx031054}.
The spin-polarized structure of the surface states was detected\,\cite{xu-14nc4566} being in good agreement with topological surface states. One of the obstacles are the low temperatures $T<4$\,K which are required to observe the surface states\,\cite{kim-13sr3150}. In a recent strain experiments it has been reported that with 0.7\%  tensile strain the surface-dominated conduction can be observed up to a temperature of 240\,K, persisting even after the strain has been removed.
Theoretically, the surface quasiparticle interference has been calculated\,\cite{baruselli-14prb205105}. Also, it has been shown that the cubic topological Kondo insulator models possess distinct topological crystalline insulating phases\,\cite{baruselli-15prl156404}.

In terms of topological toy models one can understand topological Kondo insulators in a  simplified picture: 
while the $d$ electrons form a conduction band, the $f$ electrons are typically localized and can be described as Ising or Heisenberg spins (\ie the $f$ band is (almost) flat). The Kondo hybridization which is present due to interactions features in certain crystals a symmetry such that it mimics the form of a gapless Dirac-theory. The aforementioned normal-state dispersions act as a mass-term for the gapless Dirac-theory associated with the Kondo hybridization. The result is a QSH-type insulator in 2D\,\cite{werner-13prb035113} or a strong/weak TI in 3D\,\cite{dzero-16arcmp249}. 
Similar findings were reported using the Gutzwiller wave function technique thus showing that also non-local correlation effects lead to topological Kondo insulating phases on the square lattice\,\cite{wysokinski-16prb121102}.
Also 1D versions of a topological Kondo insulator have been discussed\,\cite{alexandrov-14prb115147,zhong-17epjb147,pillay-18prb205133}.
Most works concentrate, however, on 3D and the SmB$_6$ compound. As previously mentioned, whether or not SmB$_6$ is indeed a topological Kondo insulator has been debated in the past years\,\cite{dzero-16arcmp249}.

Let us emphasize again that all the interaction induced topological phases discussed in Sec.\,\ref{sec:IITI} possess a (non-interacting) band-structure analogue. That is, one could adiabatically transform the interaction-induced phase into a topological band structure without closing of the bulk gap. That would even apply to the strongly interacting topological Kondo insulators.


\subsection{Exotic Strongly Correlated Topological Phases}
\label{sec:stronglycorr-TI}

In this section, we consider exotic quantum states of matter which are also 
 ``interaction-induced topological phases''. The major difference to the models discussed in the previous section is, however, that these systems do not have a bandstructure analog. Moreover, these states of matter merely exist due to the interplay of  non-trivial band topology and strong electron-electron interactions. Examples discussed in the following are topological Mott insulators (Sec.\,\ref{sec:TMI}), fractional Chern insulators and fractional topological insulators (Sec.\,\ref{sec:fci+fti}); topological spin liquids (not further discussed here) should also be mentioned\,\cite{balents10n199}.

\subsubsection{Topological Mott Insulators}
\label{sec:TMI}

The Topological Mott insulator (TMI) it is the prototype of an {\it interacting topological insulator}.
Originally proposed by Pesin and Balents\,\cite{pesin-10np376} for Pr$_2$Ir$_2$O$_7$, it realizes a  three-dimensional U(1) spin liquid\,\cite{wang-16prx011034}. One might intuitively think about this state of matter in the following way: due to the strong correlations the charge degrees of freedom have been stripped from the original electrons and are frozen in a Mott insulating phase; the spinons (\ie emergent quasiparticles carrying only spin degree of freedom) inherit the non-trivial band topology of the underlying TI bandstructure. The TMI exhibits spin-only Dirac cone surface states. Today, more than seven years after the proposal, there are no experimental signs or clues of the TMI phase; even the theoretical understanding is rather limited\,\cite{pesin-10np376,witczak-krempa-10prb165122,kargarian-11prb165112,cho-12njp115030,wang-16prx011034,rachel-12prb075106,young-08prb125316,scheurer-15sr8386,yoshida-14prl196404}. The relationship between interacting topological insulators and quantum spin liquids as well as their classification was discussed in Ref.\,\cite{guo-18ap244}.

The reader may be warned that the term ``Topological Mott insulator'' has also been introduced and widely used for the interaction-induced Chern or QSH insulators as discussed in Sec.\,\ref{sec:raghu}. But these systems have nothing in common with the TMI phase of Pesin and Balents: the former are adiabatically connected to non-interacting band insulators while the latter is a strongly correlated state of matter not having any bandstructure analog.

Here we will first briefly explain why the TMI phase does {\it not} exist in the KMH model or any other 2D $\mathbb{Z}_2$ TI. Then we will 
discuss the pyrochlore model of Pesin and Balents and its extensions. Eventually we will introduce a 3D generalization of the Hofstadter-Hubbard model\,\cite{scheurer-15sr8386} discussed in Sec.\,\ref{sec:TRI-HH} which provides a scenario to realize the TMI phase using ultracold quantum gases in an optical lattice setup\,\cite{scheurer-15sr8386}. 

Within slave-rotor theory\,\cite{florens-02prb165111,florens-04prb035114,zhao-07prb195101}, one introduces phase variables $\theta_j$ conjugate to the total number of charges or fermions on lattice site $j$ and fermionic auxiliary (``spinon'') operators $f_{j\sigma}$ by rewriting the original fermion operators:
\begin{equation}
c^\pd_{j\sigma} = e^{i \theta_j} f^\pd_{j\sigma}\ , \qquad c^\dag_{j\sigma} = e^{-i \theta_j} f^\dag_{j\sigma}\ .
\end{equation}
The motivation is that the electrons $c_{j\sigma}$  are represented by a collective phase degree of freedom $\theta$ (conjugate to charge) and spinons $f_{j\sigma}$ (describing spin degree of freedom). Introducing an additional quantity, the angular momentum $L_j \equiv -i \pa_{\theta_j}$ associated with a quantum O(2) rotor $\theta$, simplifies the original Hubbard term: the quartic fermionic Hubbard interaction term reduces to a bilinear $L_j^2$ in the angular momentum operators, a simple kinetic term. This simplification comes for the price that (i) bandstructure terms of the bare electrons are now quartic in the rotor and spinon variables and (ii) the Hilbert space has been enlarged, unphysical states are present. The first point can be resolved by performing all kinds of mean-field decouplings\,\cite{florens-02prb165111,florens-04prb035114}; the second point is more fundamental as it requires the introduction of a constraint,
\begin{equation}
\sum_{\sigma} f_{i\sigma}^\dag f_{i\sigma}^\pd + L_i = 1\ .
\end{equation}
Imposing this constraint guarantees to involve only physical states.

As a side remark, let us emphasize that already on this operative level one can claim that any $\mathbb{Z}_2$ topological band insulator will be stable towards weak electron-electron interactions at least up to $U \sim t$. This can be best seen by considering the Hubbard term at half filling,
\begin{equation}
U \sum_i n_{i\up}n_{i\dw} = \frac{U}{2}\sum_i \left( \sum_\sigma n^{(f)}_{i\sigma} - 1 \right)^2 = \frac{U}{2} \sum_i L_i^2\ .
\end{equation}
Apparently Hubbard-$U$ only affects the rotor sector, and by readily approximating the hopping terms in a mean-field fashion the resulting model is an XY model with its well-known phase transition from a Bose-condensed or superfluid phase $(U<t)$ to a Mott insulating phase $(U>t)$. Condensation of the rotor means that a uniform ansatz $\theta_i - \theta_j=0$ is favored. Since $\exp{(\pm i[\theta_i - \theta_j])}=1$, the auxiliary fermions are proportional to the original electrons. With other words, as long as $U<t$ the bandstructures of spinons and electrons are (up to some renormalization factors) identical -- the TI phase persists\,\cite{rachel-12prb075106}. 

Working out the slave-rotor theory quantitatively for the KMH model suggests that at $U_c(\lambda)$ a phase transition into the TMI phase occurs,
\begin{equation}
U_c(\lambda) = \left[ \frac{1}{2N_{\Lambda}} \sum_{\bs{k}'} \frac{1}{\sqrt{\xi_{\bs{k}} - {\rm min}(\xi_{\bs{k}})}} \right]^{-2}\ .
\end{equation}
Further analysis including the effect of gauge fluctuations reveals that the presence of a dynamical gauge field is inevitable. Following Polyakov the fractionalized TMI phase is not stable against gauge fluctuations (``instanton proliferation'')\,\cite{polyakov75pl82} causing an XY instability, \ie an easy-plane antiferromagnetically ordered phase\,\cite{rachel-12prb075106,hermele-08prb224413}.
Since these arguments are rather generic for 2D TR invariant systems, we conclude that the TMI phase cannot exist in 2D TIs.

\begin{figure}[t!]
\centering
\hspace{-7pt}
\includegraphics[scale=0.64]{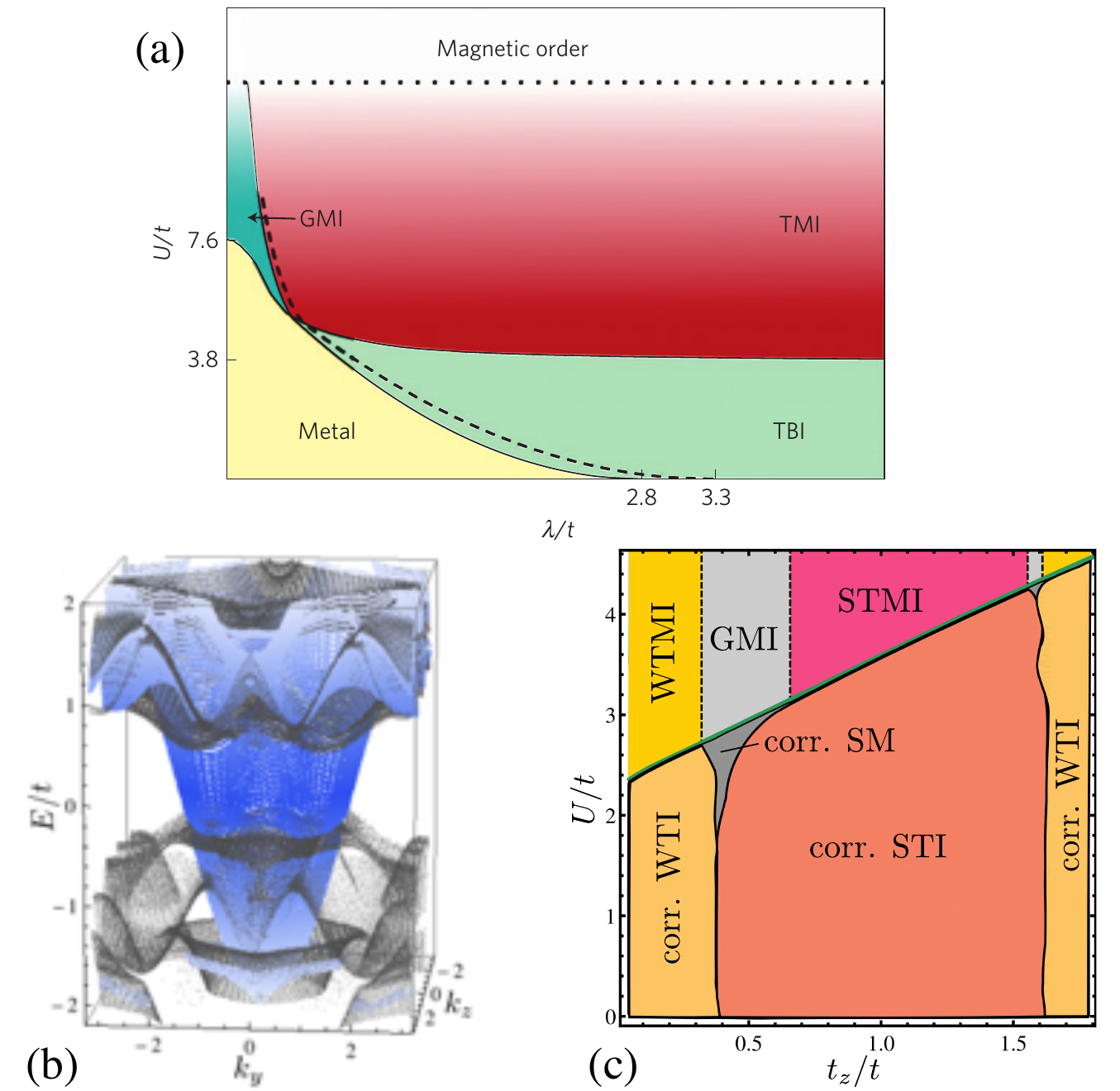}
\caption{\small  
(a) Interacting phase diagram of the topological pyrochlore model\,\cite{pesin-10np376} where the topological Mott insulator was originally proposed.
Reprinted with permission from \cite{pesin-10np376}. Copyright (2010) by Nature Springer (Nature Physics).
(b) Surface state of the STI phase within the TR invariant 3D Hofstadter-Hubbard model shown in the surface BZ with $\bs{k}_{\rm surf} = (k_y, k_z)$.
(c) Interacting phase diagram of the TR invariant 3D Hofstadter-Hubbard $U$ vs.\ $t_z$ including various topological Mott phases as explained in the text.
Panels (b), (c) are taken from\,\cite{scheurer-15sr8386}.}
\label{fig:3DHH}
\end{figure}

The original TMI proposal of Pesin and Balents is formulated as a multi-band Hubbard model on the pyrochlore lattice which should be suitable for certain Ir-based pyrochlore oxides A$_2$Ir$_2$O$_7$\,\cite{pesin-10np376}. The model is sufficiently simplified and contains as free parameters only the spin-orbit coupling $\lambda$ and local Coulomb interactions $U$; the phase diagram is shown in Fig.\,\ref{fig:3DHH}\,(a). The non-interacting system features a metallic regime for weak and a strong TI phase for strong spin-orbit coupling. While very strong electron correlations cause magnetic ordering, here the intermediate-$U$ regime is of interest. Within slave-rotor mean-field theory, a stable TMI phase can be found in addition to a gapless Mott insulating phase. The TMI phase is an exo\-tic Mott insulator with spin-charge separation. In contrast to conventional U(1) spin liquids, here the spinons (the fractionalized excitations of the spin liquid or TMI state) exhibit a topologically non-trivial behavior; in particular, in the TMI phase the system features a spinon surface state\,\cite{pesin-10np376}. When lattice distortions are incorporated, 
also weak TMI phases were suggested\,\cite{kargarian-11prb165112}.

The previously discussed TR invariant Hofstadter-Hubbard model (see Sec.\,\ref{sec:TRI-HH}) can be extended to three spatial dimensions by introducing hopping in the $z$ direction such that it picks up different Peierls phases for $\up$- and $\dw$-spins\,\cite{scheurer-15sr8386}. 
Starting at $t_z=0$ with the 2D system, we can investigate the dimensional crossover to 3D. The surface state of the STI phase at $t_z=t$ is shown in Fig.\,\ref{fig:3DHH}\,(b) proving the topological character of the STI phase.
Similar to the behavior of edge states of the Hofstadter problem in 2D (see for instance Fig.\,\ref{fig:hofstadter}\,(b)), also in 3D the diabolic point (\ie the Dirac nodal point) is buried in the bulk.

The corresponding interacting phase diagram is shown in Fig.\,\ref{fig:3DHH}\,(c). It contains correlated WTI and STI phases. At sufficiently large interaction strength $U_c(t_z)$ in the range $2.25 < U_c(t_z) < 4.4$, one finds the transition into the fractionalized phase. The semimetallic phase boundary between WTI and STI phases becomes a gapless Mott insulator (GMI) with a semimetallic spinon spectrum\,\cite{scheurer-15sr8386} being very similar to nodal spin liquid states. The correlated WTI phases become weak TMI (WTMI)\,\cite{kargarian-11prb165112} phases for $U>U_c$. Most importantly, the rather large correlated STI phase undergoes a phase transition in the strong TMI (STMI) regime for $U>U_c$, see Fig.\,\ref{fig:3DHH}\,(c).

This analysis demonstrates that the TMI phase of Pesin and Balents is {\it not} limited to the pyrochlore model\,\cite{pesin-10np376} and can be found on other lattices and for other models.
Moreover, the cold-atom setup proposed in\,\cite{scheurer-15sr8386} opens the possibility to investigate TMI phases in a controlled way: as most bandstructure parameter as well as Hubbard-$U$ can be tuned with high precision it promises to be a successful route to the observation of topological Mott insulators. 
Another interesting system involving electron fractionalization in topological band structures  was proposed to exist in heavy transition metal oxides\,\cite{maciejko-14prl016404}.

%
%
\subsubsection{Fractional Chern and Topological Insulators}
\label{sec:fci+fti}

The fractional quantum Hall effect is realized in a two-dimensional electron gas - nevertheless it represents a strongly correlated electron
system. Due to the heavily quenched kinetic energy of perfectly flat continuum Landau levels, the ratio of interactions and kinetic energy is effectively large. For a surprisingly long time there was no proposal to stabilize the fractional quantum Hall effect in a lattice system. When it was realized that (topological) bandstructures could be tuned to become sufficiently flat without the need of unphysically large hopping ranges\,\cite{tang-11prl236802,sun-11prl236803,neupert-11prl236804}, fractional Chern insulators (FCIs) became popular. Here we will briefly discuss the main ideas behind FCIs and  also sketch the TR invariant extension, fractional topo\-logical insulators (FTIs). The interested reader should consult the review articles about FCIs and FTIS\,\cite{bergholtz-13ijmpb1330017,neupert15ps014005}. Also a brief but more general review about Fractional Topological Insulators is available\,\cite{maciejko-15np385}.

The main reason why FCIs were not proposed earlier might be due to the insight that most fractional quantum Hall states require longer-ranged Coulomb interactions to be stabilized\,\cite{haldane83prl605}. Including longer-ranged Coulomb interactions on a lattice typically leads to a ground state with conventional order such as charge order ({\it cf.} discussion in Sec.\,\ref{sec:raghu}). In fact, FCI ground states do compete with charge ordered states and fine-tuning is indeed required.

The first ingredient for FCIs is a sufficiently flat band with a non-zero Chern number or finite Hall conductivity. For the continuum Landau levels the non-zero Chern number stems from the external magnetic field. The lattice Chern insulators are more elegant as they do not require any magnetic field - the bandstructure itself leads to the non-trivial topology. Including longer-ranged hoppings to the minimal Chern insulator models allows to tune the ``band flatness''\,\cite{tang-11prl236802,sun-11prl236803,neupert-11prl236804}. The second ingredient is an appropriate band filling in analogy to the fractional filling of Landau levels. For instance, in order to stabilize the lattice version of the $\nu=1/3$ Laughlin state the band filling also needs to be adjusted to 1/3. The third and final ingredient are electron--electron interactions\,\cite{regnault-11prx021014,sheng-11nc389}. In case of spinless particles, first- and second-neighbor Coulomb repulsion are already sufficient to stabilize several fractional quantum Hall states\,\cite{regnault-11prx021014}. For hardcore bosons, the $\nu=1/2$ Laughlin state forms even without the need of any non-local interactions. Flat bands on arbitrary lattices\,\cite{wu-12prb075116,hu-11prb155116} and with higher or arbitrary Chern number have been discussed\,\cite{wang-12prb201101,liu-12prl186805, trescher-12prb241111,yang-12prb241112}. FCI ground states with  Abelian and non-Abelian anyon excitations have been identified\,\cite{sterdyniak-13prb205137} including Moore-Read type states\,\cite{regnault-11prx021014}. Also a hierarchy of FCIs states has been shown to exist\,\cite{laeuchli-13prl126802} as well as multi-component fractional states\,\cite{zeng-17prb125134}. Pseudo potential descriptions in analogy to the fractional quantum Hall effect have been dervied\,\cite{wu-13prl106802,lee-13prb035101}.
Fractional many-body states have also been proposed and investigated for the Hofstadter problem (see Sec.\,\ref{sec:QHE}). In contrast to FCIs, the fractional phases in Hofstadter bands realize a lattice version of the fractional quantum Hall effect  {\it in the presence} of an external magnetic field. For both fermions and bosons fractional ground states have been shown to exist\,\cite{moeller-15prl126401}. 

All these works about FCIs have in common that they consider non-interacting bandstructures which are already topologically non-trivial. In Sec.\,\ref{sec:IITI} we have seen that interactions can also ``make'' topologically trivial bandstructures non-trivial. In a recent work, it was shown that it is even possible to directly induce an FCI phase in topologically trivial bands\,\cite{kourtis-18prb085108}: in solid-liquid composites, the fractional phase can be induced by the formation of charge order, based on the idea of symmetry-breaking topological order\,\cite{kourtis-14prl216404}.

In the following, let us consider the spinful extension of FCIs. Bernevig and Zhang proposed such a state of matter already in 2006\,\cite{bernevig-06prl106802}. Levin and Stern pointed out that it is not sufficient to ``glue'' together arbitrary fractional Chern insulator ground states; they derived a criteria for spin-conserving systems which involves the spin-Hall conductance $\sigma_{\rm sH}$ and the  elementary charge $e^\star$ in units of $e$: if $\sigma_{\rm sH}/e^\star$ is odd, the resulting state is a fractional topological insulator; if it is even, one obtains a trivial insulator instead\,\cite{levin-09prl196803}. Microsocpic models and studies based on exact diagonalization started after FCIs had been established (see above).
The construction is usually straight-forward:
the FCI states originating from a spinless bandstructure with Chern number $C=1$ and $C=-1$, respectively, will be the same but possess opposite chirality. For a topologically non-trivial $\mathbb{Z}_2$ bandstructure (in the spirit of the Kane-Mele construction) it must, hence, be possible to obtain a TR-invariant fractionalized phase where the  $\up$-spins display the chiral FCI state while the $\dw$-spins display the anti-chiral FCI state. Both combined lead to a TR invariant ground state just like the TR invariant TIs: the resulting state is a fractional topological insulator\,\cite{neupert-11prb165107,wu-12prb085129,repellin-14prb245401} given it fulfills the Levin-Stern criteria\,\cite{levin-09prl196803}. Using exact diagonalization, such constructions have explicitly been shown to exist\,\cite{neupert-11prb165107,repellin-14prb245401}.

To date there is no experimental evidence for the realization of a fractional Chern or topological insulator phase. This has inspired physicists to think about schemes which allow for the engineering of such phases: coupled-wire constructions. They can be thought of as arrays of one-dimensional systems (``wires'') which can be treated within Luttinger liquid formalism. By coupling them in a smart way, they can reproduce topological states of matter. Originally pioneered for the integer\,\cite{yakovenko91prb11353,lee-94prb10788} and fractional quantum Hall effects\,\cite{kane-02prl036401,teo-14prb085101} in two dimensions, these ideas have been generalized and used to engineer essentially all types of topological states of matter. Given that the wires are modelled as Luttinger liquids, electron-electron interactions are intrinsically embedded into these systems, thus allowing to analytically study the interacting versions of these topological systems including fractional topological insulators\,\cite{neupert-14prb205101,meng-14prb235425,sagi-15prb195137,iadecola-16prb195136,meng-15prb115152,meng-15prb241106,klinovaja-15prb085426,klinovaja-14prb115426,volpez-17prb085422}. In principle, it seems possible to experimentally build such wire networks thus underlining the importance of coupled-wire constructions.

\subsection{Interacting Surface States of Topological Insulators}
\label{sec:TIsurface}

So far we always considered electron--electron interactions affecting the total system, \ie both bulk and edges or surfaces, respectively. In this subsection, we restrict our focus onto the surface of a strong TI in 3D, ignore the bulk since it is gapped, and ask how interactions might influence the surface states.

The surface states of 3D STIs are very special with respect to different aspects:
\begin{enumerate}
\item
They are described by a two-dimensional Dirac theory and there must be an odd number of such Dirac cones; in some cases there is only a single Dirac cone per surface.
It is interesting to ask  whether or not the surface of an STI violates the {\it fermion doubling theorem} (stating that Dirac cones on the lattice always need to come in pairs)\,\cite{nielsen-81npb173}. A possible solution to this riddle is to recall that the slab geometry features on both top  {\it and} bottom surface a Dirac cone\,\cite{hasan-10rmp3045}. In total, surface Dirac cones still come in pairs.
\item
Fu and Kane pointed out that breaking of the protecting U(1)$_{\rm charge}$ symmetry, for instance by a superconducting proximity effect, leads on the 3D STI surface to an unconventional gapped $s$-wave superconductor with Majorana zero modes bound to the vortex cores\,\cite{fu-08prl096407}. 
\item
When TR symmetry is broken, such as by magnetic coating, the Dirac cone surface state acquires a gap,
resulting in a $\nu=1/2$ quantum Hall effect\,\cite{qi-08prb195424,sitte-13prb205107} without fractionalized excitations, as expected from a single massive Dirac theory.

\end{enumerate}

\begin{figure}[t!]
\centering
\includegraphics[scale=0.65]{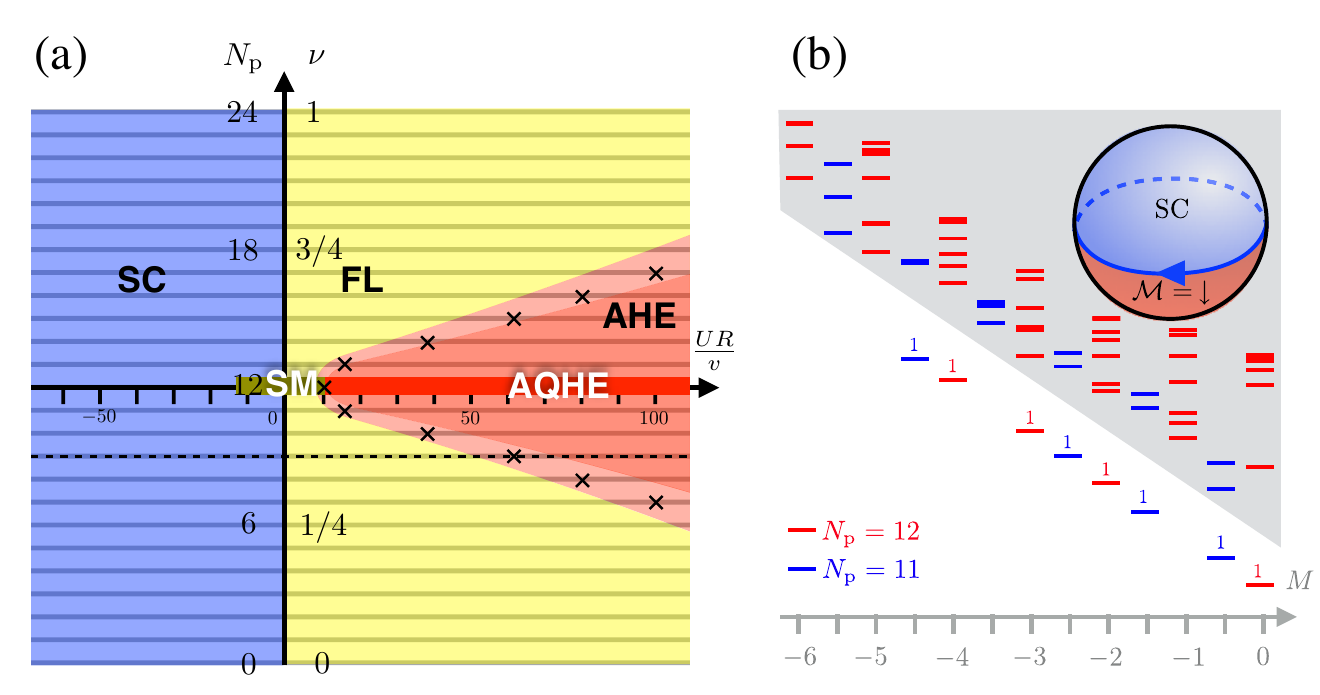}
\caption{\small (a) Phase diagram of the STI surface in the presence of contact interactions with amplitude $U$ vs.\ filling fraction $\nu$. (b) Level distribution of the finite-size spectra for the chiral Majorana mode. Figures are taken from\,\cite{neupert-15prl017001}.}
\label{fig:STIsurface}
\end{figure}

Recently, it was proposed that there is yet a third possibility to gap the single Dirac cone on the 3D STI surface, transcending mean-field scenarios but involving topological order (see Sec.\,\ref{sec:SPTvsTO}). Motivated by the study of bosonic topological insulators in 3D\,\cite{melitski-13prb035131,vishwanath-13prx011016,wang-13prb235122,burnell-14prb245122}, it was shown that such gapped surface states without breaking any symmetries are indeed possible. These works directly stimulated the analogous investigation of the fermionic 3D TIs\,\cite{melitski-15prb125111,wang-13prb115137,bonderson-13jsm09016} and showed the topological order emerging on the surface of a strong 3D TI to be of Non-Abelian type. In Ref.\,\cite{fidkowski-13prx041016} a similar idea was pushed forward for the surface of 3D topological superconductors. A non-perturbative lattice construction for SPT and for symmetry-enriched  topological phases and their gapped surfaces is provided in Ref.\,\cite{wang-17arXiv1705.06728}.

These findings are most remarkable in the following sense: as discussed previously, a 3D TI is an SPT phase which has a gapped bulk but metallic surface states which are protected as long as the protecting symmetries are intact. Apparently the presence of electron-electron interactions circumvents these strict requirements, leading to a gapped surface which fully respects the protecting symmetries; the stronlgy correlated groundstate is accompanied by topological order. 
Most of these proposals are based on field-theoretical considerations; microscopic studies of interaction effects on the STI surface are limited to an exact diagonlization study\,\cite{neupert-15prl017001}. 
Therein the STI surface has been studied in a microscopic setup by employing a spherical geometry. The Dirac cone surface states are described by a two-dimensional Dirac Hamiltonian, which is given in the limit of long wavelengths as
\begin{equation}\label{Dirac-surface}
H = v \hat{\bs{n}} \left( -i \nabla \times \bs{\sigma}\right)
\end{equation}
where $\bs{\sigma}$ is the spin operator, $\hat{\bs{n}}$ is the surface normal, and $v$ the Dirac velocity. For a spherical TI with radius $R$, \eqref{Dirac-surface} becomes\,\cite{imura-12prb235119} $H_0 = v/R (\sigma_x \Lambda_\theta + \sigma_y \Lambda_\phi )$ where $\Lambda$ is the dynamical electron angular momentum when a magnetic monopole with strength $2\pi\sigma_z$ is present.
Similar to Haldane's elegant idea to map the QHE for spinless fermions onto a sphere with a monopole in its center, here we end up with a setup where spinful electrons move on a sphere with a monopole in its center which possesses opposite sign for opposite spins to ensure TR invariance. Eigenstates in this setup are readily labeled, because Landau levels on the sphere are spanned by two mutually commuting SU(2) algebras. One is associated with the guiding center momentum $\bs{L}$ and one with the cyclotron momentum $\bs{S}$. Another advantage is the absence of an edge; keep in mind that a surface state cannot have an edge in a normal sense. The spherical geometry provides a natural way of avoiding an edge.

A straightforward exact diagonalization study with Hubbard-like density-density interactions leads at half filling to a ferromagnetically polarized half-QHE (repulsive interactions) or to an $s$-wave superconductor (attractive interactions), see Fig.\,\ref{fig:STIsurface}\,(a). These calculations show that local density-density interactions are not sufficient to stabilize the predicted topologically ordered phases\,\cite{melitski-15prb125111,wang-13prb115137,bonderson-13jsm09016}, longer-ranged and potentially more-complex interaction terms need to be added. This remains an open problem suited for future studies because {\it any} interaction term can be implemented  within the discussed framework. 

An $s$-wave superconducting termination of a 3D STI is a topological superconductor in the following sense: (i) it supports Majorana zero modes  in its vortex cores and (ii) a chiral Majorana mode at the boundary with a ferromagnetic region of the surface. Using exact diagonalization, this can be conveniently shown by considering the following modified setups: (i) a superconductor--ferromagnet domain wall and (ii) the region between two ferromagnetic domains. The corresponding level counting for the chiral Majorana mode is shown in Fig.\,\ref{fig:STIsurface}\,(b).

An interesting aspect of interacting surface states of STIs is the emergence of supersymmetry\,\cite{grover-14s280,ponte-14njp013044}. It has been predicted to occur at the quantum critical point at the end of the semimetallic (SM) line (with one Dirac fermion) where the superconducting (SC) phase evolves for $U<0$ [see Fig.\,\ref{fig:STIsurface}\,(a)]. Indeed a recent Monto Carlo study reported the observation of supersymmetric behavior\,\cite{li-17arXiv1711.04772}. An extension of this line of research led to the prediction of emergent supersymmetric electrodynamics on the surface of the topological Kondo insulator SmB$_6$ (see Sec.\,\ref{sec:TKI}) at the critical point between the semimetal (with three Dirac fermions) and the nematic pair density wave phase\,\cite{jian-17prl166802}.

%
%
\section{Relation to Magnetism in Iridates}
\label{sec:iridates}

The honeycomb iridates A$_2$IrO$_3$ (A=Na, Li) together with $\alpha$-RuCl$_3$ are today known to play a major role in the active field of {\it Kitaev materials}\,\cite{trebst17arXiv1701.07056,hermanns-18arcmp17,winter-17jpcm493002}. Here we are not reviewing the experimental and theoretical developments of the past years but focus instead on the relationship of Kitaev spin exchange and topological insulating bandstructures, following the original papers which initiated the excitement about Na$_2$IrO$_3$\,\cite{shitade-09prl256403,jackeli-09prl017205}.

\begin{figure}[b!]
\centering
\includegraphics[scale=0.5]{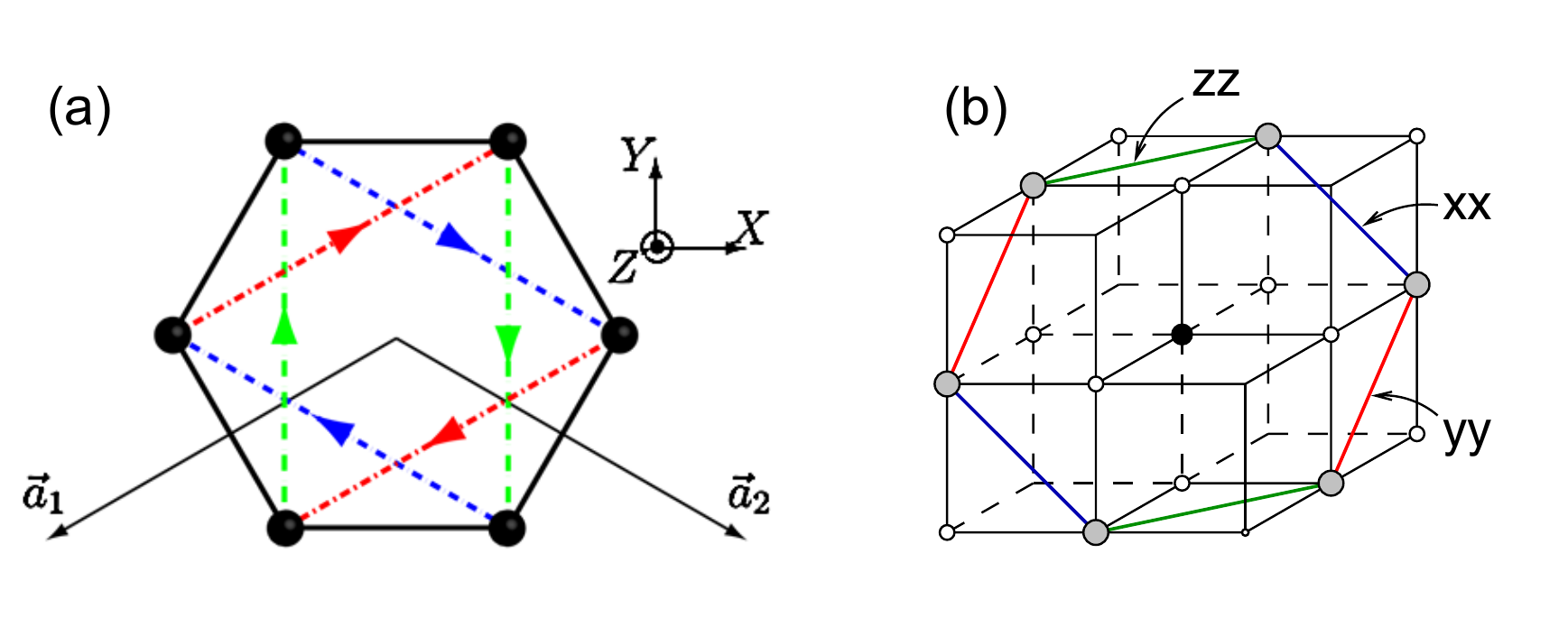}
\caption{\small
(a) Topological insulator picture. The transfer integrals on the honeycomb lattice: a black solid line shows $-t$, while blue short-dashed, red dash-dotted, and green long-dashed arrows indicate $it'\sigma_x$, $it'\sigma_y$, $it'\sigma_z$, respectively.
Reprinted with permission from \cite{shitade-09prl256403}. Copyright (2009) by the American Physical Society.
(b) Frustrated magnet picture. Hexagonal unit cell of A$_2$BO$_3$-type layered compound, in which magnetic ions (B-sites) form a honeycomb lattice. (Black dot: nonmagnetic A-site). On an $xx$-bond, the interaction is $S^x_i S^x_j$, etc. For this structure, the spin Hamiltonian contains Kitaev spin exchange.
Reprinted with permission from \cite{jackeli-09prl017205}. Copyright (2009) by the American Physical Society.
}
\label{fig:na2iro3}
\end{figure}

The Kane-Mele proposal\,\cite{kane-05prl226801} stimulated the search for other honeycomb lattice compounds aiming to find a successful material realization of the Kane-Mele model. As pointed out in Sec.\,\ref{sec:SIH}, the layered honeycomb lattice iridate Na$_2$IrO$_3$ was amongst the first proposals along this line. {\it Ab initio} calculation revealed\,\cite{shitade-09prl256403} that the relevant nearest and next-nearest neighbor hoppings integrals are similar to the Kane-Mele scenario, see Fig.\,\ref{fig:na2iro3}\,(a).
Shitade \ea\  claimed that--if Coulomb interactions are not too strong--the groundstate should be a correlated QSH insulator.	
About the same time, a competing proposal by Jackeli and Khaliullin was put forward: they claimed that 
Na$_2$IrO$_3$ is deep in a Mott phase such that charge fluctuations are frozen out and the system is well described by a spin model\,\cite{jackeli-09prl017205,chaloupka-10prl027204}. Moreover, they proposed that due to large SOC and 90 degree bonding of the edge-sharing oxygen tetrahedra, which surround the Ir atoms, the interactions are dominated by Kitaev spin exchange\,\cite{Kitaev06ap2} which is known to realize a spin liquid ground state. In Fig.\,\ref{fig:na2iro3}\,(b) the hexagonal unit cell in Na$_2$IrO$_3$ is shown with the nearest-neighbor Kitaev spin exchange.
Together with additional generic Heisenberg interactions, the resulting 
spin Hamiltonian is referred to as Kitaev-Heisenberg (or Heisenberg-Kitaev) model in the literature, given by the Hamiltonian
\begin{equation}\label{kh-ham}
\mathcal{H}_{\rm KH}(\alpha) = \alpha \sum_{\langle ij \rangle} \bs{S}_i \bs{S}_j - 2\alpha \sum_{\langle ij \rangle_\gamma} S_i^\gamma S_j^\gamma
\end{equation}
where $0\leq \alpha \leq 1$ and $\gamma=x,y,z$ depending on the direction of the bond $\langle ij \rangle$. 
The second term in \eqref{kh-ham} represents the previously mentioned compass or Kitaev interactions.
The phase diagram of the KH model hosts for dominating Heisenberg exchange antiferromagnetic Neel order ($0\leq\alpha\leq 0.4$) and for dominating Kitaev exchange the spin liquid phase ($0.8\leq \alpha \leq 1$). In between another collinear antiferromagnet, dubbed {\it stripy} antiferromagnet, is realized. Both the paper by Shitade \ea~and the one by Jackeli and Khaliullin have been very influential.

Early experiments showed that magnetic long-range order occurs below $T_N=15$\,K in Na$_2$IrO$_3$\,\cite{singh-10prb064412,singh-12prl127203}. 
As if the competing proposals of Shitade \ea~and Jackeli and Khaliullin are not controversial enough, subsequent inelastic neutron scattering experiments revealed that the magnetic order is of {\it zigzag} type\,\cite{choi-12prl127204} -- not present in the Kitaev-Heisenberg (KH) model. In the meantime, it is also known that the sister compound Li$_2$IrO$_3$ is likewise magnetically ordered with $T_N=15$\,K but the order is of incommensurate spiral type\,\cite{williams-16prb195158}. Apparently both materials are not correlated TIs -- but their magnetic groundstate is not contained in the original KH model\,\eqref{kh-ham}.
In another work, Na$_2$IrO$_3$ was suggested to realize a 3D strong TI phase\,\cite{kim-12prl106401}; most research is, however, motivated by the idea that these compounds feature Kitaev spin exchange.
As emphasized before, here we are not reviewing the manifold experimental and theoretical efforts of the past years; an overview of the current status can be found here\,\cite{trebst17arXiv1701.07056,hermanns-18arcmp17,winter-17jpcm493002}.

In the following, we will discuss the connection between the TI physics of Shitade \ea~and the KH model of Jackeli and Khaliullin. We note that the strong coupling expansion of the topological Hubbard model (the SIH model in Sec.\,\ref{sec:SIH}) results in the spin Hamiltonian Eq.\,\eqref{SI-SOC-spin}\,\cite{reuther-12prb155127}. The topological Hubbard model generates, hence, in the strong-coupling limit Kitaev spin exchange on next-nearest neighbor bonds instead on nearest neighbor bonds as suggested by Jackeli and Khaliullin. Partially motivated by this discrepancy, models were proposed containing both nearest and next-nearest neighbor Kitaev exchange (dubbed $K_1$--$K_2$ models) and rich groundstate phase diagrams derived\,\cite{reuther-14prb100405,rousochatzakis-15prx041035}.

The considerations in the previous paragraph might raise the question which bandstructure could result in nearest-neighbor Kitaev spin exchange. Defining the spin-dependent, imaginary spin-orbit term of the SIH model on nearest-neighbor bonds in addition to the standard real nearest neighbor hopping, indeed results in Kitaev and Heisenberg terms. The strong coupling expansion produces, however, another  term which results from processes where an electron hops from site $i$ to $j$ with spin-orbit amplitude $\tilde\lambda$ and hops back from $j$ to $i$ with  normal hopping amplitude $t$\,\cite{yu-13prb041107,laubach-17prb121110}. The corresponding spin exchange in the strong coupling is the symmetric of-diagonal $\Gamma$ exchange.
In order to avoid the cross term, one could replace the spin-orbit term by a spin-dependent {\it real} hopping term -- then the mixed terms cancel and the resulting spin Hamiltonian is the KH model \eqref{kh-ham}. 
Such a one-band Hubbard model was indeed  discussed and shown to have several interesting properties\,\cite{hassan-13prl037201,liang-14prb075119}. The underlying bandstructure explicitly breaks, however, TR symmetry. That immediately implies that the weak-coupling regime cannot contain any  TR invariant $\mathbb{Z}_2$ TI phase as proposed by Shitade \ea

The simplest model which captures all these aspects consists of three spinful orbitals, which one might identify with the $t_{2g}$ manifold consisting of $d_{yz}$, $d_{xz}$, and $d_{xy}$ orbitals\,\cite{rau-14prl077204,laubach-17prb121110}. The Hamiltonian is given by\footnote{Note that for an accurate first principle treatment additional terms will be present; here we restrict ourselves to discuss a minimal model.}
\begin{equation}\label{ham-3band}
H = H_{\rm kin} + H_{\rm SOC} + H_{UH}\ ,
\end{equation}
consisting of kinetic, spin-orbit, and interaction part, respectively. $H_{\rm kin}$ reads
\begin{equation}
H_{\rm kin} = \sum_{\langle ij \rangle_\gamma} \sum_{n,n',\alpha} d_{in\alpha}^\dag T^\gamma_{nn'} d^\pd_{jn'\alpha}  + {\rm H.c.}
\end{equation}
where $d_{in\alpha}$ denotes a fermionic annihilation operator on site $i$ with orbital $n\in\{yz, xz, xy\}$ and spin $\alpha\in\{\up,\dw\}$. For a $z$ bond the hopping matrix $T_{nn'}^z$ is gien by
\begin{equation}
T_{nn'}^z = \left(\begin{array}{ccc} t_1 & t_2 & 0 \\ t_2 & t_1 & 0 \\ 0 & 0 & t_1' \end{array}\right),
\end{equation}
and the other matrices $T_{nn'}^y$ and $T_{nn'}^z$ follow by cyclic permutations of the rows and columns.
 The spin-orbit term $H_{\rm SOC}$ is given by
\begin{equation}\label{3band:soc}
H_{\rm SOC} = \frac{\lambda}{2} \sum_i \sum_{n,n',\alpha,\alpha'} d_{in\alpha}^\dag \, \bs{l}_{nn'} \cdot \bs{s}_{\alpha\alpha'}\, d_{in'\alpha'}^\pd\ , 
\end{equation}
where  $\bs{l}$ is an angular momentum operator with $\bs{l}^2 = l(l+1)=2$ and $\bs{\sigma}$ is the vector of Pauli matrices. For positive $\lambda$, \eqref{3band:soc} generates a low-energy $J=1/2$ doublet, representing the subspace for the KH model.

The simplest SU(2) symmetric extension of the Hubbard repulsion, which generates anisotropic spin exchange in the $J=1/2$ subspace, has the form
\begin{equation}
H_{\rm int} = \frac{U -3J_H}{2} \sum_i (N_i - 1)^2 - 2J_H \sum_i \bs{S}_i^2
\end{equation}
with Hund's coupling $J_H$, with $N_i=\sum_{n,\alpha} d_{in\alpha}^\dag d_{in\alpha}^\pd$ and $S_i^\mu = \frac{1}{2}\sum_{n\alpha\alpha'} d_{in\alpha}^\dag s_{\alpha\alpha'}^\mu d_{in\alpha'}^\pd$.
Projecting the result of the standard strong coupling expansion ($U \gg t_{1/2}$ and $\lambda \gg t_{1/2}^2/U$) on the low-energy $J=1/2$ doublet leads to the pure KH model \eqref{kh-ham} for $t_1=t_1'$\,\cite{laubach-17prb121110}. Due to the orbital structure of Na$_2$IrO$_3$, $t_1\not=t_1'$ is more realistic leading to additional symmetric off-diagonal $\Gamma$ spin exchange, $S_i^x S_j^y + S_i^y S_j^x$.

Analysis of the 3-band model at $U=J_H=0$ yields a phase diagram containing several metallic and insulating phases\,\cite{laubach-17prb121110}. 
As a central result, at the relevant filling $1/6$ a large region of the non-interacting phase diagram is in a topological insulator phase. This topological phase is clearly adiabatically connected to the one found in Ref.\,\cite{shitade-09prl256403} based on the lesson learned from Kane and Mele that there are in 2D only two types of TR invariant insulators: trivial insulators and topological insulators.

The interacting phase diagram is parametrized by Hubbard $U$ (we assume that $J_H$ is much smaller than $U$) and some bandstructure parameters.  Thus the phase diagram is framed by the weak-coupling regime of Shitade \ea\,\cite{shitade-09prl256403} and the strong-coupling regime of the KH spin model\,\cite{jackeli-09prl017205,chaloupka-10prl027204}. Attempts in solving the intermediate-$U$ regime agree that a magnetic phase with zigzag-order is present\,\cite{laubach-17prb121110,igarashi-16jpcm026006}. It seems as if there was a natural tendency to form zigzag magnetic order when driving the topological insulator within the Na$_2$IrO$_3$ environement into the Mott phase. 
Whether or not this zigzag phase (being in the vicinity to the weak-coupling TI phase) is the same as the one which has been observed experimentally\,\cite{singh-10prb064412,singh-12prl127203,choi-12prl127204}   remains unclear.

The reader may note that there have been several attempts to explain the experimentally found zigzag phase in Na$_2$IrO$_3$ by extending the original KH spin Hamiltonian \eqref{kh-ham}: longer-ranged Heisenberg exchange\,\cite{singh-12prl127203,kimchi-11prb180407,katukuri-14njp013056}, second neighbor Kitaev exchange\,\cite{reuther-14prb100405,rousochatzakis-15prx041035}, additional Ising terms\,\cite{kimchi-15prb245134}, a negative-sign version of the original KH model\,\cite{chaloupka-13prl097204}, significant $\Gamma$ exchange\,\cite{rau-14prl077204,catuneanu-17arXiv1701.07837,lampen-KelleyarXiv1803.04871} and a combination of $\Gamma$ and longer-ranged isotropic exchange\,\cite{winter-16prb214431}.
{\it Ab initio} based works also suggested a more itinerant explanation for the phenomenology of Na$_2$IrO$_3$ such as the formation of quasimolecular orbitals\,\cite{mazin-12prl197201,foyevtsova-13prb035107}; in particular, the magnitude of Coulomb interactions, spin-orbit coupling and hopping strength was found to be of similar size which would support the above discussed scenario of the three-band Hubbard model\,\cite{laubach-17prb121110} where charge fluctuations are responsible for the magnetic zigzag order. 

As a final remark, we emphasize that a magnetically ordered zigzag phase which is close to the weak-coupling TI phase might display remnants of the TI physics. In Ref.\,\cite{laubach-17prb121110} topologically trivial edge states were observed in the single-particle spectral function. Indeed, recently the observation of metallic surface states were reported in an angle-resolved photoemission spectroscopy (ARPES) measurement\,\cite{alidoust-16prb245132}. In a more recent ARPES and x-ray absorption study the relevance of the surface termination was pointed out as well as the role of charge transfer from Na atoms to Ir-derived states\,\cite{moreschini-17prb161116}.

%
%
\section{Conclusion and Outlook}
\label{sec:conclusion}

In this review various aspects of electron-electron interactions applied to topological insulators are discussed. The consequences are manifold and include many different fields of contemporary condensed matter physics: from quantum Hall effects to quantum magnets and spin liquids, from one-dimensional Luttinger liquids to three-dimensional Hubbard models, from homotopy groups to renormalization group methods, and from SmB$_6$ and its long history to the most recent developments of ultracold quantum gases. 
Each of this topics clearly deserves its own review article; here only their important aspects for interacting topological insulators are considered.

The central theme is the effect of electron-electron interactions in topological bandstructures. TR invariant topological insulators, prototypes of symmetry protected topological phases, are typically stable up to moderate interaction strength. Also the metallic edge states, which are protected against single-particle backscattering are stable towards interactions as long as the bulk phase remains intact. Stronger interactions then often induce phase transitions into more conventional  phases such as antiferromagnetically or charge-ordered states. In certain situations, the competition between strong interactions and non-trivial topology at half filling leads to  exotic states of matter such as the topological Mott insulator phase and fractional Chern insulators.
In addition, it is shown that there are several interesting proposals where topologically non-trivial bandstructures can be driven into a topological insulator phase by virtue of interactions. For the class of the honeycomb iridates A$_2$IrO$_3$ it is shown that the frustrated magnetism of these materials is intertwined with the underlying topological insulating bandstructures.

We wish to conclude with an outlook about possible future directions. Currently a lot of interest has been attracted by topological Weyl semimetals which feature nodal points in the bulk spectrum\,\cite{Volovik03,murakami07njp356,wan-11prb205101,burkov-11prl127205}. Since they possess topologically protected Fermi arc surface states these systems are sometimes regarded as ``gapless topological insulators''. Several of these materials are known to be superconducting\,\cite{qi-16nc11038} or magnetic\,\cite{wang-16arXiv:1603.00479} implying that electron--electron interactions play a substantial role. 
Consequently, the process of systematically understanding the effect of interactions in Weyl semimetals has begun\,\cite{jian-15prl237001,wei-12prl196403,wang-13prb161107,maciejko-14prb035126,witzcak-14prl136402,wang-16prb075115,laubach-16prb241102}.
Even more interesting, in the recently predicted and discovered ``type-II Weyl semimetals''\,\cite{xu-15prl265304,soluyanov-15n495} where the Weyl nodes are overtilted, instabilities are expected to play a more dominant role due to the drastically enhanced density of states at the Fermi level.

Another exciting idea is to search for magnon modes in quantum antiferromagnets which exhibit topological properties. For instance, the presence of topological linear band crossings of magnon modes in antiferromagnets provide the analog of Weyl fermions in electron systems (thus dubbed ``Weyl magnons'')\,\cite{li1602.04288}. In another recent example, magnon bands with finite Chern number have been predicted\,\cite{laurell-1609.03612}. One might wonder to what extent  other topological phases exist in the excitation spectra of quantum magnets.

\section{Acknowledgments}
I am particularly grateful to Karyn Le Hur and Matthias Vojta for countless discussions and for their continued support in the past years.
I would like to thank my friends, colleagues, and co-authors I had the pleasure to work with in the context of interacting topological insulators:
M.\ Buchhold,
D.\ Cocks,
M.\ Ezawa,
L.\ Glazman,
M.\ Greiter,
W.\ Hofstetter,
M.\ Laubach,
W.-M.\ Liu,
T.\ Neupert,
F.\ v.\ Oppen,
P.\ P.\ Orth,
J.\ Reuther,
A.\ Rod,
M.\ Scheurer,
T.\ L.\ Schmidt,
R.\ Thomale,
and W.\ Wu.
For interesting and fruitful discussions
I further want to acknowledge 
D.\ Abanin,
D.\ P.\ Arovas,
F.\ F.\ Assaad,
B.\ A.\ Bernevig,
P.\ Brydon,
R.\ Claessen,
M.\ Daghofer,
L.\ Fritz,
P.\ Gegenwart,
M.\ Hohenadler,
G.\ Jackeli,
J.\ Knolle,
S.\ Kourtis,
T.\ C.\ Lang,
Y.\ Iqbal,
J.\ Maciejko,
R.\ Moessner,
D.\ K.\ Morr,
C.\ Platt,
F.\ Pollmann,
N.\ Regnault,
N.\ Perkins,
E.\ Prodan,
I.\ Rousochatzakis,
C.\ Timm,
S.\ Trebst,
R.\ Valenti,
W.\ Witczak-Krempa,
and P.\ W\"olfle.
Part of this work was supported by the German Research Foundation (DFG) through the priority program ``Topological Insulators'' SPP 1666 and through the Collaborative Research Center SFB 1143. Hospitality of the KITP St.\ Barbara is acknowledged.

\bibliography{interacting_ti_v03}

\end{document}